\documentclass[journal]{IEEEtran}
\usepackage{graphicx}
\usepackage[justification=centering]{caption}
\ifCLASSINFOpdf
\else
\fi
\usepackage{amsmath,amsfonts,amssymb}
\usepackage{bbm}
\usepackage{amsfonts}
\usepackage{color}

\usepackage{algorithm}
\usepackage{algpseudocode}

\usepackage{diagbox}

\usepackage[numbers,sort&compress]{natbib}

\begin{document}
%
\title{Privacy Attack in Federated Learning is Not Easy: An Experimental Study}
%
%
%

\author{Hangyu~Zhu,
        Liyuan~Huang,
        Zhenping~Xie
\thanks{H. Zhu, L. Huang, and Z. Xie are with the School of Artificial Intelligence and Computer Science, Jiangnan University, No. 1800, Lihu Avenue, Wuxi, 214122, P. R. China. E-mail: hangyu.zhu@jiangnan.edu.cn; 6223110002@stu.jiangnan.edu.cn; xiezp@jiangnan.edu.cn.}
\thanks{}
\thanks{Manuscript received April 19, 2005; revised August 26, 2015.}}

%
%

\markboth{Journal of \LaTeX\ Class Files,~Vol.~14, No.~8, August~2015}%
{Shell \MakeLowercase{\textit{et al.}}: Bare Demo of IEEEtran.cls for IEEE Journals}
%



\maketitle

\begin{abstract}
Federated learning (FL) is an emerging distributed machine learning paradigm proposed for privacy preservation. Unlike traditional centralized learning approaches, FL enables multiple users to collaboratively train a shared global model without disclosing their own data, thereby significantly reducing the potential risk of privacy leakage. However, recent studies have indicated that FL cannot entirely guarantee privacy protection, and attackers may still be able to extract users' private data through the communicated model gradients. Although numerous privacy attack FL algorithms have been developed, most are designed to reconstruct private data from a single step of calculated gradients. It remains uncertain whether these methods are effective in realistic federated environments or if they have other limitations. In this paper, we aim to help researchers better understand and evaluate the effectiveness of privacy attacks on FL. We analyze and discuss recent research papers on this topic and conduct experiments in a real FL environment to compare the performance of various attack methods. Our experimental results reveal that none of the existing state-of-the-art privacy attack algorithms can effectively breach private client data in realistic FL settings, even in the absence of defense strategies. This suggests that privacy attacks in FL are more challenging than initially anticipated.

\end{abstract}


%
\IEEEpeerreviewmaketitle

\section{Introduction}
%
%
%
%
\IEEEPARstart{M}{achine} learning has witnessed significant success in the fields of computer vision \cite{voulodimos2018deep}, natural language processing \cite{chowdhary2020natural} and so on. For parametric machine learning models, especially deep neural networks \cite{lecun2015deep} containing numerous learnable parameters, it is generally agreed that more training data leads to greater potential performance of the model. And if one were able to collect data from multiple distributed devices, the model trained on such a dataset might exhibit superior learning performance \cite{kaissis2020secure}. This form of data 'centralization' has indeed emerged as the predominant training method for decades.


Nonetheless, the aggregation of distributed data onto a single device markedly compromises user privacy. The pioneering research of Fredrikson et al \cite{fredrikson2014privacy} was among the earliest to address the leakage of private information. Richard et al. \cite{mcpherson2016defeating} further demonstrated that the protected data can be vulnerable to the recovery of hidden information. These findings underscore the necessity of exploring alternative learning paradigms. Moreover, recent regulatory developments, such as the enactment of the General Data Protection Regulation (GDPR) \cite{regulation2016regulation} by the European Union, have heightened concerns regarding data security and privacy.

To address the above mentioned privacy issue, federated learning (FL) \cite{mcmahan2017communication} has emerged as an effective method to preserve data privacy and reduce the substantial transfer costs associated with data collection \cite{yang2019federated}. Unlike traditional centralized machine learning approaches, FL retains data locally, with each client collaborating to train a joint global model. For a typical horizontal FL framework, the central server solely receives and aggregates the model parameters or gradients from the clients to derive a global model, thereby benefiting from the distributed learning. Subsequently, the updated global model is transmitted back to the clients, facilitating knowledge sharing among them. This approach allows clients to retain their training data within the device, thus safeguarding user privacy to a certain extent.

However, some research work indicate that FL remains susceptible to adversarial attacks from malicious parties. For instance, model poisoning attacks \cite{fang2020local,bhagoji2019analyzing,Cao_2022_CVPR,fi13030073,8975792,NEURIPS2021_692baebe,pmlr-v151-panda22a,Shejwalkar2021ManipulatingTB} compromise the integrity of the global model by tampering with local datasets or introducing backdoor inputs \cite{wang2020attack, bagdasaryan2020backdoor}. Even worse, the design of existing FL protocols still harbors flaws in data privacy. These vulnerabilities could potentially be exploited by attackers to identify weaknesses or gain unauthorized access to sensitive user data \cite{orekondy2018gradient,nasr2019comprehensive,luo2021feature}. 
Studies have highlighted that models trained using FL inadvertently memorize the privacy-sensitive training dataset and subsequently disclose sensitive information \cite{pustozerova2020information,carlini2021extracting,jin2021cafe,wu2023learning}.Any participating client in fact acts as a potential data provider whose intermediate updates can be observed by the central server. And adversarial attackers might manipulate them as features of the decentralized preparation process, allowing malicious server or client to adjust interactions with minimal constraints.

Specifically, several attack algorithms have been developed to reconstruct dummy images that closely resemble the original private images, including DLG \cite{zhu2019deep}, iDLG \cite{zhao2020idlg}, Inverting Gradients \cite{geiping2020inverting} from the server side and Deep Models under the GAN (DMGAN) \cite{hitaj2017deep} from the client side. However, there is a notable absence of experimental studies that systematically evaluate the strengths and weaknesses of these attack methods. Given their inherent limitations, these approaches may not be fully effective in real-world FL environments. For example, some studies \cite{geiping2020inverting,kariyappa2023cocktail,yin2021see} conduct experiments by applying attacks to a single gradient derived from batch training data, which diverges significantly from the local model updates or differences typically encountered in FL. Moreover, some methods \cite{kariyappa2023cocktail,zhao2023secure} necessitate modifications to the global model's architecture, may resulting in significant FL performance degradation.

Therefore, in this paper, we aim to conduct a comprehensive investigation into the various privacy attacks executed within a real FL environment. We specifically focus on assessing the extent to which these attacks can compromise user privacy and evaluate their effectiveness in revealing sensitive information. Nine representative attack algorithms are evaluated, and the experimental results provide valuable insights for the future development of cybersecurity-related FL research. To this end, To this end, our concise and clear code implementation is publicly available \footnote{https://github.com/hangyuzhu/leakage-attack-in-federated-learning}, enabling researchers to easily experiment with different privacy attack strategies and extend them in their research.

Based on the results of our extensive studies, we have identified several key findings. Most notably, we found that many attack algorithms are capable of reconstructing high-quality dummy images from gradients corresponding to single data points or multiple averaged data. However, in more complex federated learning environments, where averaged gradients from batch data are computed locally and updated multiple times, these algorithms perform poorly. Among the evaluated methods, Robbing the Fed (RTF) demonstrated the best attack performance. Nevertheless, it requires the insertion of an Imprint module (comprising two fully connected layers) before the learning model, which significantly degrades the training performance of federated learning.

Our contributions are summarized as follows:

\begin{enumerate}
    \item We provide a systematic overview of existing privacy attacks within federated learning (FL), discussing their underlying algorithms in detail and analyzing the advantages and disadvantages of each approach.
    \item We conduct extensive experimental studies on 9 representative privacy attack methods in a more realistic FL environment, including DLG \cite{zhu2019deep}, iDLG \cite{zhao2020idlg}, Inverting Gradients \cite{geiping2020inverting}, GGL \cite{li2022auditing}, GRNN \cite{ren2022grnn}, CPA \cite{kariyappa2023cocktail}, DLF \cite{geng2021towards}, and RTF \cite{fowl2022robbing} for server-side attacks, and DMGAN \cite{hitaj2017deep} for the client-side attack. Our experimental results reveal that most of these approaches are ineffective in scenarios where averaged gradients of batch data are locally computed and updated multiple times on each client.
\end{enumerate}

The remainder of this experimental study is organized as follows. Section \ref{Federated learning} provides an overview of FL. In section \ref{attacks in federated learning}, we introduce different types of privacy attack algorithms in FL. Section \ref{experiments} then presents the experimental results of 9 selected attack algorithms, followed by the conclusion in Section \ref{conclusion}.

\section{Overview of Federated Learning}
\label{Federated learning}

Early in 2016, Google introduced the concept of FL \cite{mcmahan2017communication}, a decentralized machine learning paradigm designed to collaboratively train a shared global model using distributed data. This approach ensures that no external parties have access to local private data, thereby significantly protecting user privacy. A typical FL problem aims to optimize the aggregated loss function $\mathcal{L}(\mathbf{W})$ as shown below:
\begin{equation}
    \mathcal{L}(\mathbf{W})=\sum_{k=1}^{K} \frac{n^{k}}{n}\mathcal{L}^{k}(\mathbf{W}) \quad  \mathcal{L}^{k}(\mathbf{W})=\frac{1}{n^{k}} \sum_{i \in |\mathcal{D}^{k}|}\mathcal{L}_{i}(\mathbf{W}) 
\end{equation}
where $\mathbf{W}$ represents the parameters of the shared global model, $K$ is the total number of clients, $n^{k}$ indicates the amount of local data on client $k$, and $\mathcal{D}^{k}_{i}=\left( \mathbf{x}_{i}^{k}, \mathbf{y}_{i}^{k} \right)$ denotes the $i$-th data sample on client $k$. Then, for each communication round $t$, the training procedure of FL primarily consists of the following three steps:




\begin{enumerate}
    \item \textbf{Download:} The central server sends the global model parameters $\mathbf{W}_{t}$ to each client $k$.
    \item \textbf{Local training:} Each client $k$ uses local training data $\mathcal{D}^{k}$ to train received model parameters $\mathbf{W}_{t}$. Subsequently, the client transmits the updated model parameters (or gradients)  $\mathbf{W}^{k}$ back to the server.
    \item \textbf{Aggregation:} The server aggregates the received updates to refine the global model $\mathbf{W}_{t+1}=\sum_{k=1}^{K}\frac{n^{k}}{n}\mathbf{W}^{k}$, which is then distributed to each client in the subsequent communication round $t+1$.
\end{enumerate}

It is important to note that the last two steps can be performed either synchronously or asynchronously, and the global model typically converges gradually over multiple communication rounds. Yang et al. \cite{yang2019federated} redefined this procedure as horizontal federated learning (HFL) and further expanded and categorized the concept into 
HFL, vertical federated learning (VFL), and federated transfer learning (FTL) \cite{9076003,saha2021federated} based on the varying data partitioning characteristics among distributed participants.

\begin{enumerate}
    \item \textbf{Horizontal Federated Learning (HFL)}, also known as sample-based FL, is characterized by having the same data feature space but different sample spaces among clients during joint training. For example, banks in various regions may share the same user characteristics, such as account, user name, age, account balance, etc., but their user groups may differ.  Typical applications of HFL include Google's Gboard \cite{hard2018federated} and WeBank's FedVision \cite{liu2020fedvision}.
    \item \textbf{Vertical Federated Learning (VFL)}, also known as characteristic-based FL, is characterized by having the same sample space but different data feature spaces among clients. Unlike HFL, where each client independently performs training, VFL requires each client to collaboratively make inferences. The need for VFL has emerged in recent years due to the limited amount of data owned by companies and institutions \cite{li2021survey}. Many enterprises have adopted VFL to enhance finance, advertising, and multimodal tasks \cite{kang2022privacy,liang2021Self}. 
    \item \textbf{Federated Transfer Learning (FTL)} addresses a significant limitation of both HFL and VFL: the requirement for consistency in sample or data feature spaces. Unlike HFL and VFL, FTL does not necessitate feature space or user space consistency \cite{saha2021federated}. For instance, if a hospital and a bank from different cities collaborate in training, they may have different samples and data feature spaces. In such scenarios, HFL and VFL are not applicable, but FTL can effectively address model training challenges. FTL aims to develop an effective model for a target domain by leveraging knowledge from other domains. It first trains a model using the source domain's sample and feature space, then adapts this model for the target domain, applying insights gained from the source domain's features to enhance performance in the target domain.
\end{enumerate}

And this paper mainly focuses on exploring of privacy attack in HFL, as HFL is the most commonly encountered use case. For brevity, we use FL to refer to HFL unless specified otherwise in the following manuscript.

\section{Privacy attacks in federated learning}\label{attacks in federated learning}

\begin{figure}[!htbp]
    \centering
    \includegraphics[width=0.46\textwidth]{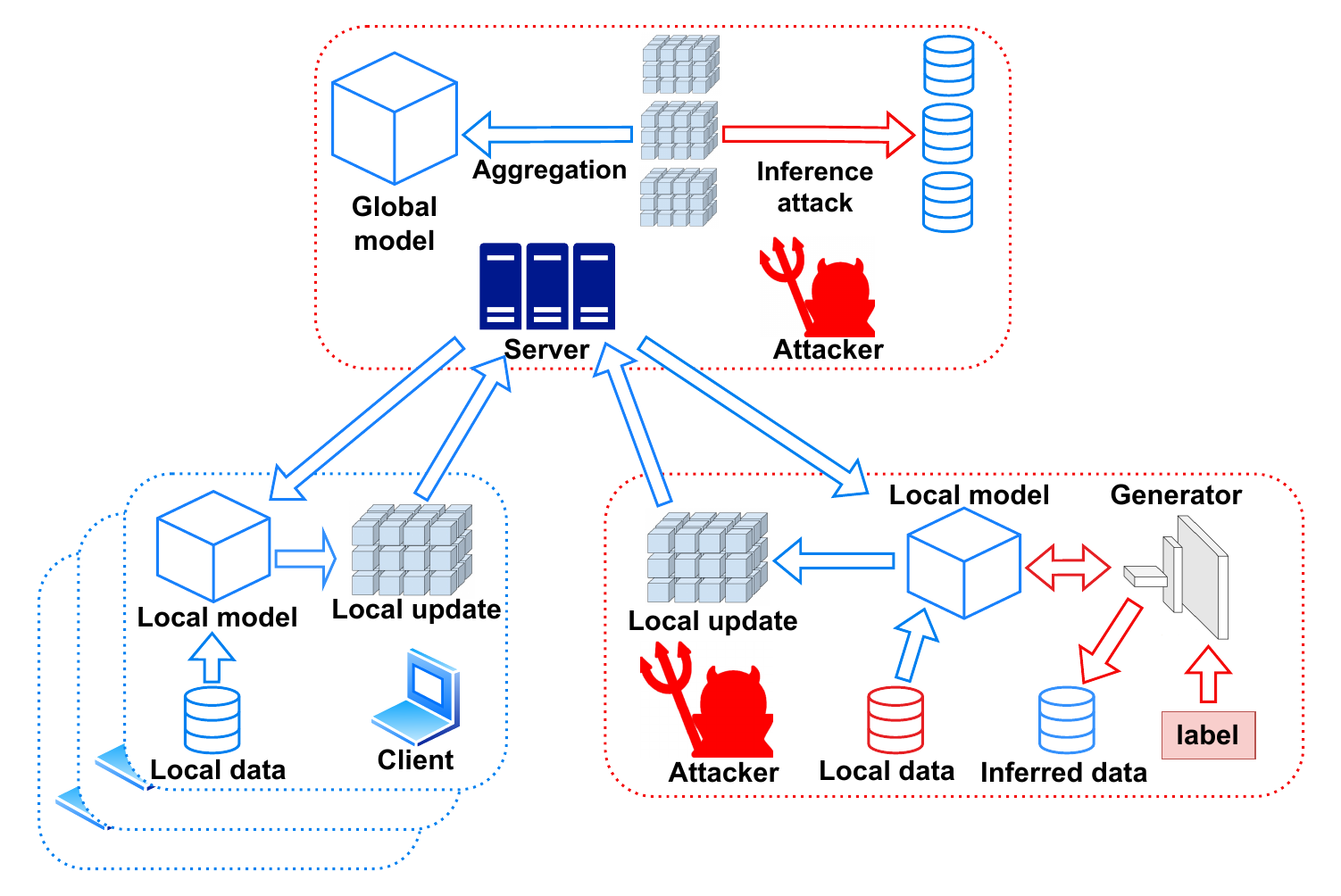}
    \hfill
    \caption{A simple example of privacy attacks in FL. Attacks can be performed on both server-side and client-side.}
    \label{fig:opahfl}
\end{figure}

Recent studies have indicated that FL cannot fully guarantee data privacy. As shown in Fig. \ref{fig:opahfl}, attackers may still deduce private data to some extent at both server-side and client-side through communicated model parameters. And based on different attack targets, we introduce and discuss privacy attacks from two perspectives: label inference attack and input reconstruction attack. For each type of attack, existing methods are systematically reviewed and discussed, along with their advantages and disadvantages.


\subsection{Label Inference}
Adversarial attackers may implicitly deduce the private labels of other parties through the analysis of communicated model gradients. And a timeline highlighting parts of milestones in recent label inference algorithms is summarized in Fig. \ref{fig:milelabel}. Li et al. \cite{li2021label} demonstrated that there are two methods by which one party can accurately recover the ground-truth labels owned by the other party in a two-party split learning scenario. Zhao et al. \cite{zhao2020idlg} proposed an analytical approach to extract the ground-truth labels by exploiting the direction of the gradients computed using the cross-entropy loss, as shown below:

\begin{figure*}[!htbp]
    \centering
    \includegraphics[width=0.92\textwidth]{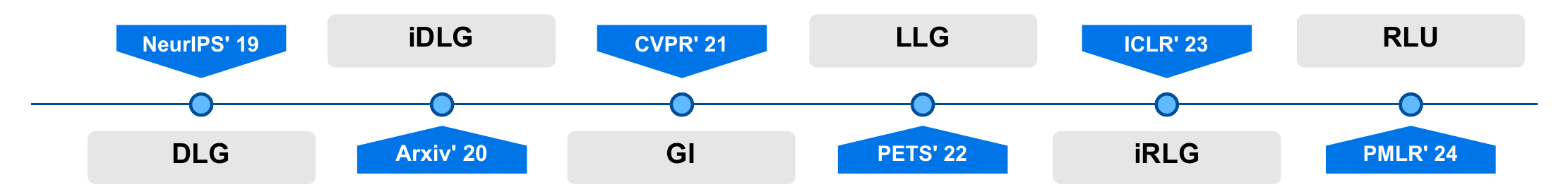}
    \caption{Some recent label inference attack methods.}
    \label{fig:milelabel}
\end{figure*}

\begin{equation}
\begin{aligned}
    &\mathbf{p} = \textit{softmax}(\hat{\mathbf{y}}) = \frac{\text{exp}(\hat{\mathbf{y}})}{\textstyle \sum_{{c}'} \text{exp}(\hat{\mathbf{y}}_{{c}'})}  \\
    &\mathcal{L}(\hat{\mathbf{y}}, \mathbf{y}) = - \sum_{{c}'} \mathbf{y}_{{c}'} \log \mathbf{ p}_{{c}'} 
\end{aligned}
\end{equation}
where $\mathbf{y}$ is the one-hot encoding vector of the ground-truth label $c \in \left[ C \right]$, $\hat{\mathbf{ y}}$ is the model output logits of \emph{a single input $\mathbf{x}$}, and $\mathbf{p}_{{c}'}$ denotes the predicted score for the ${c}'$-th label class. The gradient $d\hat{\mathbf{y}}_{{c}'}$ of the loss $ \mathcal{L}(\hat{\mathbf{y}}, \mathbf{y})$ with respect to any slot ${c}'$ of the prediction vector $\hat{\mathbf{y}}$ can be defined as: 
\begin{equation}
\label{eq:dyhat}
\begin{aligned}
    d\hat{\mathbf{y}}_{{c}'} &= \frac{\partial \mathcal{L}(\hat{\mathbf{y}}, \mathbf{y})} {\partial \hat{\mathbf{y}}_{{c}'}} \\ &= - \frac{\partial \sum_{{c}'} \mathbf{y}_{{c}'} \log \frac{\text{exp}(\hat{\mathbf{y}}_{{c}'})}{\sum_{{c}''} \text{exp}(\hat{\mathbf{y}}_{{c}''})} }{\partial \hat{\mathbf{y}}_{{c}'}} \\ &=\mathbf{p}_{{c}'}-\mathbf{y}_{{c}'}
\end{aligned}
\end{equation}

Since the prediction $\mathbf{p}_{{c}'}$ ranges from 0 to 1, it is easy to find that $d\hat{\mathbf{y}}_{{c}'} \in (-1,0)$ when the label slot ${c}'=c$, and $d\hat{\mathbf{y}}_{c} \in (0,1)$ otherwise. Therefore, the true label class can be easily deduced through the gradient $d\hat{\mathbf{y}}_{{c}'}$ with the negative value.

Even in scenarios where $d\hat{\mathbf{y}}_{{c}'}$ is unavailable, such as missing model biases in the final fully connected layer, inference of the private label remains feasible through analysis of the model weights. Specifically, the gradients $\nabla \mathbf{W}^{L}_{{c}'} \in \mathbb{R}^{n_{L-1}}$ with respect to ${c}'$-th row of the model weights $\mathbf{W}^{L} \in \mathbb{R}^{n_{L} \times n_{L-1}}$ of the last layer $L$ can be formulated as:

\begin{equation}
\label{eq:dW}
\begin{split}
    \nabla \mathbf{W}_{{c}'}^{L} &= \frac{\partial \mathcal{L}(\hat{\mathbf{y}}, \mathbf{y})}{\partial \hat{\mathbf{y}}_{{c}'}} \cdot \frac{\partial \hat{\mathbf{y}}_{{c}'}}{\partial \mathbf{W}_{{c}'}^{L}}\\
    &= d\hat{\mathbf{y}}_{{c}'} \cdot \frac{\partial {(\mathbf{W}_{{c}'}^{L}}\mathbf{a}^{L-1} + \mathbf{b}_{{c}'}^{L})}{\partial \mathbf{W}_{{c}'}^{L}}\\
    &= (\mathbf{p}_{{c}'}-\mathbf{y}_{{c}'}) \cdot \mathbf{a}^{L-1}
\end{split}
\end{equation}
where $L$ is the total number of model layers, $\mathbf{b}_{{c}'}^{L} \in \mathbb{R}$ is the ${c}'$-th model bias of layer $L$, $\mathbf{a}^{L-1} \in \mathbb{R}^{n_{L-1}}$ denotes a single activation output of layer $L-1$. When the network employs non-negative activation functions, such as commonly used Relu, the positivity of $\mathbf{a}^{L-1}$ ensures. Consequently, $\nabla \mathbf{W}^{L}_{{c}'}$ and $d\hat{\mathbf{y}}_{{c}'}$ share a concordant sign. Thus, a straightforward deduction emerges: the negativity of the gradient vector $\nabla \mathbf{W}^{L}_{{c}'}$ reliably indicates the ground-truth label $c$.
\begin{equation}
\label{eq:idlg}
    c = {c}', \quad s.t. \quad \left \langle \nabla \mathbf{W}_{{c}''}^{L}, \nabla \mathbf{ W}_{{c}'}^{L} \right \rangle \le 0, \quad \forall {c}' \ne {c}''
\end{equation}

However, this approach can only successfully infer private label of the gradients calculated by \emph{a single data point} which is not representative of actual FL situations in the real world applications. To address this challenge, Yin et al. \cite{yin2021see} proposed the first method to extract labels from gradients with respect to a batch of multiple images. While this is only valid when there are no duplicate samples within the batch. Aidmar et al. \cite{wainakh2022user} investigated an attack to extract the labels of the users' local training data from the shared gradients by exploiting the direction and magnitude of the gradients. Based on the previous work, Geng et al. \cite{geng2021towards} proposed a simple zero-shot approximation method to restore labels from training batch containing data with repeated label classes. According to Eq. \eqref{eq:dW}, the sum of values in $\nabla{\mathbf{W}}_{{c}'}^{L}$ along $l$-th column ($l \in [n_{L-1}]$) for averaged batch data $\Delta{\mathbf{w}}_{{c}'}^{L} \in \mathbb{R}$ is shown in Eq. \eqref{eq:dWsumavg}:

\begin{equation}
\label{eq:dWsumavg}
\begin{split}
    \Delta{\mathbf{w}}_{{c}'}^{L}&=\frac{1}{|B|}\sum_{i}\sum_{l}(\mathbf{P}_{i,{c}'}-\mathbf{Y}_{i,{c}'}) \cdot \mathbf{A}_{i, l}^{L-1} \\ &=\frac{1}{|B|}\sum_{i}(\mathbf{P}_{i,{c}'}-\mathbf{Y}_{i,{c}'}) \sum_{l}\mathbf{A}_{i, l}^{L-1} \\
\end{split}
\end{equation}
where $B$ is a set of batch data indices, $i \in B$ is the sample index, $\mathbf{Y} \in \mathbb{R}^{|B| \times n_{L}}$ represents batch labels, $\mathbf{P} \in \mathbb{R}^{|B| \times n_{L}}$ and $\mathbf{A} \in \mathbb{R}^{|B| \times n_{L-1}}$ are the batch outputs of layer $L$ and $L-1$, respectively. If we assume the sum of $i$-th data sample $\sum_{l}\mathbf{A}_{i, l}^{L-1}=\mathbf{A}_{i}^{L-1}\mathbf{1}_{l} \in \mathbb{R}$ can be approximated to the batch mean value $\bar{a}^{L-1} \approx \sum_{i}\mathbf{A}_{i}^{L-1}\mathbf{1}_{l} / |B|$, then $\Delta{\mathbf{w}}_{{c}'}^{L} \approx \frac{1}{|B|}\sum_{i}(\mathbf{P}_{i, {c}'}-\mathbf{Y}_{i, {c}'})\bar{a}^{L-1}$. Consequently, the number of label counts for class ${c}'$ can be calculated by the following equation:

\begin{equation}
\label{eq:labelcounts}
\sum_{i}\mathbf{Y}_{i,{c}'} \approx \sum_{i}\mathbf{P}_{i, {c}'}-\frac{|B| \Delta{\mathbf{w}}_{{c}'}^{L}}{\bar{a}^{L-1}}
\end{equation}
where $\mathbf{P}_{i, {c}'}$ and $\bar{a}$ ($L-1$ omitted for brevity)  are estimated by feeding $|B|$ dummy data $\mathbf{X}_{d}$ into the neural network. Furthermore, Dimitrov et al. \cite{dimitrov2022data} extended Eq. \eqref{eq:labelcounts} to let it suited in the scheme of multiple local batch training in FedAvg. Given that $\mathbf{\bar{p}_{{c}'}}=\sum_{i}\mathbf{P}_{i, {c}'}/ |B|$, the feed forward mean approximations for the dummy inputs are $\mathbf{\bar{p}_{{c}'}}^{s}$ and $\bar{a}^{s}$ at the global model weights $\mathbf{W}^{s}$, and $\mathbf{\bar{p}_{{c}'}}^{k}$ and $\bar{a}^{k}$ at the client $k$'s model weights $\mathbf{W}^{k}$, respectively. The estimates between $\mathbf{W}^{s}$ and $\mathbf{W}^{k}$ then be linearly interpolated for $U^{k}$ local training steps using the following equations:

\begin{equation}
\label{eq:pinter}
\begin{aligned}
    &\mathbf{\widetilde{p}}_{{c}', u}^{k}=\frac{u}{U^{k}}\mathbf{\bar{p}_{{c}'}}^{s}+\frac{U^{k}-u}{U^{k}}\mathbf{\bar{p}}_{{c}'}^{k} \\
    &\widetilde{a}_{u}^{k}=\frac{u}{U^{k}}\bar{a}^{s}+\frac{U^{k}-u}{U^{k}}\bar{a}^{k} 
\end{aligned}
\end{equation}
where $U^{k}=\left \lceil \left | \mathcal{D}^{k} \right | / |B| \right \rceil \times E$, $\mathcal{D}^{k}$ represents the entire dataset on client $k$, $E$ is the number of local training epochs, $u \in [U^{k}]$ represents every local update step. Thus, the label count $\widetilde{\mathbf{\lambda}}_{{c}'}^{k}$ for any specific label class ${c}'$ can be calculated by:

\begin{equation}
\label{eq:llc}
\begin{aligned}
    &\widetilde{\lambda}_{{c}', u}^{k}=\left | \mathcal{D}^{k} \right | \cdot \widetilde{\mathbf{p}}_{{c}', u}^{k}-\frac{\left | \mathcal{D}^{k} \right | \cdot \Delta{\mathbf{W}}_{{c}'}^{k}}{\widetilde{a}_{u}^{k}} \\
    &\widetilde{\lambda}_{{c}'}=\frac{1}{E}\sum_{u=1}^{U^{k}}\widetilde{\lambda}_{{c}', u}^{k}
\end{aligned}
\end{equation}
where $\widetilde{\lambda}_{{c}', u}^{k}$ is the label count for every training step $u$ on client $k$, $\Delta{\mathbf{W}}_{i}^{k}=(\mathbf{W}^{s, L}-\mathbf{W}^{k, L}) / \eta$ is the quotient of $i$-th row of weights difference and the learning rate $\eta$. It should be noticed that $\widetilde{\lambda}_{i}$ should be further adjusted to ensure that it satisfies $\left | \mathcal{D}^{k} \right |=\sum_{i} \widetilde{\lambda}_{i}$.

Recently, Ma et al. \cite{ma2022instance} introduced an iLRG method to restore instance-wise labels via shared batch-averaged client gradients, demonstrating promising attack performance particularly on untrained models. This work is to further extend Eq. \eqref{eq:dyhat} to a batch of $B$ samples belonging to ${c}'$ as shown below:

\begin{equation}
    \sum_{i}d\hat{\mathbf{Y}}_{i, {c}'} = \sum_{i}\frac{\partial \mathcal{L}_{i}}{\partial \mathbf{b}_{{c}'}}=\sum_{i}\left (\mathbf{P}_{i, {c}'}-\mathbf{Y}_{i, {c}'} \right )
\end{equation}

After adjusting the order, the number of label counts for any ${c}'$ can be easily derived as follows:

\begin{equation}
\label{eq:lbd}
\begin{aligned}
    \sum_{i}\mathbf{Y}_{i, {c}'}&=\sum_{i}\mathbf{P}_{i, {c}'}-\sum_{i}\frac{\partial \mathcal{L}_{i}}{\partial \mathbf{b}_{{c}'}} \\
    &=\sum_{i}\mathbf{P}_{i, {c}'}-\sum_{i}\nabla{\mathbf{b}}_{i, {c}'}
\end{aligned}
\end{equation}

Consequently, the key factor for achieving valid $\sum_{i}\mathbf{Y}_{i, {c}'}$ in Eq. \eqref{eq:lbd} is to approximate accurate $\sum_{i}\mathbf{P}_{i, {c}'}$ which is not directly accessible to the attacker. Thus, let $N_{{c}'}$ denotes the number of ${c}'$-class data samples, $\sum_{i}\mathbf{P}_{i}$ can be factorized into $\sum_{{c}'}N_{{c}'}\mathbf{\bar{P}}_{B_{{c}'}}$, and the average predictions  $\mathbf{\bar{P}}_{B_{{c}'}}$ are approximated as follows:

\begin{equation}
\begin{aligned}
    \mathbf{\bar{P}}_{B_{{c}'}}&=\frac{1}{|B_{{c}'}|}\sum_{i \in B_{{c}'}}\textit{softmax}\left( \mathbf{W}^{L}\mathbf{A}_{i}^{L-1}+\mathbf{b}^{L} \right) \\
    &\approx \textit{softmax}\left( \mathbf{W}^{L}\mathbf{\bar{A}}^{L-1}_{B_{{c}'}}+\mathbf{b}^{L} \right)
\end{aligned}
\end{equation}
where the average outputs of the last hidden layer $\mathbf{\bar{A}}^{L-1}_{B_{{c}'}}=\frac{1}{|B_{{c}'}|} \sum_{i \in B_{{c}'}}\mathbf{A}^{L-1}_{i}$ can be achieved by inter-class approximation $\mathbf{\bar{A}}^{L-1}_{B_{{c}'}} \approx \overline{\frac{\mathcal{L}_{B_{{c}'}}}{\mathbf{b}}}^{-1} \times \overline{\frac{\mathcal{L}_{B_{{c}'}}}{\mathbf{W}^{L}}}^{\intercal}$. Since $N_{{c}'}$ is still unknown to attackers, additional approximation techniques are required to deduce the label counts and the performance of the above iRLG drastically deteriorates as the model’s accuracy improves.

To address this limitation, Chen and Vikalo \cite{chen2024recovering} introduced a novel RLU label recovery method that demonstrates strong performance on both well-trained and untrained models. Given that  $s_{{c}'}(\mathbf{x})=\mathbf{p}_{{c}'}=\frac{\text{exp}(\hat{\mathbf{y}}_{{c}'})}{\sum_{{c}''} \text{exp}(\hat{\mathbf{y}}_{{c}''})}$, the authors provided the concept of the expected erroneous confidence $\mathcal{S}_{c, {c}'}$ for class ${c}'$ on any data sample with the true label $c \neq {c}'$ as shown in Eq. \eqref{eq:S}:

\begin{equation}
\label{eq:S}
    \mathcal{S}_{c, {c}'}=\mathbb{E}_{ \left ( \mathbf{x}, \mathbf{y}  \right ) \sim \mathcal{D}_{c}}\left [ s_{{c}'}\left ( \mathbf{x} \right ) \right ]
\end{equation}
where $\mathcal{D}_{c}$ contains data samples with label $c$ and $\left ( \mathbf{x}, \mathbf{y}  \right )$ represents data pairs sampled from $\mathcal{D}_{c}$. Specifically for the condition that the local training epoch $E=1$, the local update with respect to bias $\mathbf{b}_{{c}'}$ is $\Delta{\mathbf{b}_{{c}'}}=-\frac{\eta}{|B|}\sum_{i=1}^{|B|}\nabla{\mathbf{b}_{i, {c}'}}$. Then the expectation of $\Delta{\mathbf{b}_{{c}'}}$ can be defined as follows:

\begin{equation}
\label{eq:expsb}
\begin{split}
    \mathbb{E}\left [ \Delta{\mathbf{b}_{{c}'}} \right ]&=-\frac{\eta}{|B|}\sum_{i=1}^{|B|}\mathbb{E}\left[ \nabla{\mathbf{b}_{i, {c}'}} \right ] \\
    &=\frac{\eta}{|B|}\left ( N_{{c}'}\sum_{c \neq {c}'}\mathcal{S}_{{c}', c}-\sum_{c \neq {c}'}N_{c}\mathcal{S}_{c, {c}'} \right )
\end{split}
\end{equation}
where $N_{{c}'}$ represents the number of data points in $B$ with label class ${c}'$. Then, $N_{{c}'}$ can be estimated by minimizing the following loss function:

\begin{equation}
\label{eq:optz}
    \min_{\mathbf{z} \in \mathbb{R}^{C}}\left \| \mathbf{H}\mathbf{z} - \mathbf{v} \right \|_{2}^{2}
\end{equation}
where vector $\mathbf{v}=\Delta{\mathbf{b}} / \eta \in \mathbb{R}^{C}$, $\mathbf{z} $ follows a discrete probability distribution with $0 < \mathbf{z}_{{c}'} < 1, \left \| \mathbf{z} \right \|_{1}=1$, and $\mathbf{H}$ is a coefficient matrix whose diagonal element is $\mathbf{H}_{{c}', {c}'}=\sum_{c \neq {c}'}\mathcal{S}_{{c}', c}$ and the $\left( c, {c}' \right)$ off-diagonal element is $\mathbf{H}_{c, {c}'}=-\mathcal{S}_{c, {c}'}$. Since the prediction $\mathbf{p}_{{c}'}$ used to calculate $\mathcal{S}_{c, {c}'}$ is unavailable, the authors made an assumption that the server contains an auxiliary dataset $\mathcal{D}^{A}$ to be processed through the model weights $\mathbf{W}$. And the resulting output logits $\hat{\mathbf{Y}}_{i}, i \in \mathcal{D}^{A}_{c}$ are used to calculate the mean $\bar{\mu}_{c}=\sum_{i \in \mathcal{D}^{A}_{c}}\hat{\mathbf{Y}}_{i}/|\mathcal{D}^{A}_{c}|$ and covariance $\bar{\Sigma}_{c}\left ( c, {c}' \right )=\frac{1}{|\mathcal{D}^{A}_{c}|}\sum_{i \in \mathcal{D}^{A}_{c}} \left ( \hat{\mathbf{Y}}_{i, c}-\bar{\mu}_{c, c} \right )\left (  \hat{\mathbf{Y}}_{i, {c}'}-\bar{\mu}_{c, {c}'} \right ) $, respectively. Given $\bar{\mu}_{c}$ and $\bar{\Sigma}_{c}$, $\mathcal{S}_{c, {c}'}$ can be indirectly inferred by $M$ points sampled from $\hat{\mathbf{Y}}_{i} \sim \mathcal{N}\left ( \bar{\mu}_{c}, \bar{\Sigma}_{c} \right )$:

\begin{equation}
\label{eq:approxScc}
    \mathcal{S}_{c, {c}'} \approx \frac{1}{M}\sum_{i=1}^{M}\frac{\text{exp}(\hat{\mathbf{Y}}_{i, {c}'})}{\sum_{{c}''}\text{exp}(\hat{\mathbf{Y}}_{i, {c}''})}
\end{equation}
As $\mathcal{S}_{c, {c}'}$ and the corresponding $\mathbf{H}$ can be approximated, the optimal solution $\mathbf{z}^{*}$ in Eq. \eqref{eq:optz} allows for the estimation of $N_{{c}'}$ can be easily derived by computing $\bar{N}_{{c}'}=\left \lfloor |B| \cdot \mathbf{z}^{*}_{{c}'} \right \rceil$.

While for more realistic FL scenarios that $E > 1$, the expectation of the local update at $t$-th communication round becomes:

\begin{equation}
\label{eq:mapproxscc}
    \mathbb{E}\left [ \Delta{\mathbf{b}_{{c}'}^{t}} \right ]=\frac{\eta}{|B|}\sum_{e=1}^{E}\left ( N_{{c}'}^{t, e}\sum_{c \neq {c}'}\mathcal{S}_{{c}', c}^{t, e}-\sum_{c \neq {c}'}N_{c}^{t, e}\mathcal{S}_{c, {c}'}^{t, e} \right )
\end{equation}
where both $\mathcal{S}_{c, {c}'}^{t, 1}$ and $\mathcal{S}_{c, {c}'}^{t, E}$ are inferred by Eq. \eqref{eq:approxScc} using the global model parameters $\mathbf{W}^{t}$ and updated local model parameters $\mathbf{W}^{E}$, respectively. While any intermediate erroneous confidence $\mathcal{S}_{c, {c}'}^{t, e+1}$ from $\mathcal{S}_{c, {c}'}^{t, 2}$ to $\mathcal{S}_{c, {c}'}^{t, E-1}$ can be indirectly estimated by $\mu_{c,{c}'}^{t, e+1}=\mu_{c,{c}'}^{t, e} + \Delta{\mu}_{c,{c}'}^{t, e}$, where $\Delta{\mu}_{c,{c}'}^{t, e}$ is approximated by the change of output logits shown below:

\begin{equation}
\begin{split}
    \Delta{\mu}_{c,{c}'}^{t, e}&=\mathbb{E}_{i \sim \mathcal{D}_{c}^{A}}\left[ \Delta{\hat{\mathbf{Y}}_{i, {c}'}} \right] \\
    &=\mathbb{E}\left[ \Delta{\mathbf{b}}_{{c}'}^{t, e} \right] \cdot \bar{\mathbf{a}}^{\intercal}\bar{\mathbf{a}} \\
    &=\frac{\eta}{|B|}\left ( N_{{c}'}^{t, e}\sum_{c \neq {c}'}\mathcal{S}_{{c}', c}^{t, e}-\sum_{c \neq {c}'}N_{c}^{t, e}\mathcal{S}_{c, {c}'}^{t, e} \right ) \cdot \bar{\mathbf{a}}^{\intercal}\bar{\mathbf{a}}
\end{split}
\end{equation}
where $\bar{\mathbf{a}}$ is the average outputs of the last hidden layer which can be approximated by equation $\bar{\mathbf{a}} \approx \Delta{\mathbf{W}} / \Delta{\mathbf{b}}$. Regardless of the inference errors arising from approximation processes, RLU algorithm assumes that each iteration uses the same batch of data for weight updates according to Eq. \eqref{eq:mapproxscc}. This assumption may lead to inaccurate label count deductions for the attacked client, thereby further misdirecting the gradient inversion optimization.

Until now, label inference attacks have not performed well in FL scenarios involving multiple local epochs with slightly larger batch size (e.g. more than 50). And incorrect label count deduction significantly deteriorates the reconstruction performance of private data on the server side. Furthermore, the accuracy of inferred dummy labels can significantly impact the effectiveness of input reconstruction attacks.


\subsection{Input Reconstruction}

\begin{figure*}
    \centering
    \includegraphics[width=0.92\textwidth]{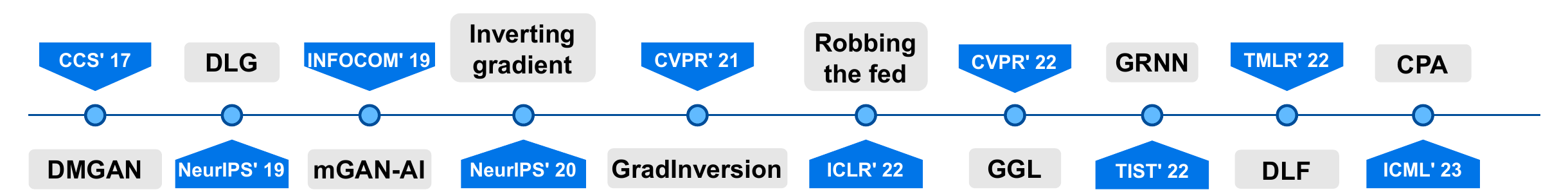}
    \caption{Some input reconstruction attack methods.}
    \label{fig:mileinput}
\end{figure*}

Ganju et al. \cite{ganju2018property} have demonstrated that the attacker can infer properties of the training data by analyzing the shared model parameters of neural networks. This category of privacy attack, commonly referred to as input reconstruction on the server-side (shown in Fig. \ref{fig:opahfl}), can be categorized within FL into text reconstruction and image reconstruction. A timeline of some recent input reconstruction algorithms is summarized in Fig. \ref{fig:mileinput}. Note that, text reconstruction is generally more difficult than image reconstruction in FL, due to the discrete, sparse, and high-level semantic nature of the text data. In this section, we focus primarily on image reconstruction.

\subsubsection{Text Reconstruction}
text data, typically represented as sequences of discrete tokens, is characterized by sparse and complex semantic structures. The relationships between words (context) are crucial, and these dependencies add complexity to gradient interpretation. Unlike continuous data, the discrete nature of text makes gradients less directly interpretable \cite{CARTUYVELS2021143}, and this inherent robustness to small gradient changes further complicates precise text reconstruction without additional linguistic knowledge. In \cite{melis2019exploiting}, the adversary knows all non-sensitive attributes of a given record and analyzes the gradients generated by different words in a shared global model to infer sensitive attributes used by other clients during training. Based on this work, Gupta et al.\cite{gupta2022recovering} proposed an attack targeting the recovery of full sentences from large language models. Song et al. \cite{song2020information} discovered that embeddings can leak information about input data, revealing sensitive attributes inherent in the inputs. Lyu et al. \cite{lyu2021novel} systematically investigated attribute inference attacks. In their study, the attacker first generates a multinomial distribution for the attribute of interest and computes the predicted values of these attributes. These predicted sensitive attributes, along with other non-sensitive attributes, are then fed into the global model to obtain gradients. Finally, the adversary optimizes the predicted attributes by aligning the virtual gradients with the real gradients.


\subsubsection{Image Reconstruction}
on the other hand, image data, being continuous and high-dimensional, relies heavily on the spatial relationships between pixels, with each pixel serving as a distinct feature. The structured nature of image features, such as edges and textures, allows attackers to exploit these patterns, making reconstruction more straightforward. Additionally, the continuous nature of images means that small gradient changes can significantly alter the image, further facilitating reconstruction. Here, we will explore image reconstruction upon the shared gradients in FL from two aspects: optimization-based attacks and analytic attacks. Additionally, approaches based on GAN \cite{goodfellow2014generative} will be examined in a separate subsection.

\begin{figure}[!htbp]
    \centering
    \includegraphics[width=0.46\textwidth]{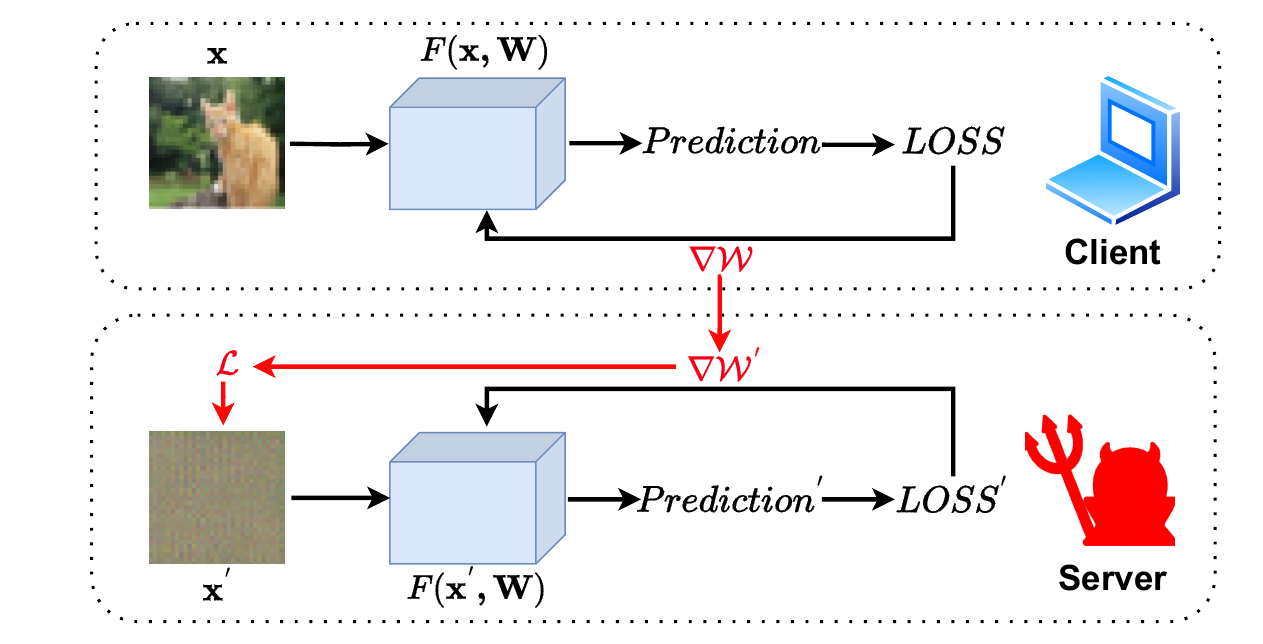}
    \caption{An example of optimization-based attack}
    \label{fig:optattack}
\end{figure}

\paragraph{Optimization-Based Attack}
this is achieved by minimizing the distance between the gradients of the randomly generated dummy image and the ground-truth gradients, typically using L1-norm \cite{deng-etal-2021-tag-gradient} or L2-norm \cite{zhu2019deep} as the optimization loss function. And they can be applied to the gradients of even unconverged models \cite{pan2020theory, li2021deep, luo2022effective} by forcing a dummy image $\hat{\mathbf{x}}$ to approximate the clients' privacy data as shown below: 

\begin{equation}
\label{eq:aialoss}
    {\hat{\mathbf{x}}}' = \mathop{\arg\min}\limits_{\hat{\mathbf{x}}}\mathcal{L} (\hat{\mathbf{x}}; \mathbf{W}, \nabla \mathbf{W}_{g})+\mathcal{R}(\hat{\mathbf{x}})
\end{equation}
where $\mathbf{W}$ represents the current model parameters, $\nabla \mathbf{W}_{g}$ is the gradients of the ground-truth data, and $\mathcal{R}(\hat{\mathbf{x}})$ is the auxiliary regularization term. Note that, in FL, image reconstruction attacks are usually performed on the server side, and $\nabla \mathbf{W}_{g}^{k}$ of client $k$ can be derived by calculating $\mathbf{W}^{t-1}-\mathbf{W}^{k}$, where $\mathbf{W}^{t-1}$ is the parameters of the global model at the previous communication round $t-1$ and $\mathbf{W}^{k}$ is the current parameters from client $k$ trained based on $\mathbf{W}^{t-1}$. A more straightforward example is also provided in Fig. \ref{fig:optattack} to offer a clearer, intuitive understanding of the process involved in this optimization-based attack.

Early in 2019, Zhu et al. \cite{zhu2019deep} first proposed the gradient inversion attack, named Deep Leakage Gradients (DLG), to retrieve both input features and labels of the training data. By optimizing the dummy image $\hat{\mathbf{x}}$ in Eq. \eqref{eq:aialoss}, the L2-distance between the gradients of the ground-truth data $\nabla \mathbf{W}_{g}$ and dummy data $\nabla \mathbf{W}_{d}$ would be shrunk, making $\hat{\mathbf{x}}$ resembles the corresponding ground-truth image $\mathbf{x}$. The image reconstruction process of DLG is shown in Algorithm \ref{alg:dlg}, where $\eta$ is the learning rate, and $E$ is the number of reconstruction epochs.

\begin{algorithm}[!htbp]
\renewcommand{\algorithmicrequire}{\textbf{Input:}}
\renewcommand{\algorithmicensure}{\textbf{Output:}}
\caption{Deep Leakage from Gradients (DLG)} \label{alg:dlg}
\begin{algorithmic}[1]
\Require current model parameters $\mathbf{W}$, the gradients $\nabla \mathbf{W}_{g}$ computed on the ground-truth data $(\mathbf{x}_{g}, y_{g})$
\Ensure reconstructed dummy data pairs $(\mathbf{x}_{d}, y_{d})$
\State $\mathbf{x}_{d} \leftarrow \mathcal{N}(0,1), y_{d} \leftarrow \mathcal{N}(0,1)$
\For{each epoch $e=0,1, \ldots E-1$}
\State $\nabla \mathbf{W}_{d}^{e} \leftarrow \partial \mathcal{L}(f(\mathbf{x}_{d}; \mathbf{W}), y_{d}) / \partial \mathbf{W}$
\State $\mathcal{L} \leftarrow \left \| \nabla{\mathbf{W}}_{d}^{e} - \nabla{\mathbf{W}}_{g} \right \|^{2} $
\State $\mathbf{x}_{d} \leftarrow \mathbf{x}_{d} - \eta \nabla_{\mathbf{x}_{d}}\mathcal{L}$
\State $y_{d} \leftarrow y_{d} - \eta \nabla_{y_{d}}\mathcal{L}$
\EndFor
\State \textbf{return} $\mathbf{x}_{d}, y_{d}$
\end{algorithmic}
\end{algorithm}

\begin{figure}[!htbp]
\centering
\includegraphics[width=0.46\textwidth]{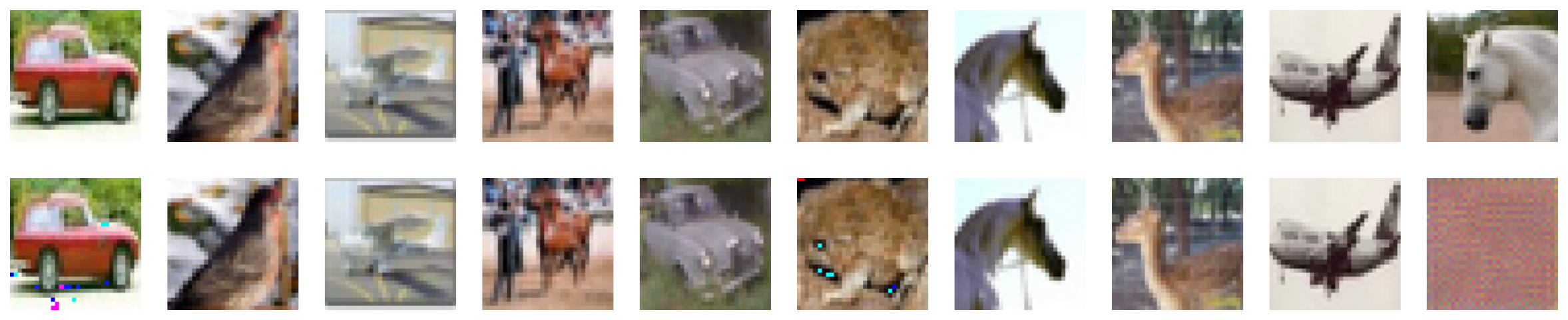}
\caption{The outcomes of gradient attack by DLG.}
\label{fig:dlg}
\end{figure}

Since then, many gradient-based attack approaches have focused on further improving the performance of DLG. Qian et al. \cite{qian2020can} considered the prior knowledge of auxiliary data to initialize the dummy data with uniform distribution, enhancing the quality of reconstructed images in DLG. Moreover, Zhao et al. \cite{zhao2020idlg} found that DLG method frequently generates incorrect labels during the optimization period, misleading the convergence direction of image recovery \cite{geng2021towards}. The simulation results of DLG attack performed on gradients with respect to a single CIFAR-10 image \cite{krizhevsky2009learning} for 10 repeated runs (each run only reconstructs one dummy image) are shown in Fig. \ref{fig:dlg}. The first row and the second row contain 10 ground-truth data and the corresponding reconstructed dummy data for 10 separate simulations, respectively. It is evident that the white horse image located at the last column fails to be retrieved.

To address this issue, improved DLG (iDLG) were proposed as shown in Algorithm \ref{alg:idlg} to directly infer the private label according to Eq. \eqref{eq:idlg} other than relying on direct optimization (line 6 in Algorithm \ref{alg:dlg}).

\begin{algorithm}[!htbp]
\renewcommand{\algorithmicrequire}{\textbf{Input:}}
\renewcommand{\algorithmicensure}{\textbf{Output:}}
\caption{Improved Deep Leakage from Gradients (iDLG)} \label{alg:idlg}
\begin{algorithmic}
\Require current model parameters $\mathbf{W}$, the gradients $\nabla \mathbf{W}_{g}$ computed on the ground-truth data $(\mathbf{x}_{g}, y_{g})$
\Ensure dummy data pairs $(\mathbf{x}_{d}, y_{d})$
\State $\mathbf{x}_{d} \leftarrow \mathcal{N}(0,1)$
\State Deduce $y_{d}$ using Eq. \eqref{eq:idlg} \Comment{Different from DLG}
\For{each epoch $e=0,1, \ldots E-1$}
\State $\nabla \mathbf{W}_{d}^{e} \leftarrow \partial \mathcal{L}(f(\mathbf{x}_{d}; \mathbf{W}), y_{d}) / \partial \mathbf{W}$
\State $\mathcal{L} \leftarrow \left \| \nabla{\mathbf{W}}_{d}^{e} - \nabla{\mathbf{W}}_{g} \right \|^{2} $
\State $\mathbf{x}_{d} \leftarrow \mathbf{x}_{d} - \eta \nabla_{\mathbf{ x}_{d}}\mathcal{L}$
\State $y_{d} \leftarrow y_{d} - \eta \nabla_{y_{d}}\mathcal{L}$
\EndFor
\State \textbf{return} $\mathbf{x}_{d}, y_{d}$
\end{algorithmic}
\end{algorithm}

\begin{figure}[!htbp]
\centering
\includegraphics[width=0.46\textwidth]{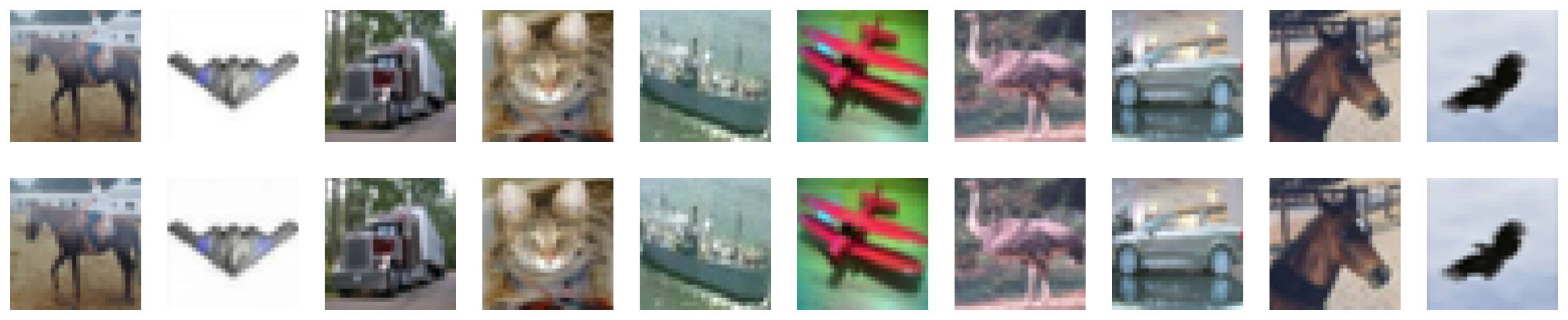}
\caption{The outcomes of gradient attack by iDLG.}
\label{fig:idlg}
\end{figure}

The reconstructed outcomes by iDLG attack performed on gradients with respect to one image for 10 separate runs are shown in Fig. \ref{fig:idlg}, where the first row contains 10 ground-truth images and the second row contains the respective retrieved dummy images. It is obvious to find that the quality of the reconstructed images from iDLG is superior to those from DLG, due to the correct inference of dummy labels.

\begin{figure}[!htbp]
    \centering
    \includegraphics[width=0.46\textwidth]{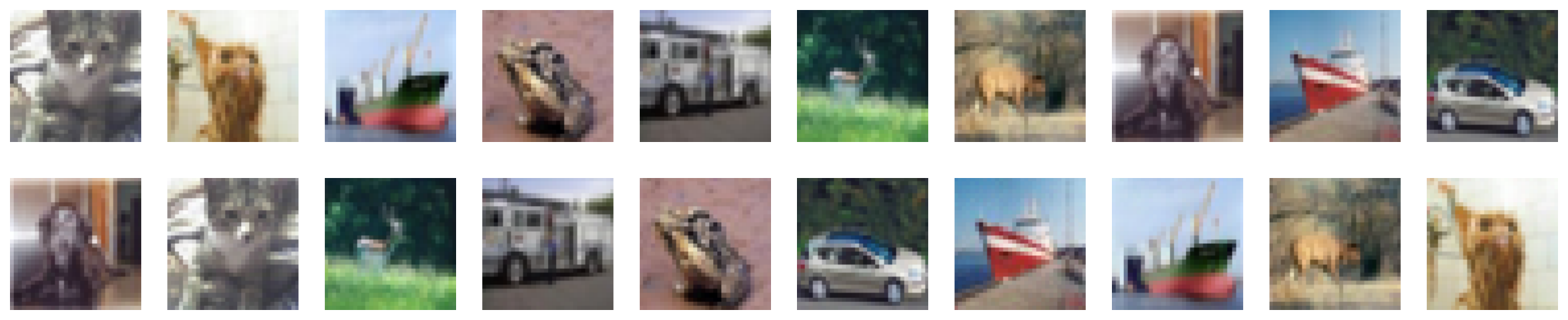}
    \caption{The outcomes of gradient attack by Inverting Gradients.}
    \label{fig:ig_result}
\end{figure}

However, iDLG has an inherent limitation that it can only recover one image at a time, as the label inference attack described in the Eq. \eqref{eq:idlg} is applicable to the gradients of a single data. In addition, both DLG and iDLG are sensitive to the weight distribution of the inferred model, and cannot recover images under a normal weight initialization \cite{wang2020sapag}. Moreover, the L2-norm distance utilized cannot convey sufficient information of the high-dimensional direction of the gradients. For instance, the angle between two data points quantifies the change in prediction at one data point when taking a gradient step towards another, a critical factor overlooked by the Euclidean distance function. Therefore, Geiping et al. (Inverting Gradients) \cite{geiping2020inverting} replaced the Euclidean distance with cosine similarity to capture the significant information in the high-dimensional direction of the gradients and improve the stability of DLG-based methods. The equation of cosine similarity loss is shown below:

\begin{equation}
\label{eq:ig}
    \mathop{\arg\min}\limits_{\hat{\mathbf{x}}}1-\frac{\left \langle \nabla_{\mathbf{ W}}\mathcal{L}\left ( \hat{\mathbf{x}},\hat{y} \right ), \nabla_{\mathbf{ W}}\mathcal{L}\left ( \mathbf{x}, y \right )  \right \rangle}{\left \| \nabla_{\mathbf{ W}}\mathcal{L}\left ( \hat{\mathbf{x}},\hat{y} \right ) \right \| \left \| \nabla_{\mathbf{W}}\mathcal{L}\left ( \mathbf{x}, y \right )  \right \|}+\alpha \text{TV}(\hat{\mathbf{x}})
\end{equation}
where TV$(\hat{\mathbf{x}})$ represents total variation \cite{RUDIN1992259} of the dummy data $\hat{\mathbf{x}}$. Except that, He et al. \cite{he2023fast} proposed a method that improved the stability of the inference attack and the reliability of the reconstructed label by changing the distance metric from Euclidean distance to the Wasserstein distance \cite{vaserstein1969markov} to calculate the loss between the dummy gradients and the ground-truth gradients. The corresponding reconstructed outcomes for batch gradients with respect to 10 image data are presented in Fig. \ref{fig:ig_result}, where the second row represents the recovered images. And even for batch gradient attack, Inverting Gradient still show promising attack performance, generating high-quality dummy images that closely resemble the ground-truth images.

Furthermore, Jeon et al. \cite{jeon2021gradient} optimized the objective function by incorporating auxiliary prior knowledge to improve reconstruction performance. Hatamizadeh et al. \cite{hatamizadeh2022gradvit} used the mean and variance of the inputs captured by the batch normalization layers as priors to enhance the quality of retrieved images when models containing batch normalization layers  are selected for training. Yin et al. \cite{yin2021see} proposed a group consistency regularization term \cite{10.5555/2969033.2969202} that employs multiple independent optimization processes and statistical information from batch normalization layers to enable consistent improvements across evaluation indicators, thereby narrowing the gap between the reconstructed and ground-truth data. However, the method proposed by Yin et al. does not align with the real FL scenarios, as it requires clients to train only one batch of data per iteration and then upload updates immediately.

\begin{figure}[!htbp]
    \centering
    \includegraphics[width=0.46\textwidth]{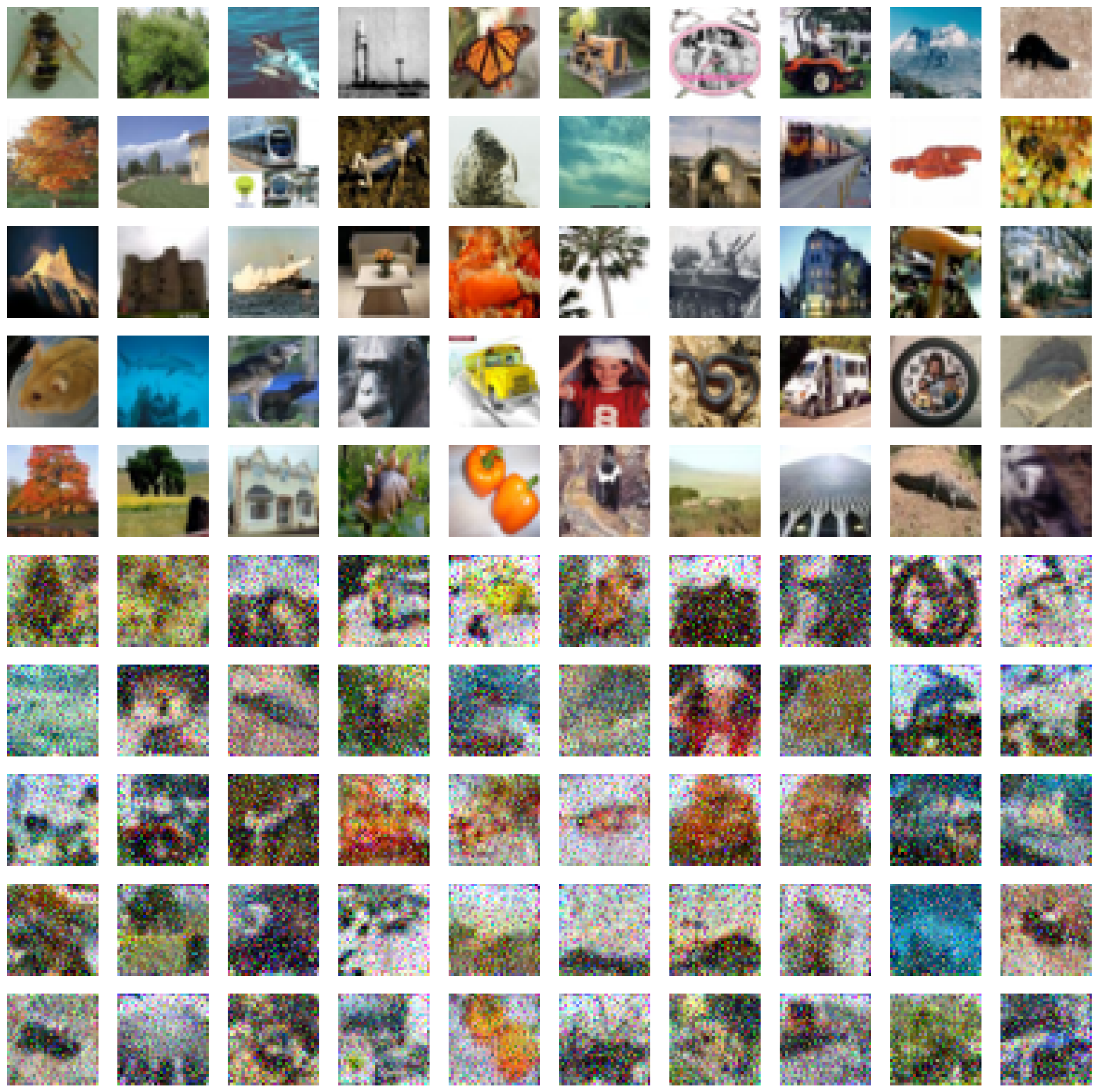}
    \caption{The outcomes of DLF attack for 50 images.}
    \label{fig:dlfsmall}
\end{figure}

To address the aforementioned issue and enhance the practicality of gradient leakage attacks, Dimitrov et al. (DLF) \cite{dimitrov2022data} proposed a novel approach that combines a simulation-based reconstruction loss with an epoch order-invariant prior. This method effectively recovers private images from model differences computed through multiple epochs of local batch training. Additionally, the authors highlighted the significance of label inference counts (as introduced in Eq. \eqref{eq:pinter}\eqref{eq:llc}) in influencing the success of image recovery. The simulation results of DLF for 50 images, with a batch size of 10 and 10 local training epochs, are presented in Fig. \ref{fig:dlfsmall}, where the last five rows are reconstructed 50 dummy images. Moreover, Yang et al. \cite{yang2022using} recovered data from highly compressed gradients by isolating the gradients of the final layer of the neural network model. Acknowledging Wei et al.'s observation \cite{wei2020framework} that different initialization methods impact the performance of gradient-based attacks, Zhao et al. \cite{pmlr-v234-zhao24b} modified the objective function to eliminate dependencies on the learning rate, thereby addressing the challenge of high initialization requirements. Furthermore, Sun et al. \cite{10445924} employed anomaly detection techniques to enhance attack effectiveness with minimal auxiliary data.

\begin{figure}[!htbp]
    \centering
    \includegraphics[width=0.46\textwidth]{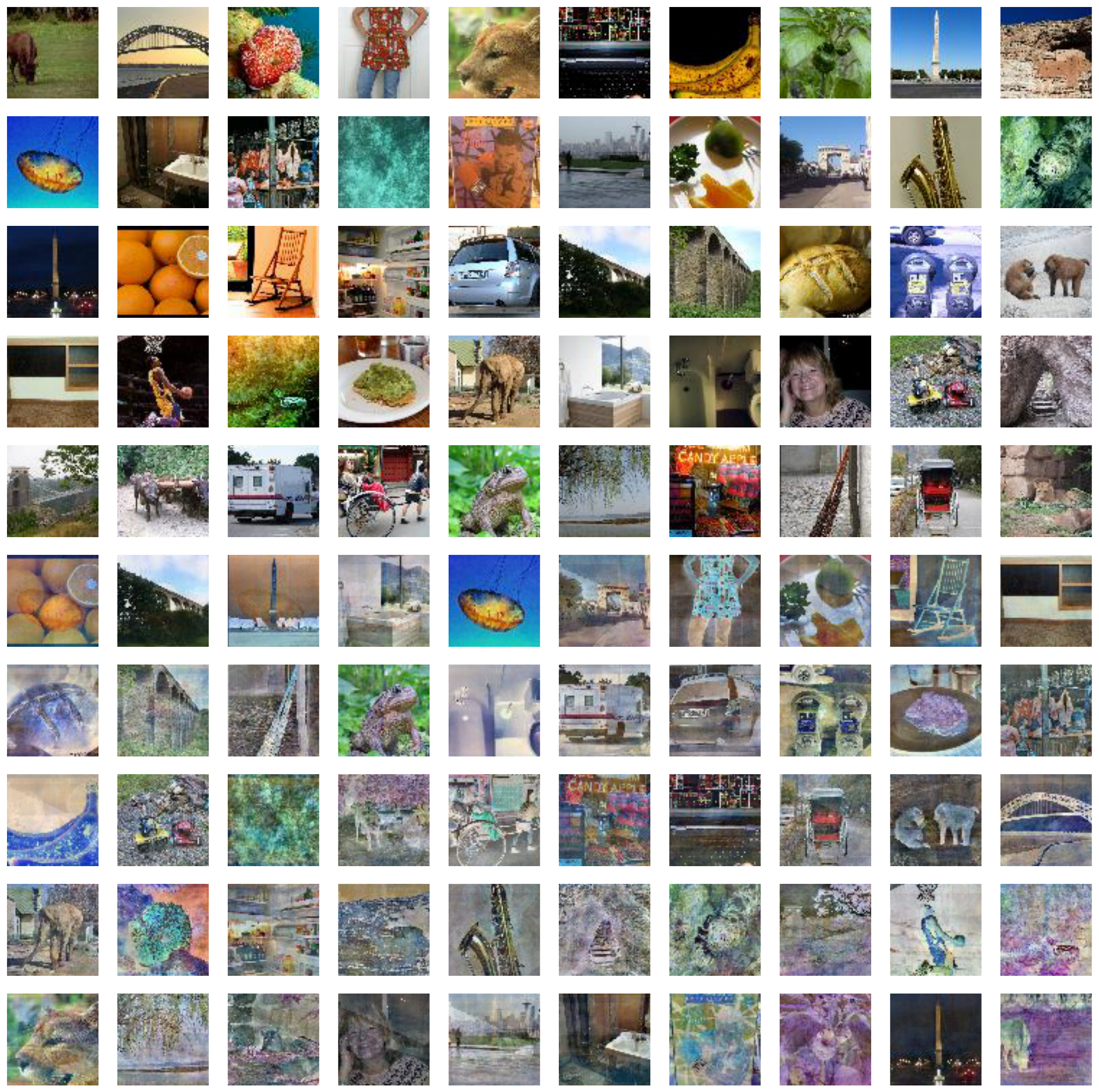}
    \caption{The outcomes of CPA attack for 50 images.}
    \label{fig:cpafc2}
\end{figure}

More recently, inspired by \cite{yin2020dreaming}, Kariyappa et al. proposed the Cocktail Party Attack (CPA) \cite{kariyappa2023cocktail} which recovers private data from gradients aggregated over a large batch size. This approach is enabled by the novel insight that the aggregated gradient for a fully connected layer is a linear combination of its inputs as shown below:

\begin{equation}
\label{eq:bss}
    \mathbb{E}\left( \nabla{\mathbf{W}}_{d,l} \right)=\frac{1}{|B|}\sum_{i \in B}\frac{\partial \mathcal{L}_{i}}{\partial \mathbf{Y}_{i,l}}\mathbf{X}_{i,d}^{\intercal}
\end{equation}
where $\mathbf{Y}_{i,l}$ is the layer output of $l$-th neuron for input sample $\mathbf{X}_{i}$ and $\mathbf{X}_{i,d}$ is the $d$-th input data feature. Recovering the input $\mathbf{X}_{i}$ from $\mathbb{E}\left( \nabla{\mathbf{W}}_{d,l} \right)$ can therefore be formulated as a blind source separation (BSS) problem. If we reformulate Eq. \eqref{eq:bss} into a matrix multiplication operation $\mathbf{G}=\mathbf{C}\mathbf{X}$ (where $\mathbf{X} \in \mathbb{R}^{|B| \times n_{0}}$ denote $|B|$ data inputs, $\mathbf{C} \in \mathbb{R}^{|B| \times |B|}$ represents the coefficients of the linear combinations, and $\mathbf{G} \in \mathcal{R}^{|B| \times n_{0}}$ is the whitened gradients), the matrix $\hat{\mathbf{X}}$ can be estimated by the following equation:

\begin{equation}
    \hat{\mathbf{X}}=\mathbf{C}^{-1}\mathbf{G}=\mathbf{U}\mathbf{G}
\end{equation}
where each row of $\hat{\mathbf{X}}$ represents a single reconstructed image and $\mathbf{U} \in \mathbb{R}^{|B| \times |B|}$ is defined as an unmixing matrix. Since the whitened gradients $\mathbf{G}$ are available during training, recovering $\hat{\mathbf{X}}$ can be reformulated as the optimization problem of finding the optimal unmixing matrix $\mathbf{U}$. For attacking a multi-layer perceptron (MLP) neural network trained on image data, the following optimization function is employed:

\begin{equation}
\label{eq:cpa1}
    \mathbf{U}=\mathop{\arg\max}\limits_{\mathbf{U}^{*}}\mathbb{E}_{i}\left[ J\left( \mathbf{U}_{i}^{*}\mathbf{G} \right)-\lambda_{tv}\mathcal{R}_{tv}\left( \mathbf{U}_{i}^{*}\mathbf{G} \right) \right]-\lambda_{MI}\mathcal{R}_{MI}
\end{equation}
where $J\left( \mathbf{U}_{i}^{*}\mathbf{G} \right)=J(\mathbf{X}_{i}^{*})=\mathbb{E} \left[ \frac{1}{a^{2}} \log \cosh^{2} (a\mathbf{X}_{i}) \right]$ is the negentropy metric \cite{HYVARINEN2000411} measuring non-Gaussianity, $\mathcal{R}_{tv}$ denotes the total variation prior \cite{RUDIN1992259}, $\mathcal{R}_{MI}=\mathbb{E}_{i \neq {i}'}\text{exp}\left(T \left| CS \left( \mathbf{U}_{i}^{*}, \mathbf{U}_{{i}'}^{*} \right) \right| \right)$ represents mutual independence loss, and $\lambda_{tv}$ and $\lambda_{MI}$ are two hyperparameters. And the simulation outcomes of CPA attack for gradients with respect to a batch of 50 Tiny-ImageNet data \cite{le2015tiny} on a MLP neural network with a single hidden layer are presented in Fig. \ref{fig:cpafc2}, where the last 5 rows are reconstructed images. It is evident that the CPA is capable of successfully recovering more complex images (64 $\times$ 64) even with significantly larger batch size.

\begin{figure}[!htbp]
    \centering
    \includegraphics[width=0.46\textwidth]{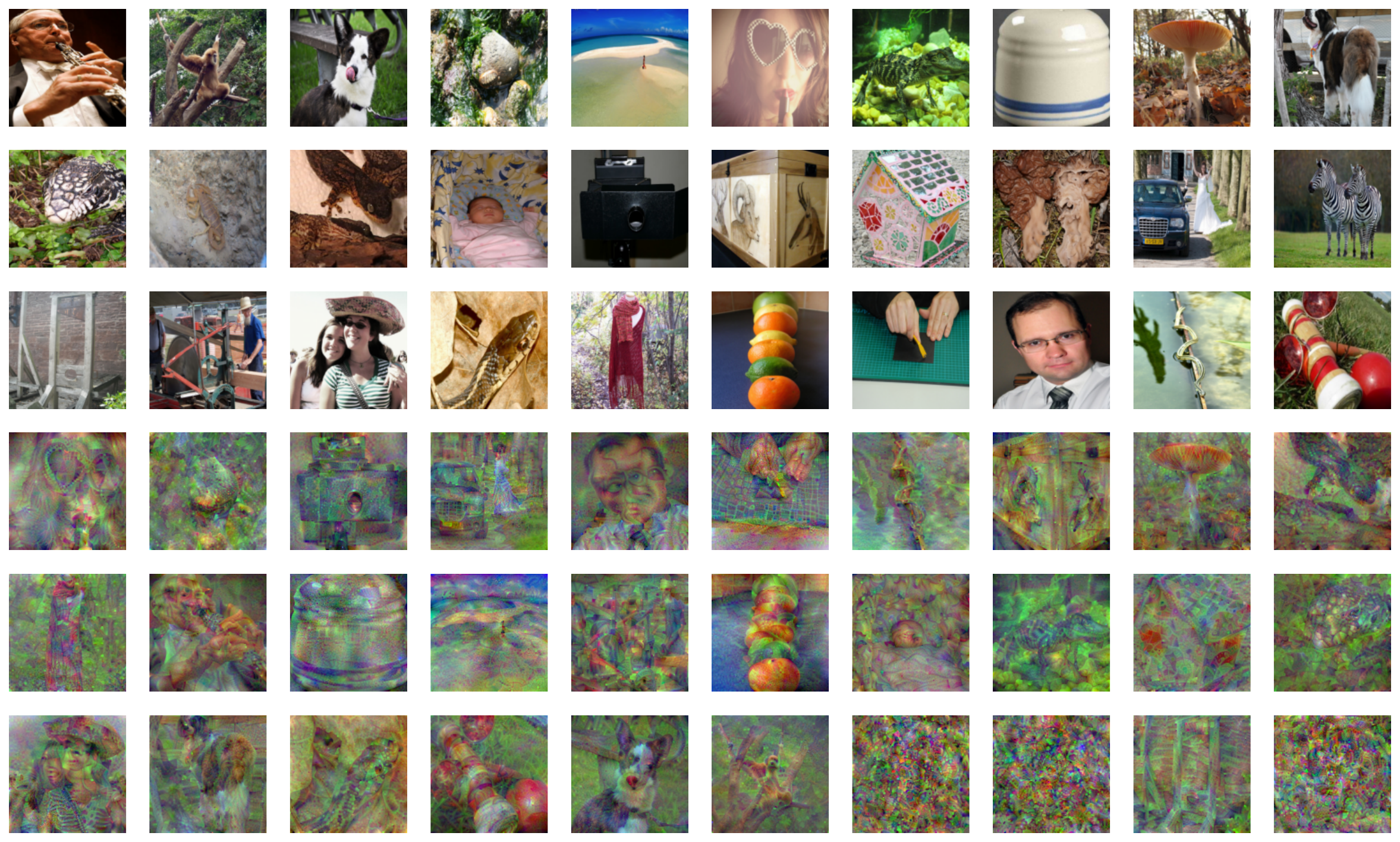}
    \caption{The outcomes of CPA-FI attack for 30 images.}
    \label{fig:cpafi}
\end{figure}

However, CPA may not be directly applicable to convolutional neural networks (CNNs), which are commonly used for image classification tasks. To address this limitation, the authors propose using CPA to recover the embedding vector $\mathbf{z}$, generated just before the first fully connected (FC) layer of the CNN model, and then applying feature inversion (FI) techniques \cite{mahendran2015understanding,ulyanov2018deep} to reconstruct the images from $\mathbf{z}$ as shown in Eq. \eqref{eq:fi}:

\begin{equation}
    \label{eq:fi}
    \hat{\mathbf{x}}=\arg\max_{\mathbf{x}^{*}}CS(f(\mathbf{x}^{*}), \mathbf{z})-\lambda_{tv}\mathcal{R}_{tv}(\mathbf{x}^{*})
\end{equation}
where $CS$ is the cosine similarity. The reconstructed results of CPA-FI on a pretrained VGG16 model \footnote{https://www.flyai.com/m/vgg16-397923af.pth} \cite{simonyan2014very} for gradients with respect to a batch of 30 ImageNet images are shown in Fig. \ref{fig:cpafi}, where the last 3 rows are reconstructed images. It is easy to find that the quality of recovered dummy images is suboptimal, with only a portion of the images being recognizable, even when the attacked VGG16 model is pretrained.

In summary, optimization-based attack approaches reconstruct images by minimizing the distance between the gradients of the ground-truth images and those of randomly initialized dummy images. However, in real FL systems, gradients with respect to the local data are not directly accessible, and gradient inversion attacks may struggle to perform effectively with larger images.

\paragraph{Analytic Attack}
unlike optimization-based attack methods, this type of approach focuses on tampering with the model architecture and decomposing gradients through the observation of multiple communication rounds in FL. Phong et al. \cite{phong2017privacy} were the first to discuss analytic attacks, leveraging the key insight that the input of a learnable affine function can be recovered by dividing the gradients of the weights and biases. Geiping et al. \cite{geiping2020inverting} conducted a theoretical analysis to demonstrate that the input to a neural network can be uniquely reconstructed from the gradients of the fully connected layer. Enthoven et al. \cite{enthoven2022fidel} proposed a mechanism to fully reveal privacy in a MLP neural network trained on a singular sample.

As introduced in \cite{phong2017privacy, geiping2020inverting, qian2020minimal}, the feed-forward pass of the first fully connected layer of a MLP neural network can be defined as:

\begin{equation}
    \mathbf{Y}_{i} = \mathbf{X}_{i}\mathbf{W} + \mathbf{b}
\end{equation}
where $\mathbf{X}_{i} \in \mathbb{R}^{n_{0}}$ is the $i$-th input data sample,  $\mathbf{W} \in \mathbb{R}^{n_{0} \times n_{1}}$ and $\mathbf{b} \in \mathbb{R}^{n_{1}}$ are the weights and bias of the network layer, respectively. Then, the derivative of the training loss $\mathcal{L}$ with respect to the $l$-th column weights $\mathbf{W}_{l}$ can be calculated as:
\begin{equation}
     \frac{\partial{\mathcal{L}}}{\partial{\mathbf{W}_{l}}} = \frac{\partial{\mathcal{L}}}{\partial{\mathbf{Y}_{i,l}}} \cdot \frac{\partial{\mathbf{Y}_{i,l}}}{\partial{\mathbf{W}_{l}}}= \frac{\partial{\mathcal{L}}}{\partial{\mathbf{Y}_{i,l}}} \cdot \mathbf{X}_{i}
\end{equation}
wherein $\frac{\partial{\mathcal{L}}}{\partial{\mathbf{Y}_{i,l}}}$ can be easily achieved by the following equation: 
\begin{equation}
    \frac{\partial{\mathcal{L}}}{\partial{\mathbf{Y}_{i,l}}} = \frac{\partial{\mathcal{L}}}{\partial{\mathbf{Y}_{i,l}}} \cdot \frac{\partial{\mathbf{Y}_{i,l}}}{\partial{\mathbf{b}_{l}}} = \frac{\partial{\mathcal{L}}}{\partial{\mathbf{b}_{l}}}    
\end{equation}
Therefore, the input sample $\mathbf{X}_{i}$ of the first fully connected layer can be recovered as:
\begin{equation}
\label{eq:analyticattack}
    \mathbf{X}_{i} = \frac{\partial{\mathcal{L}}}{\partial{\mathbf{W}_{l}}} \cdot {(\frac{\partial{\mathcal{L}}}{\partial{\mathbf{Y}_{i,l}}})}^{-1} = \frac{\partial{\mathcal{L}}}{\partial{\mathbf{W}_{l}}} \cdot {(\frac{\partial{\mathcal{L}}}{\partial{\mathbf{b}_{l}}})}^{-1}
\end{equation}

However, the aforementioned analytic attack approach is limited to MLP neural networks with one hidden layer and requires the presence of a non-zero bias. Additionally, it is only applicable when the batch size is restricted to one. Fan et al. \cite{fan2020rethinking} extended the previous attack from MLP neural networks to Secret Polarization networks. Zhu et al. \cite{zhu2020r} introduced R-GAP, employing refined rank analysis to explain attack performance and identify network architectures that support full recovery, while also proposing its use to modify these architectures. However, R-GAP does not address batch input size limitations. To overcome this, Wen et al. \cite{wen2022fishing} developed a strategy that accommodates arbitrarily large batch sizes by adjusting model parameters to amplify the gradient contributions of target data while reducing those of other data.

\begin{figure}[!htbp]
    \centering
    \includegraphics[width=0.46\textwidth]{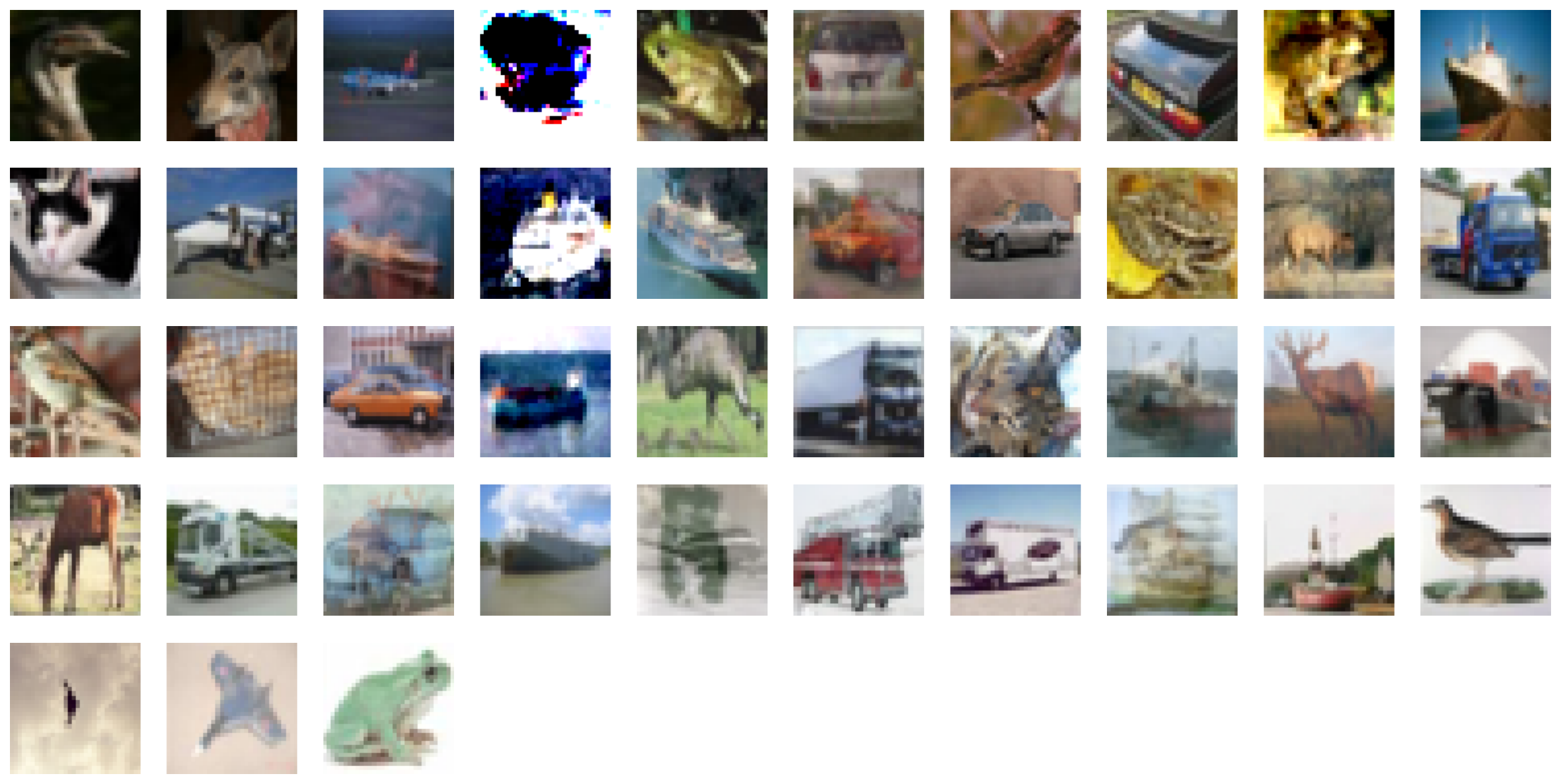}
    \caption{The reconstructed images of RTF attack.}
    \label{fig:rtf}
\end{figure}

Fowl et al. (Robbing the Fed, RTF) \cite{fowl2022robbing} further modified the model architecture in FL by inserting an additional imprint module before the original learning model. This imprint module consists of a single fully connected layer followed by a ReLU activation function, and its feed-forward computation for a single image $\mathbf{X}_{i}$ is computed as follows:

\begin{equation}
    M(\mathbf{X}_{i})=\sigma \left(\mathbf{X}_{i}\mathbf{W} + \mathbf{b} \right)
\end{equation}
where $\sigma$ represents ReLU activation function. For the $l$-th column of imprint model weights $\mathbf{W}$ and $l$-th entry of model bias $\mathbf{b}$, the linear combination (brightness for image data) can be defined as $h(\mathbf{X}_{i})=\left\langle \mathbf{X}_{i}, \mathbf{W}_{l} \right\rangle$. Given that $h(\mathbf{X}_{i})$ follows a Gaussian distribution, the bias of the imprint module is determined by the inverse of the standard Gaussian CDF $\Phi^{-1}$:

\begin{equation}
    \mathbf{b}_{l}=-\Phi^{-1} \left( \frac{l}{n_{1}} \right)
\end{equation}
where $n_{1}$ is the number of bins (neurons) of the first hidden layer. This is designed to separate individual images from the aggregated gradients of the model weights. Due to the characteristics of the ReLU activation function, the gradients of the model weights are non-zero only when the brightness $h(\mathbf{X}_{i}) > -\mathbf{b}_{l}$ is satisfied. Then, some specific images, e.g. $\mathbf{X}_{{i}'}$, with brightness $-\mathbf{b}_{l} \le h(\mathbf{X}_{{i}'}) \le -\mathbf{b}_{l+1}$ can be recovered as below:

\begin{equation}
\label{eq:rtf}
\begin{aligned}
    \mathbf{X}_{{i}'}&=\left( \frac{\partial{\mathcal{L}}}{\partial{\mathbf{W}_{l}}}-\frac{\partial{\mathcal{L}}}{\partial{\mathbf{W}_{l+1}}} \right) \oslash \left( \frac{\partial{\mathcal{L}}}{\partial{\mathbf{b}_{l}}} - \frac{\partial{\mathcal{L}}}{\partial{\mathbf{b}_{l+1}}} \right) \\
    &=\mathbf{X}_{{i}'}+\sum_{i \in -{i}'}\mathbf{X}_{i}-\sum_{i \in -{i}'}\mathbf{X}_{i}
\end{aligned}
\end{equation}
where $\sum_{i \in -{i}'}\mathbf{X}_{i}$ is the summation of the images from the batch with brightness $> -\mathbf{b}_{l}$. Note that, this approach does not require label recovery, and the number of reconstructed images (including failed recovery attempts) is equal to the number of bins $n_{1}$ regardless of both the training batch size and the reconstruction batch size. The reconstructed dummy images by RTF for gradient attack are shown in Fig. \ref{fig:rtf}.

\begin{figure}[!htbp]
    \centering
    \includegraphics[width=0.46\textwidth]{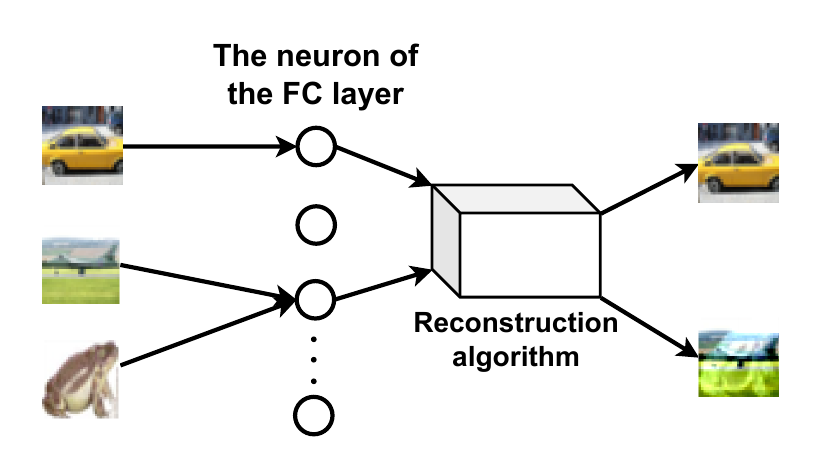}
    \caption{An example of linear layer leakage.}
    \label{fig:linear_layer_leakage}
\end{figure}

Moreover, Fowl et al. \cite{fowl2022decepticons} also employ the idea of RTF to recover the sentences from the shared gradients. However, data can only be accurately reconstructed when a single input activates a neuron in the fully-connected layer. As shown in Fig. \ref{fig:linear_layer_leakage}, when multiple inputs activate the same neuron, the reconstruction becomes a combination of these inputs, causing the attack to fail. To address this issue, Zhao et al. \cite{zhao2023loki,zhao2023secure} modified the attack module inserted at the beginning of the model. By introducing a convolutional scaling factor, they reduced the likelihood of multiple inputs simultaneously activating a neuron. Additionally, the size of the first fully-connected layer was adjusted based on the batch size or data size to further mitigate this problem.

Nevertheless, such attacks may raise client awareness due to the consistency of the modifications. To address this, Boenisch et al. \cite{boenisch2023curious} introduced the concept of trap weights, which modify the shared model weights to amplify the inherent data leakage from gradients. Lam et al. \cite{lam2021gradient} proposed a method to isolate individual updates by seperating the aggregated gradients. Similarly, Pasquini et al. \cite{pasquini2022eluding} introduced an attack in which the adversary sends different models to each client, ensuring that the shared gradients correspond to individual updates rather than aggregated ones.

In summary, analytic attacks demonstrate superior performance in generating high-quality dummy images without relying on label inference. And the simulation results presented in Section \ref{sec:serverattack} can also empirically confirm that, even in realistic FL scenarios where averaged gradients of batch data are locally computed and updated multiple times, the RTF attack remains effective in recovering high-quality dummy images. However, analytic approaches tend to perform poorly on neural network models that deviate from the MLP structure or modify the model architecture, leading to degraded FL performance.

\begin{figure}[!htbp]
    \centering
    \includegraphics[width=0.46\textwidth]{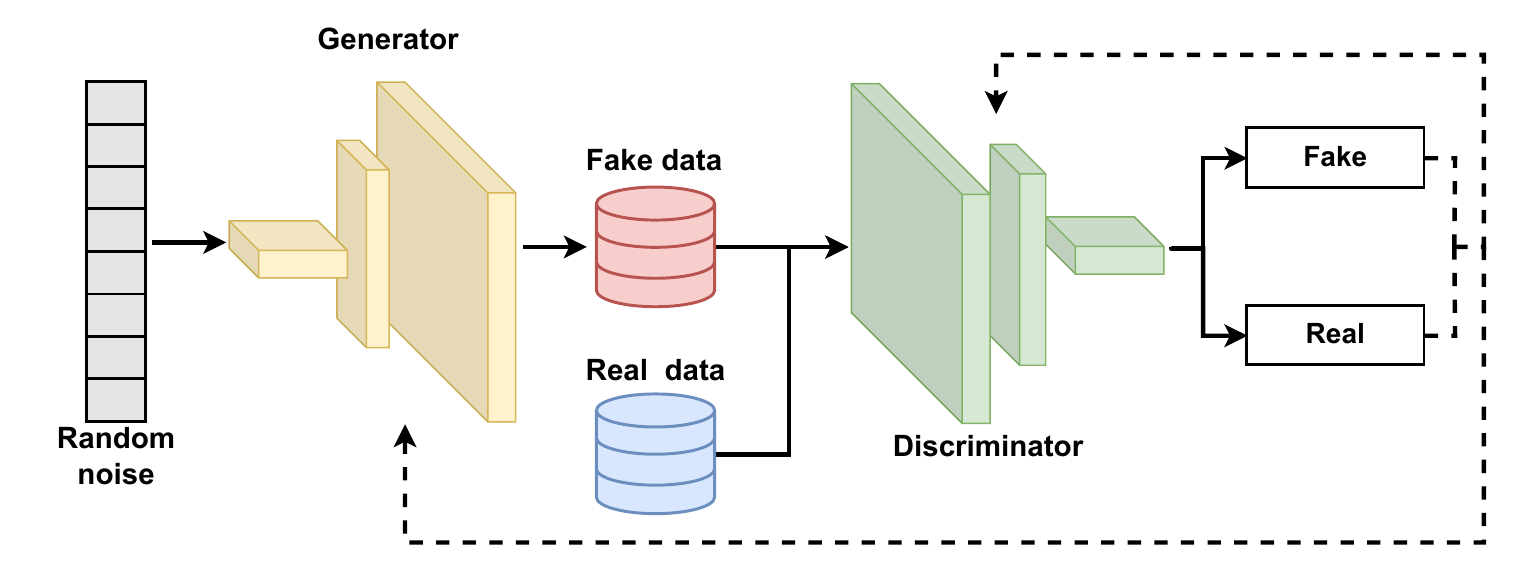}
    \caption{The structure and the training procedure of GAN}
    \label{fig:gan}
\end{figure}

\paragraph{GAN-based Attack}
Generative Adversarial Networks (GANs), first proposed by Goodfellow \cite{goodfellow2014generative}, consist of two distinct neural networks: a generator that creates data and a discriminator that evaluates it. During the GAN training process in Fig. \ref{fig:gan}, these two networks engage in a competitive dynamic, with both improving iteratively until the generator produces outputs that are virtually indistinguishable from the real data.

A typical loss function for the GAN training is described in the following equation:
\begin{equation}
\label{eq:GANtrain}
\min_{G}\max_{D}\mathcal{L}(D,G) = \mathbbm{E}\left [\log{D(\mathbf{x})}\right ] +\mathbbm{E}\left [ \log{(1-D(G(\mathbf{z})))}  \right ]
\end{equation}
where $\mathbf{x}$ is the real image, $\mathbf{z}$ is the random noise, $D$ is the discriminator and $G$ is the generator. In recent years, GAN-based privacy attacks have advanced significantly in centralized machine learning scenarios \cite{hayes2017logan, ha2022inference}. Inspired by the work proposed by Yang et al. \cite{yang2019adversarial}, Zhang et al. introduced the novel Generative Model Inference (GMI) attack \cite{zhang2020secret}, leveraging a GAN trained on auxiliary datasets to recover sensitive regions in images. The images generated by the GAN are processed by two distinct discriminators to compute the prior loss and identity loss. These losses, combined with the corrupted image, are then used as inputs for the generator to refine and produce subsequent iterations of the reconstructed image. And then, Salem et al. \cite{salem2020updates} proposed a hybrid generative model that incorporates reconstructive loss to target black-box machine learning models.

In addition, this generator-discriminator structure of GAN can also be leveraged to launch privacy attacks in FL. Each client can covertly construct its own generator to infer private data from other participants, exploiting the shared model parameters to generate synthetic reconstructions of sensitive information. Hitaj et al. \cite{hitaj2017deep} conducted one of the earliest explorations of client-side attacks using GAN-based methods, successfully recovering targeted images from other participating clients in FL. In their approach, the shared global model acts as the discriminator, with modifications made to the model structure to include an additional class for recovering fake images by the locally build generator. The corresponding adversarial FL training steps are shown in Algorithm \ref{alg:dmgan}.


\begin{algorithm}[!htbp]
\renewcommand{\algorithmicrequire}{\textbf{Input:}}
\renewcommand{\algorithmicensure}{\textbf{Output:}}
\caption{Deep Models Under the GAN} \label{alg:dmgan}
\begin{algorithmic}
\State \textbf{Server Side:}
\State Initialize the global model $\mathbf{W}_{0}$ \Comment{Discriminator $D$}
\For{each communication round $r=0,1,...,R-1$}
\For{each Client $k=1,2,...,K$ \textbf{in parallel}}
\State Download  $\mathbf{W}_{r}$ to Client $k$
\If{Client $k$ is the attacker}
\State Set tracked label class  ${c}'$
\State Perform GAN training using Eq. \eqref{eq:GANtrain}
\State Generate dummy images $\hat{\mathbf{X}}$ by the generator $G$
\EndIf
\State Client $k$ performs local training
\State Upload trained local model $\mathbf{W}^{k}$ to the server
\EndFor
\State $\mathbf{W}_{r+1} \leftarrow \sum_{k=1}^{K}\frac{n^{k}}{n} \mathbf{W}^{k}$ \Comment{FedAvg}
\EndFor
\end{algorithmic}
\end{algorithm}

\begin{figure}[!htbp]
    \centering
    \includegraphics[width=0.46\textwidth]{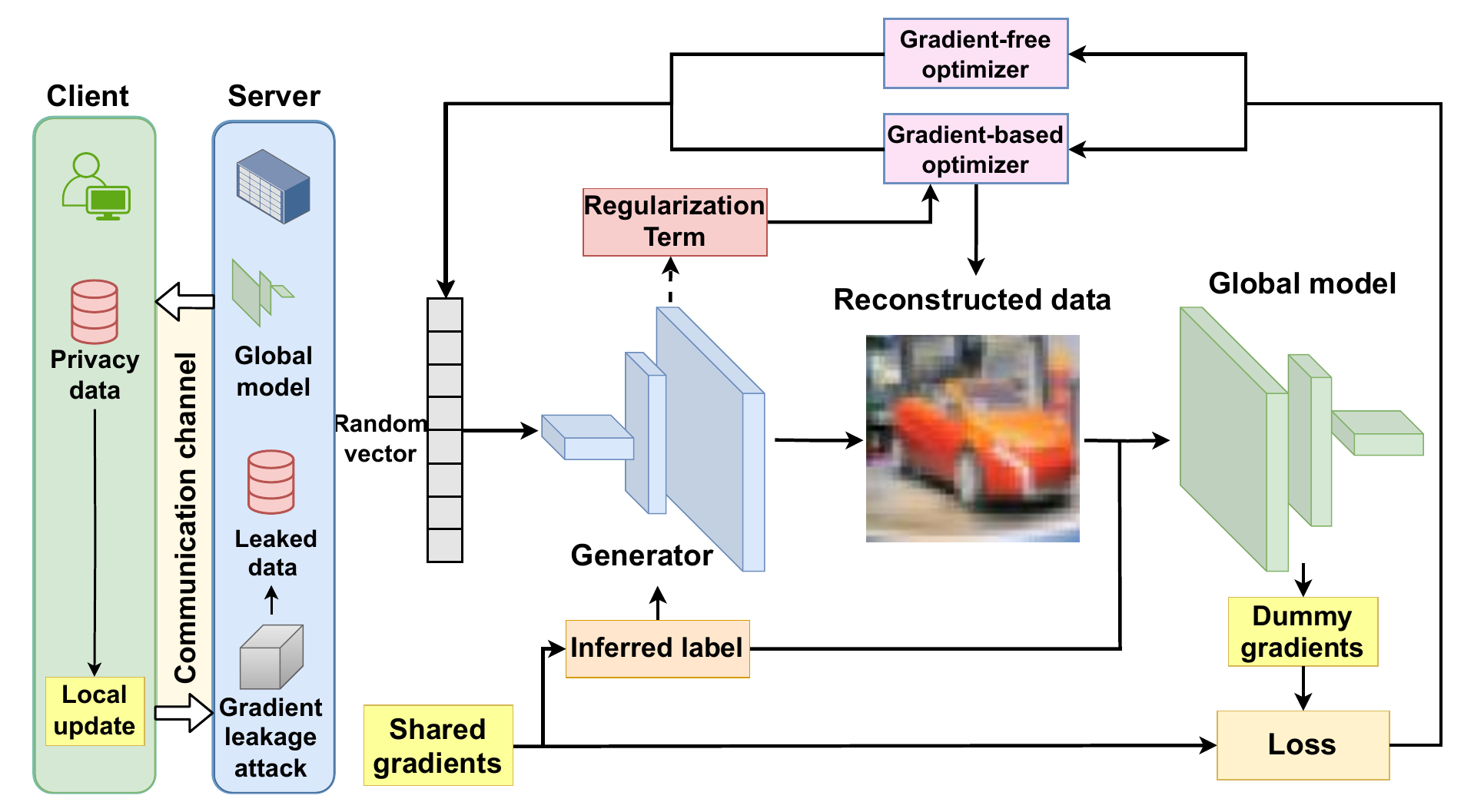}
    \caption{Illustration of GGL}
    \label{fig:ggl}
\end{figure}

Building upon the aforementioned method, Wang et al. \cite{wang2019beyond} proposed a mGAN-AI method, which employs a multitask discriminator on server capable of simultaneously discriminating category, reality, and client identity of input samples. The novel incorporation of client identity discrimination enables the generator to recover private data specific to the targeted user. In addition to utilizing the GAN's generator to indirectly compromise user data for privacy attacks, GAN-based methods can be combined with optimization-based attack techniques to further amplify their effectiveness \cite{jeon2021gradient}. Building on this concept, Li et al. \cite{li2022auditing} proposed the Generative Gradient Leakage (GGL) algorithm, which leverages GANs in a novel way. Unlike DLG, which directly optimizes dummy images, GGL optimizes lower-dimensional random noise input to a pretrained GAN generator. This approach effectively reduces the search space, leading to a higher success rate in recovering sensitive data. As shown in Fig. \ref{fig:ggl}, GGL also adopts gradient-free optimizer, such as Adaptation Evolution Strategy (CMA-ES) \cite{hansen2016cma} or Trust Region Bayesian Optimization (TuRBO) \cite{eriksson2019scalable}, to avoid convergence to local minima.

\begin{figure}[!htbp]
    \centering
    \includegraphics[width=0.46\textwidth]{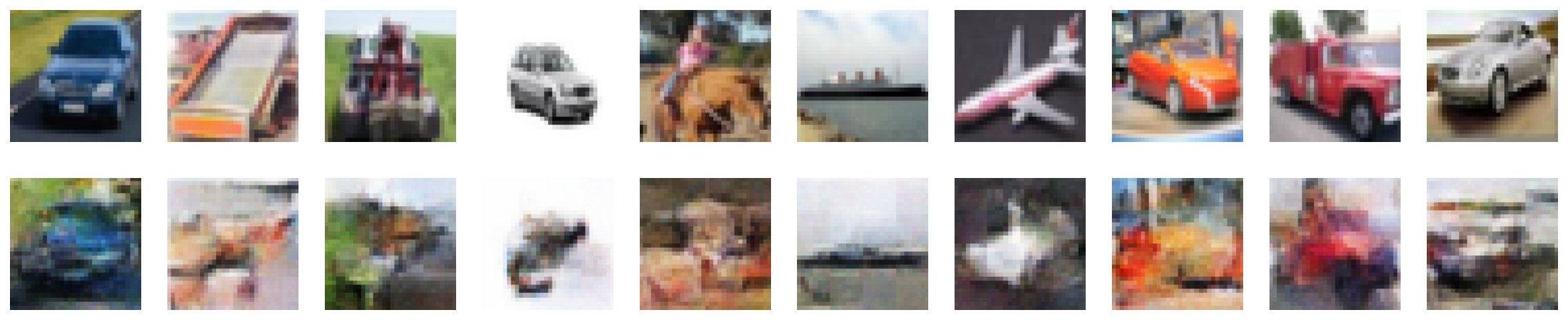}
    \caption{The outcomes of gradient attack by GGL}
    \label{fig:gglgradout}
\end{figure}

The simulation outcomes of GGL for single data gradient attack are shown in Fig. \ref{fig:gglgradout}, where the first row represents 10 original images and the second row is the corresponding dummy images for 10 runs. It is noteworthy that the reconstructed images exhibit some distortion compared to the ground-truth images, likely due to the influence of the internal mechanisms of the pretrained generator during the recovery process. 

\begin{figure}[!htbp]
    \centering
    \includegraphics[width=0.46\textwidth]{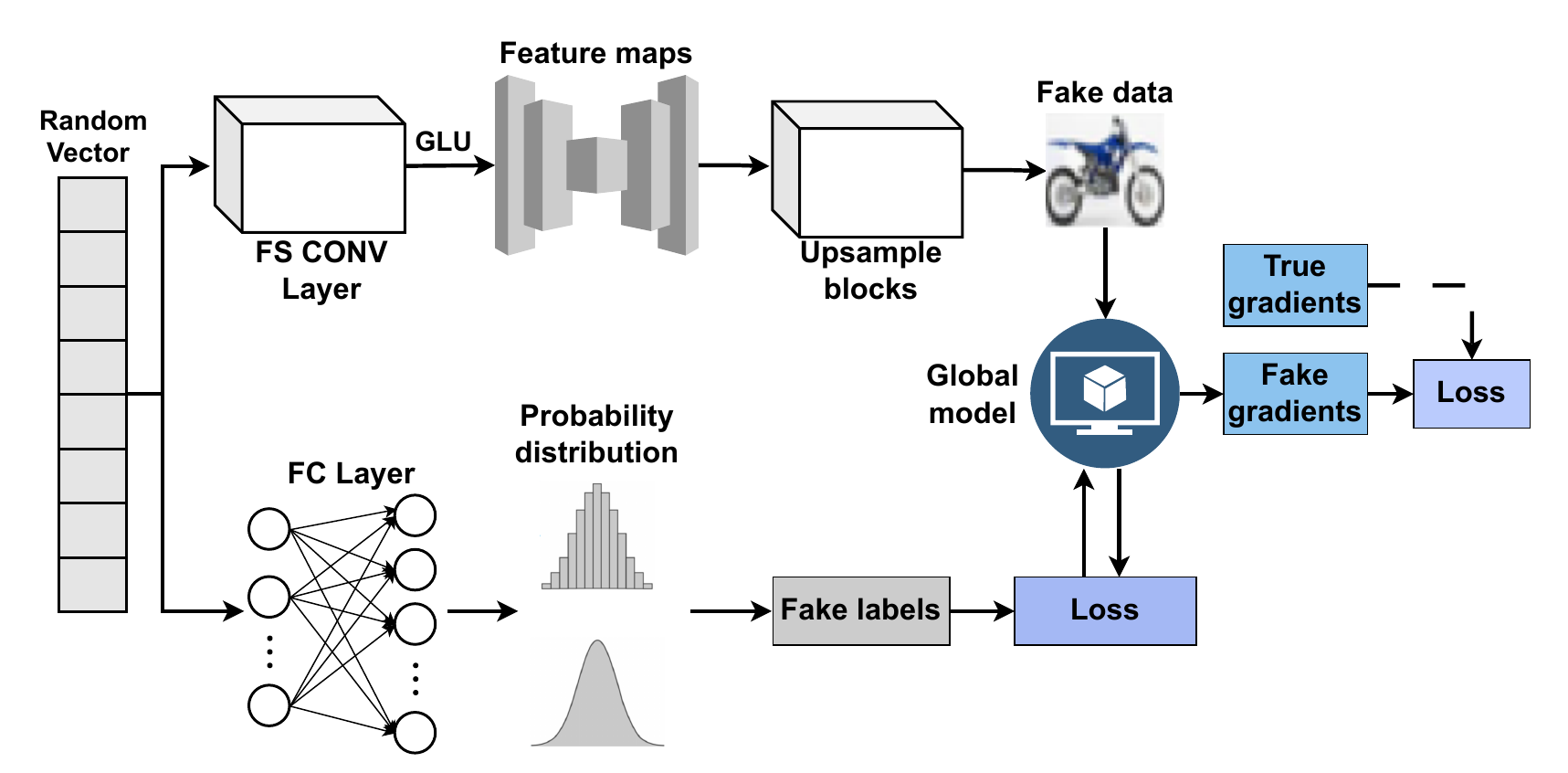}
    \caption{Training process of GRNN.}
    \label{fig:grnn}
\end{figure}

To address this issue, Ren et al. \cite{ren2022grnn} proposed the Generative Regression Neural Network (GRNN) algorithm, which generates both the image prior and corresponding label using a generative network to find inputs that satisfy the gradient-matching objective. As depicted in Fig. \ref{fig:grnn}, GRNN, optimizes the generator's model parameters rather than the input noise, eliminating the need for the generator to be pretrained. A brief training procedure of GRNN for gradient attack is also presented in Algorithm \ref{alg:grnn}, where the gradients $\nabla \mathbf{W}_{g}$ with respect to the global model weights $\textbf{W}$ are computed upon the ground-truth data pair $(\mathbf{x}, y)$. And the loss function $\mathcal{L}_{\mathbf{W}}$ includes the L2-norm used in DLG, the Wasserstein distance loss, and the total variation loss.

\begin{algorithm}[!htbp]
\renewcommand{\algorithmicrequire}{\textbf{Input:}}
\renewcommand{\algorithmicensure}{\textbf{Output:}}
\caption{GRNN} \label{alg:grnn}
\begin{algorithmic}
\Require the global model weights $\mathbf{W}$, model weights of the generator $G$ $\mathbf{W}_{G}$, the total training iteration $T$ 
\Ensure dummy data pair $(\hat{\mathbf{x}}, \hat{y})$
\State Receive the gradients of the ground-truth data $\nabla \mathbf{W}_{g}$
\State Initialize the input vector: $\mathbf{z} \leftarrow \mathcal{N}(0,1)$
\For{$t = 1$ to $T$}
\State Generate dummy data: $(\hat{\mathbf{x}}^{t}, \hat{y}^{t}) \leftarrow G \left(\mathbf{z}|\mathbf{W}^{t}_{G} \right)$
\State Compute gradients: $\nabla \mathbf{W}_{d}^{t} \leftarrow \partial \mathcal{L}_{\mathbf{W}}(\hat{\mathbf{x}}^{t}, \hat{y}^{t}) / \partial \mathbf{W}$
\State Update $G$: $\mathbf{W}_{G}^{t} \leftarrow \mathbf{W}_{G}^{t} - \eta \frac{\partial \mathcal{L}_{\mathbf{W}} \left( \nabla \mathbf{W}_{d}^{t}, \nabla \mathbf{W}_{g} \right)}{\partial \mathbf{W}_{G}^{t}}$
\EndFor
\State \textbf{return} dummy data pair $(\hat{\mathbf{x}}, \hat{y}) \leftarrow G \left(\mathbf{z}|\mathbf{W}^{T}_{G} \right)$
\end{algorithmic}
\end{algorithm}

\begin{figure}[!htbp]
    \centering
    \includegraphics[width=0.46\textwidth]{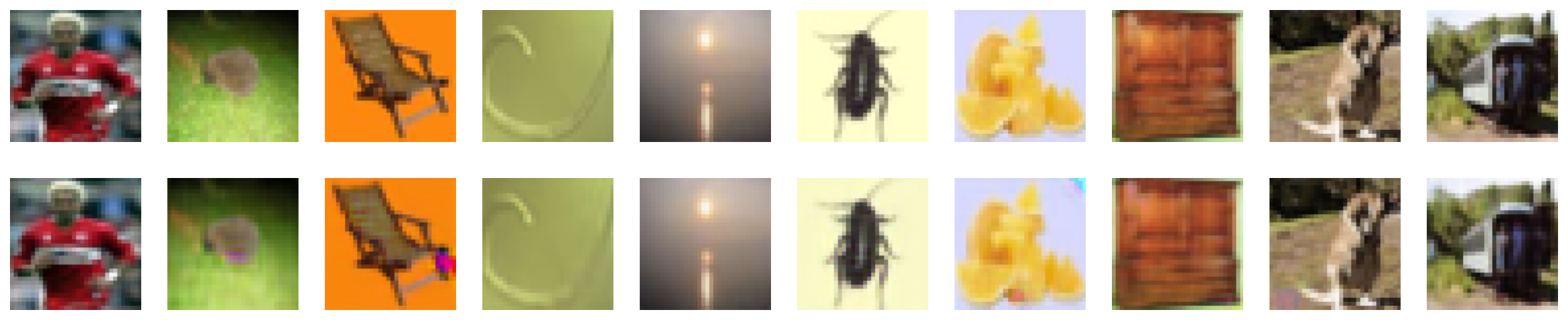}
    \caption{The outcomes of gradient attack by GRNN}
    \label{fig:grnnoutcomes}
\end{figure}

The respective simulation outcomes for 10 repeated single gradient attack by GRNN are shown in Fig. \ref{fig:grnnoutcomes}, with the second row displaying the reconstructed images.  It is important to note that the quality of these recovered images is highly sensitive to data normalization, and the results in Fig. \ref{fig:grnnoutcomes} using unnormalized data are unrealistic for normal model training.

To sum up, there are two primary approaches to utilizing GANs for privacy attacks. The first involves treating the global model as the discriminator, with the attacker constructing a generator to mimic and extract private data from the discriminator. However, this method requires modifying the global model by adding an additional output neuron, which is impractical in FL systems. The second approach integrates the GAN's generator with optimization-based attack methods, aiming to reduce the search space and improve the attack's success rate. Nonetheless, this approach often relies on pretrained generators trained on related images, which is also unrealistic in practical FL scenarios.

\section{Experiments}\label{experiments}
The purpose of this experimental study conducted here is to empirically assess the extent to which current representative privacy attack methods can recover private client images in a real FL environment. Unlike previously described single gradient attack, the uploaded model parameters from clients are computed over multiple steps of updates by averaged gradients with respect to varied batch data. And in this section, we first introduce the configurations and implementation of the privacy attacks. Then, we present the experimental results and corresponding analysis.

\subsection{Datasets}
Our experiments utilize five image classification datasets: MNIST \cite{deng2012mnist}, CIFAR-10 \cite{krizhevsky2009learning}, CIFAR-100, Tiny ImageNet \cite{le2015tiny} and ImageNet ((ILSVRC2012)) \cite{ILSVRC15}. The key attributes of these datasets are summarized in Table \ref{tab:dataset}.

MNIST is a gray-scale image dataset containing 60000 training and 10000 testing 28x28 handwritten digits. And CIFAR-10 is a RGB image dataset consisting of 50000 training and 10000 testing 32x32x3 color images with 10 different types of classification objects. Therefore, CIFAR-10 is a more challenging dataset compared to MNIST. Similarly, CIFAR-100 consists of 50,000 training images and 10,000 testing images, each 32x32x3 pixels in RGB format, featuring 100 distinct object classes for classification. Note that, from the perspective of image recovery attacks, CIFAR-10 and CIFAR-100 present similar levels of difficulty due to their comparable image resolution and structure, despite CIFAR-100 having more classification categories.

Tiny ImageNet dataset is a more complex image dataset consisting of 100,000 training, 10,000 validation and 10,000 testing 64x64x3 images with 200 different kinds of classification objects. ImageNet is long-standing landmark in computer vision, containing more than 10 million large images designed for classifying objects into 1,000 categories with minimal error. While the images vary in size, they are typically resized to 224x224x3 pixels for inputs into the learning model. It is noteworthy that larger-sized images, such as those in the ImageNet dataset, are more difficult to be recovered by privacy attacks compared to smaller images due to their increased complexity and higher resolution.

\begin{table}[ht]
\caption{Description of the Dataset}
\centering
\label{tab:dataset}
\resizebox{0.46\textwidth}{!}{
\begin{tabular}{llllll}
    \hline
        Dataset & MNIST & CIFAR-10 & CIFAR-100 & Tiny ImageNet & ImageNet \\ 
    \hline
        Channel & 1 & 3 & 3 & 3 & 3 \\
        Class & 10 & 10 & 100 & 200 & 1000 \\
        Size & 70,000 & 60,000 & 60,000 & 120,000 & 1,431,167 \\ 
        Categories & digits  & objects & objects & objects & objects \\
    \hline
\end{tabular}}
\end{table}

To simulate data distribution in FL experiments, all training data are evenly and randomly allocated to participating clients without overlap in IID scenarios. For Non-IID cases, each client is allocated a portion of the training data for each label class based on a Dirichlet distribution, $p_{c}\sim \text{Dir}_{k}\left( \beta \right)$, where $\beta=0.5$ is the concentration parameter. A smaller $\beta$ value results in a more unbalanced data partition, and vice versa. Additionally, 20\% of the allocated data on each client is used for testing, with the remaining 80\% used for training in all experiments.

\subsection{Models}
Several types of neural networks are used in this experimental study. One is a CNN model consisting of two $3 \times 3$ convolutional kernels with 32 and 64 output channels, respectively, followed by a hidden fully connected layer with 512 neurons. This model is specifically applied to CIFAR datasets and is capable of achieving reasonable learning performance in FL. Unless otherwise specified, most of the tested attack algorithms are evaluated using this model. Another one is a simple MLP neural network containing one hidden layer with 256 neurons, designed for conducting CPA attack to recover inputs in a blind source separation (BSS) problem. VGG16 and ResNet18 are also deployed in this study, with VGG16 used for feature inversion attacks and ResNet18 applied for linear leakage attacks (RTF).

\subsection{FL Settings}
To construct an environment that closely simulates privacy attacks in a realistic Federated Learning (FL) setting, the following FL parameters are listed as follows: 

\begin{itemize}
    \item Total number of clients: 10 \& 100
    \item Number of local epochs: 1, 2 \& 5
    \item Local batch size: 10 \& 50
    \item Local learning rate: 0.1 \& 0.004
    \item Local learning rate decay: 0.95
\end{itemize}

Notably, both local batch size and the number of local training epochs are particularly sensitive to privacy attacks. For simplicity, the number of local epochs is set to 1, thereby reducing the difficulty of mounting a successful attack. However, even under these conditions, most attack methods exhibit limited effectiveness for relatively small local batch size (10). Additionally, increasing the number of clients reduces the local data size per client, which facilitates easier image reconstruction but more difficult distributed model training.

\subsubsection{Evaluated Algorithms}
Nine representative attack algorithms, including 8 server-side attacks and 1 client-side attack, are evaluated and compared in our experiments.

Eight server-side privacy attack:
\begin{enumerate}
    \item \textbf{DLG \cite{zhu2019deep}}: The earliest gradient inversion algorithm which adopts L2-norm as the optimization loss function and L-BFGS \cite{fletcher2000practical} as the optimizer.
    \item \textbf{iDLG \cite{zhao2020idlg}}: Similar to DLG, iDLG also utilizes the L2-norm as the loss function. However, the label is inferred separately using Eq. \eqref{eq:idlg}.
    \item \textbf{Inverting Gradients \cite{geiping2020inverting}}: Different from DLG-based methods, Inverting Gradients adopts cosine similarity as the loss function.
    \item \textbf{GGL \cite{li2022auditing}}: A GAN-based attack approach leverages a pretrained generator to reduce the dimensionality of the reconstruction search space. Gradient-free optimizers, such as CMA-ES and TuRBO, are applied in this algorithm.
    \item \textbf{GRNN \cite{ren2022grnn}}: Different from GGL, GRNN directly optimizes the model parameters of the generator for image recovery. In addition to the L2-norm, Wasserstein distance and total variation loss are also applied.
    \item \textbf{CPA \cite{kariyappa2023cocktail}}: This pioneering work amis to recover high-resolution images from aggregated gradients by optimizing the unmixing matrix. In addition, feature inversion attack is also applied for image recovery on deep CNNs.
    \item \textbf{DLF \cite{geng2021towards}}: This method highlights the significance of correct label inference and considers the scenarios that the model uploads are computed through multiple steps of local batch training.
    \item \textbf{RTF \cite{fowl2022robbing}}: A representative analytical attack recovers images using Eq. \eqref{eq:rtf} without the need to construct an optimization loss, making the process more efficient and direct.
\end{enumerate}

One client-side privacy attack:
\begin{enumerate}
    \item \textbf{DMGAN \cite{hitaj2017deep}}: This is the first work to introduce a gradient attack using GANs, where each client locally constructs a private generator to mimic images from other clients.
\end{enumerate}

\subsection{All Other Settings}
The reconstruction hyperparameters, including reconstruction learning rate and reconstruction epochs, follow the settings of their original papers which are listed in Table \ref{tab:rechyper}. Note that, the reconstruction batch size is set to be the same as the local batch size and larger batch size makes the privacy attack process more challenging.
Other hyperparameters, such as the regularization factor of total variation loss, decorrelation weight, and etc, are also set to be the same as the corresponding papers. You can also easily find them in our open source repository \footnote{https://github.com/hangyuzhu/leakage-attack-in-federated-learning}.

\begin{table}[htbp]
    \centering
     \caption{Some Reconstruction Hyperparameters}
    \begin{tabular}{|c|c|c|c|}
    \hline
    Attack  &  Batch Size &  Learning Rate &  Epochs\\
    \hline
    DLG &  10 & 1.0 & 300 \\
    \hline
    iDLG &  1 & 1.0 & 300 \\
    \hline
    Inverting Gradients & 10 \& 50 & 1.0  & 24000\\
    \hline
    DLF &  10 \& 50 & 0.1 & 400 \\
    \hline
    GGL   & 1 & / & 25000 \\
    \hline
    GRNN   & 10 & 0.0001 & 1000\\
    \hline
    CPA & 10 & 0.001 & 25000\\
    \hline
    RTF   & / & / & /\\
    \hline
    DMGAN  & 1 & / & 10\\
    \hline
    \end{tabular}
    \label{tab:rechyper}
\end{table}

\subsection{Server-Side Privacy Attack} \label{sec:serverattack}
In this section, we systematically evaluate the aforementioned eight server-side privacy attack methods within a more realistic FL environment. For each method, we begin by selecting the easier settings with larger number of total clients and smaller local batch size. If these attacks demonstrate promising reconstruction performance under these conditions, we proceed to evaluate them under more challenging settings with smaller number of total clients and larger local batch size. Conversely, if the attacks perform poorly in the easier settings, further testing under the harder settings is deemed unnecessary. Moreover, all participating client data are set to be non-IID to simulate a more realistic FL environment. This configuration makes reconstruction attacks easier, as the data distribution becomes more concentrated within each client, allowing attackers to exploit the reduced diversity in the local datasets. 

\begin{figure}[!htbp]
    \centering
    \includegraphics[width=0.46\textwidth]{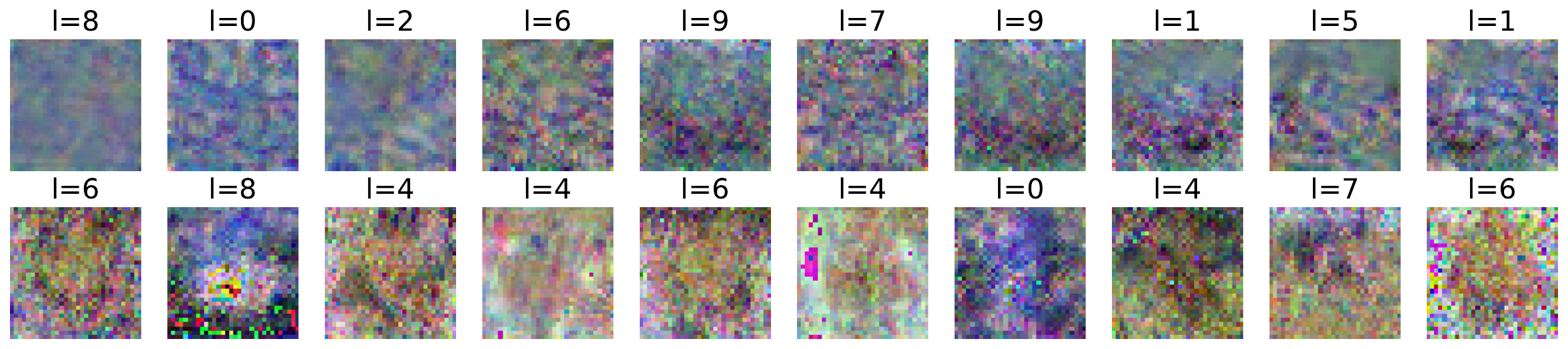}
    \caption{Reconstructed images by DLG in FL}
    \label{fig:dlg_result_fedavg}
\end{figure}

For DLG attack, CIFAR-10 dataset is utilized for simulations, with 100 total clients, resulting in each client owning approximately 500 images. The reconstruction batch size is set to 10, matching the local training batch size. The attack outcomes of a randomly selected client after 2 communication rounds using DLG are illustrated in Fig. \ref{fig:dlg_result_fedavg}, where the top of each image ('l=*') is its respective dummy label. It is evident that the quality of the reconstructed images is significantly worse than those of the gradient attack shown in Fig. \ref{fig:dlg}, making it difficult to recognize the image semantics based on the dummy labels. This is because the uploaded model parameters from each client are computed through multiple updates of averaged local gradients. And the calculated model difference (the global model parameters subtract the updated local model) diverges significantly from the gradients of the batch data.As a result, optimizing the dummy inputs to minimize the distance between the dummy gradients and the model difference leads to the generation of disordered images.

\begin{figure}[!htbp]
    \centering
    \includegraphics[width=0.46\textwidth]{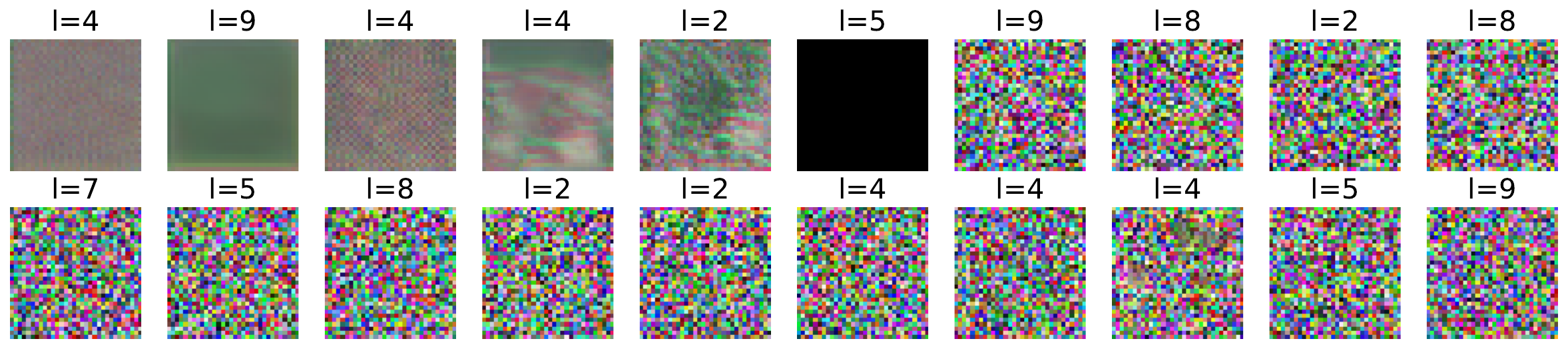}
    \caption{Reconstructed images by iDLG in FL}
    \label{fig:idlg_result_fedavg}
\end{figure}

A similar reconstruction setting is also employed in iDLG attack. However, unlike the direct label optimization used in DLG, iDLG employs label inference (Eq. \eqref{eq:idlg}) to recover the label of a single data point, which limits its ability to simultaneously deduce labels for multiple images. Consequently, the reconstruction batch size is restricted to 1. And the reconstructed images over 20 communication rounds are depicted in Fig. \ref{fig:idlg_result_fedavg}.  In this scenario, the quality of the generated dummy images degrades further, with some images failing to be recovered entirely (e.g., the sixth dark image in the first row of Fig. \ref{fig:idlg_result_fedavg}). This decline in quality is attributed to the limitations of the label inference technique, which is unsuitable for recovering labels from model differences that involve information of multiple local batch images. Thus, incorrect label inference must misguide the direction of the dummy image optimization process.

\begin{figure}[!htbp]
    \centering
    \includegraphics[width=0.46\textwidth]{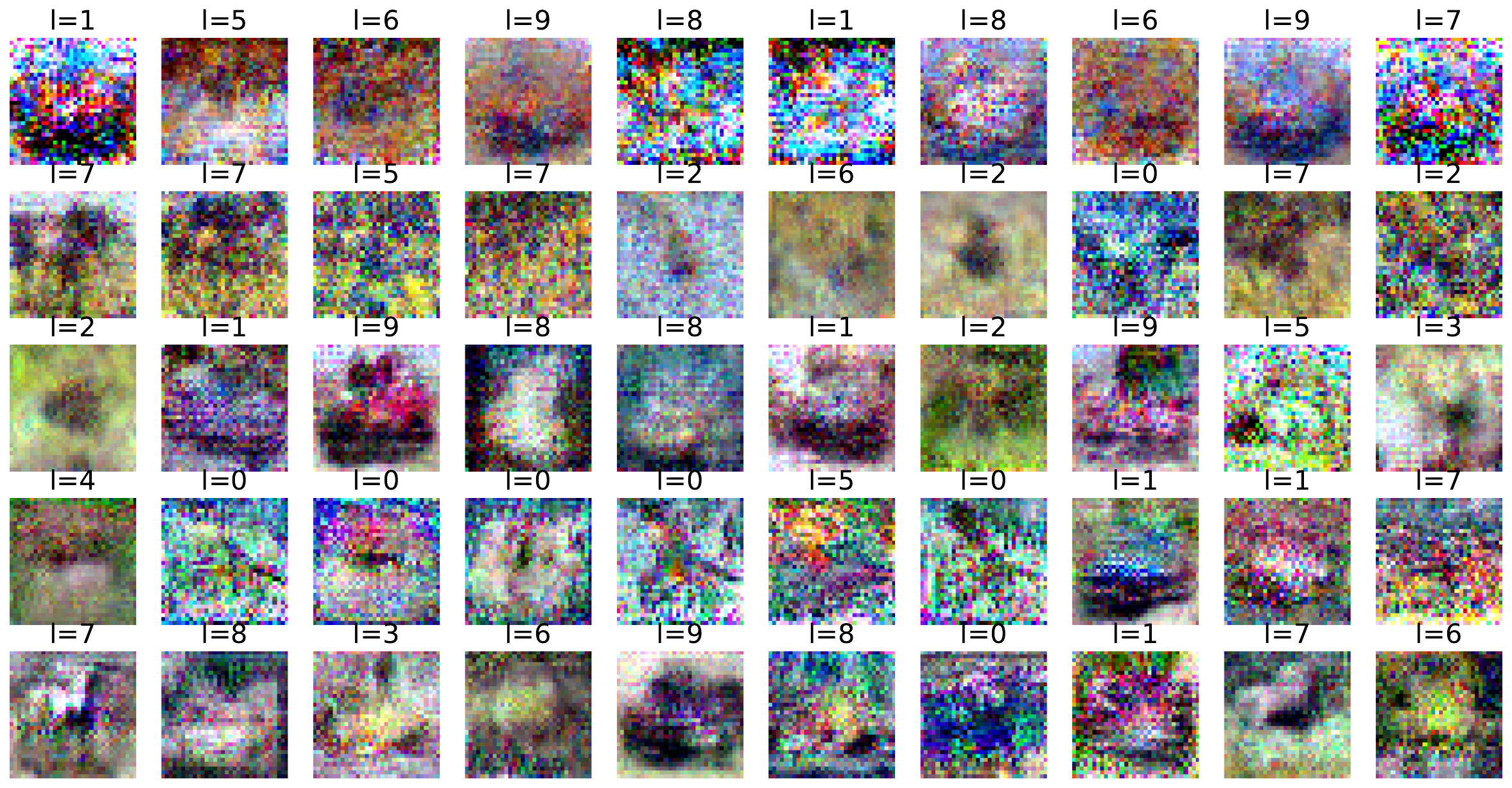}
    \caption{Reconstructed images by Inverting Gradients in FL}
    \label{fig:ig_result_fedavg}
\end{figure}

Unlike DLG-based methods, which perform on gradients from a single image or batch, the Inverting Gradients approach simulates gradient updates across multiple batches of dummy data to capture the corresponding model differences. For a batch size of 10 and 100 participating clients, the simulation results of Inverting Gradients over 5 communication rounds are presented in Fig. \ref{fig:ig_result_fedavg}, where each row represents the reconstructed dummy images at one communication round. The results suggest that the Inverting Gradients attack significantly outperforms the two previous methods, with some reconstructed images being vaguely recognizable. This indicates that Inverting Gradients may represent a leading optimization-based attack approach, and the use of the cosine similarity loss function appears to provide an intrinsic advantage. Consequently, we will evaluate this algorithm under more challenging FL settings in the following section.

\begin{figure}[!htbp]
    \centering
    \includegraphics[width=0.46\textwidth]{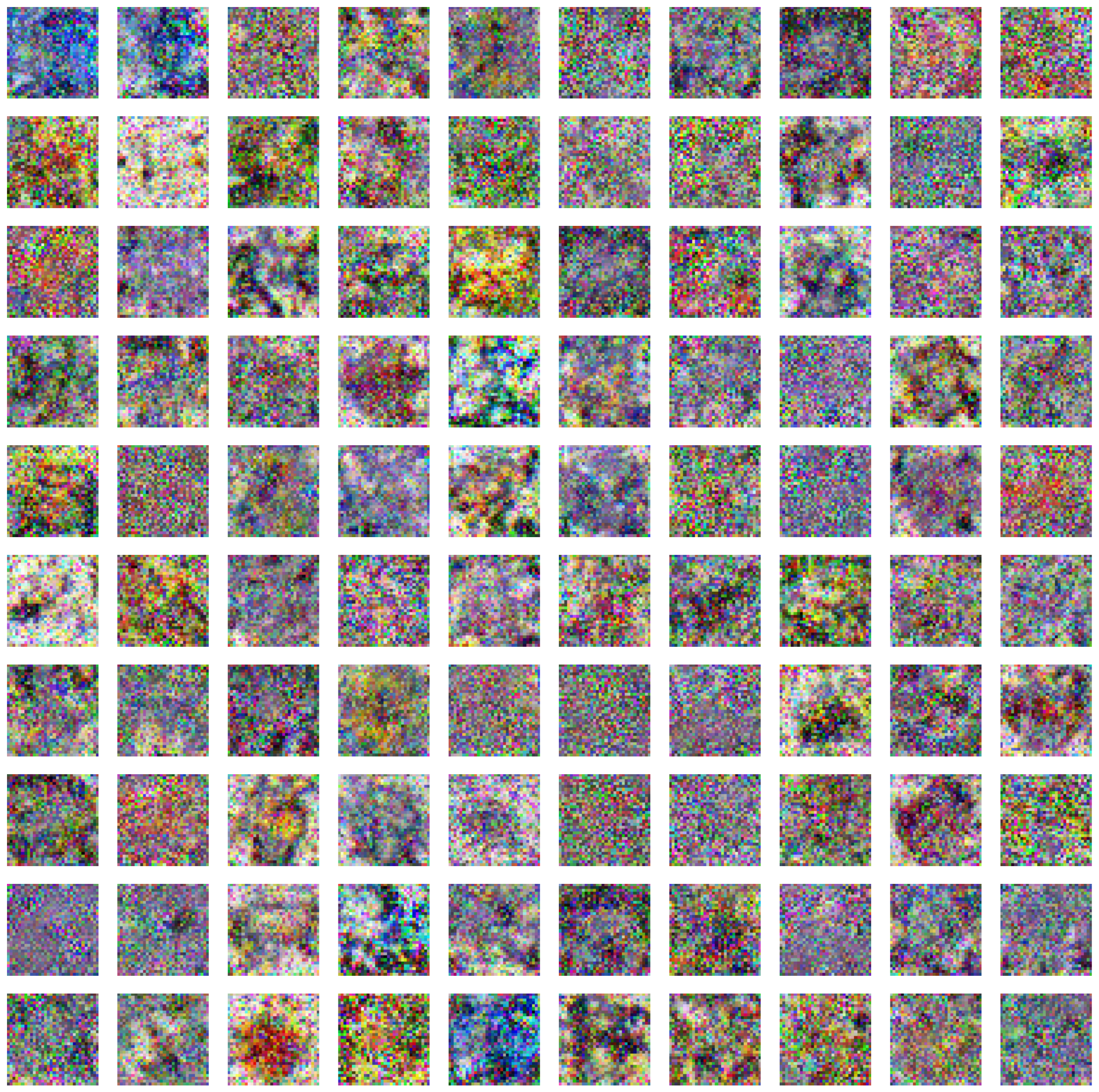}
    \caption{Reconstructed images by DLF in FL}
    \label{fig:dlf_result_fedavg}
\end{figure}

DLF also addresses the issue of multiple local updates in FL. Unlike Inverting Gradients, it employs a linear interpolation technique (in Eq. \eqref{eq:pinter} and Eq. \eqref{eq:llc}) to approximate the label counts across all local training data. However, this technique has inherent limitations, allowing a very small local learning rate (0.004) to be used, and larger learning rate, such as 0.1, would cause running bugs for label count algorithm. The reconstructed images by DLF for 10 batch size and one local epoch on CIFAR-100 are presented in Fig. \ref{fig:dlf_result_fedavg}, where DLF attempts to recover the entire set of local training data on the selected client, but only 100 randomly chosen dummy images are displayed for aesthetic purposes. 
It is surprising to observe that the quality of the reconstructed images is significantly worse than those in Fig. \ref{fig:dlfsmall}, despite both using the same batch size. A potential reason for this discrepancy could be that the DLF algorithm is sensitive to the local data size and the number of local training epochs. Larger client datasets may let the introduced label count inference become less accurate. We will exploit this issue afterwards.

\begin{figure}[!htbp]
    \centering
    \includegraphics[width=0.46\textwidth]{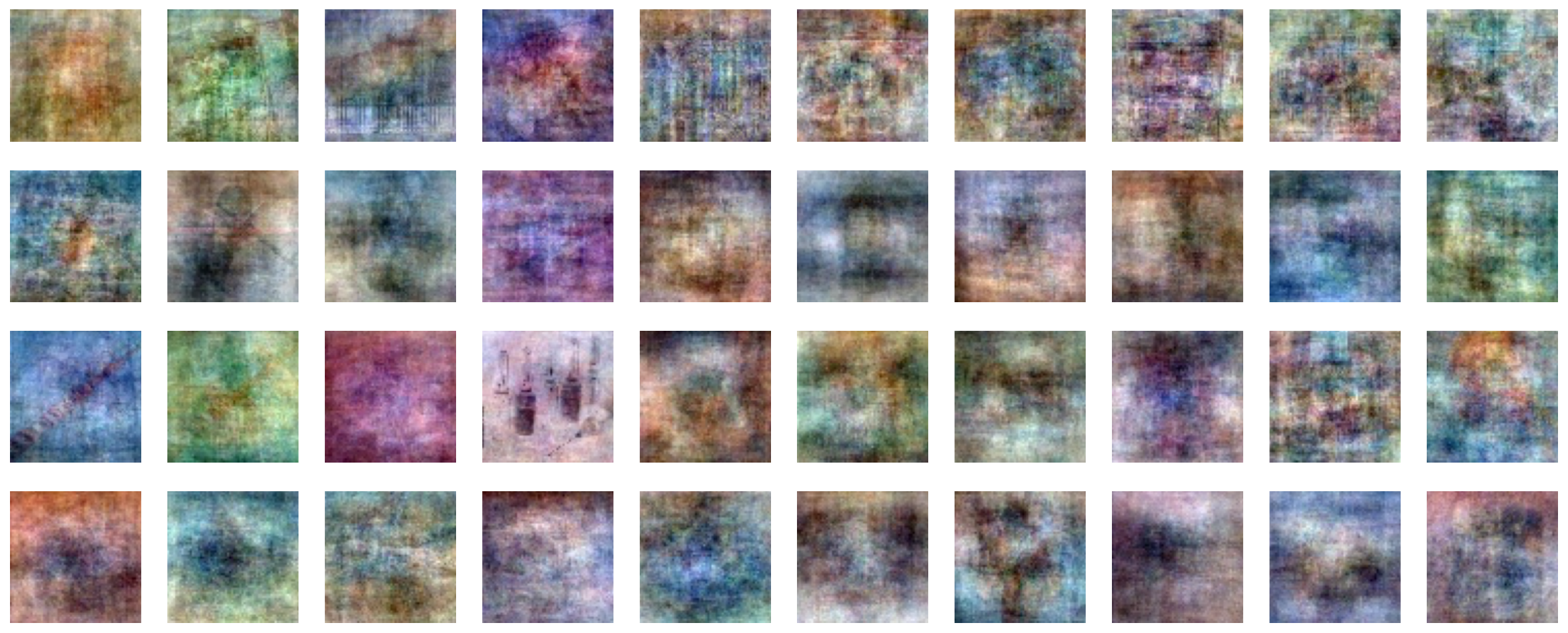}
    \caption{Reconstructed images by CPA in FL}
    \label{fig:cpa_result_fedavg}
\end{figure}

Except that, CPA attempts to use independent component analysis (ICA) to recover larger input images from aggregated gradients. The experimental evaluations of CPA are divided into two parts. The first involves attacking an MLP neural network with a single hidden layer containing 256 neurons by optimizing the unmixing matrix $U$ in the loss function described in Eq. \eqref{eq:cpa1}. The corresponding reconstructed images for 4 communication rounds on Tiny ImageNet are shown in Fig. \ref{fig:cpa_result_fedavg}, where each row is 10 batch dummy images for one round. Since CPA is originally designed for aggregated gradients with respect to a batch of data, it cannot successfully separate and recover individual images from multiple local model updates.

\begin{figure}[!htbp]
    \centering
    \includegraphics[width=0.46\textwidth]{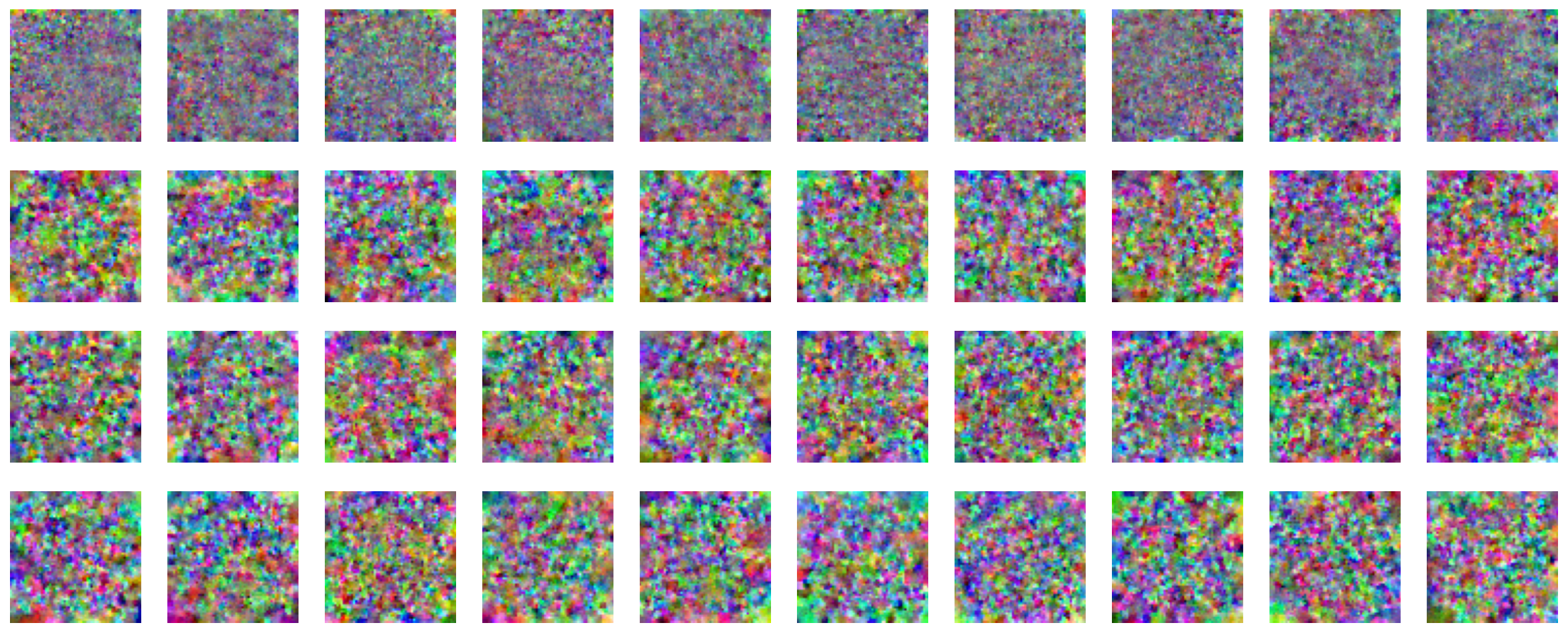}
    \caption{Reconstructed images by CPA-FI in FL}
    \label{fig:cpafi_result_fedavg}
\end{figure}

The second part focuses on attacking a deep CNN model (specifically VGG16) by applying ICA to recover the embedding inputs of the fully connected layers. This is followed by the use of feature inversion techniques to reconstruct the original inputs from these embeddings. The reconstructed images by this CPA-FI method are shown in Fig. \ref{fig:cpafi_result_fedavg}, where each row represents the inverted dummy images for one communication round. Unlike the results of the gradient attack shown in Fig. \ref{fig:cpafi}, the VGG model in this case is not pretrained, and the recovered images appear to be a complete failure, offering no useful information. This empirically inidicates that attacking more complex neural networks trained on high-resolution, large-scale images is significantly more challenging. As a result, privacy attack methods may become less effective in such scenarios.

\begin{figure}[!htbp]
    \centering
    \includegraphics[width=0.46\textwidth]{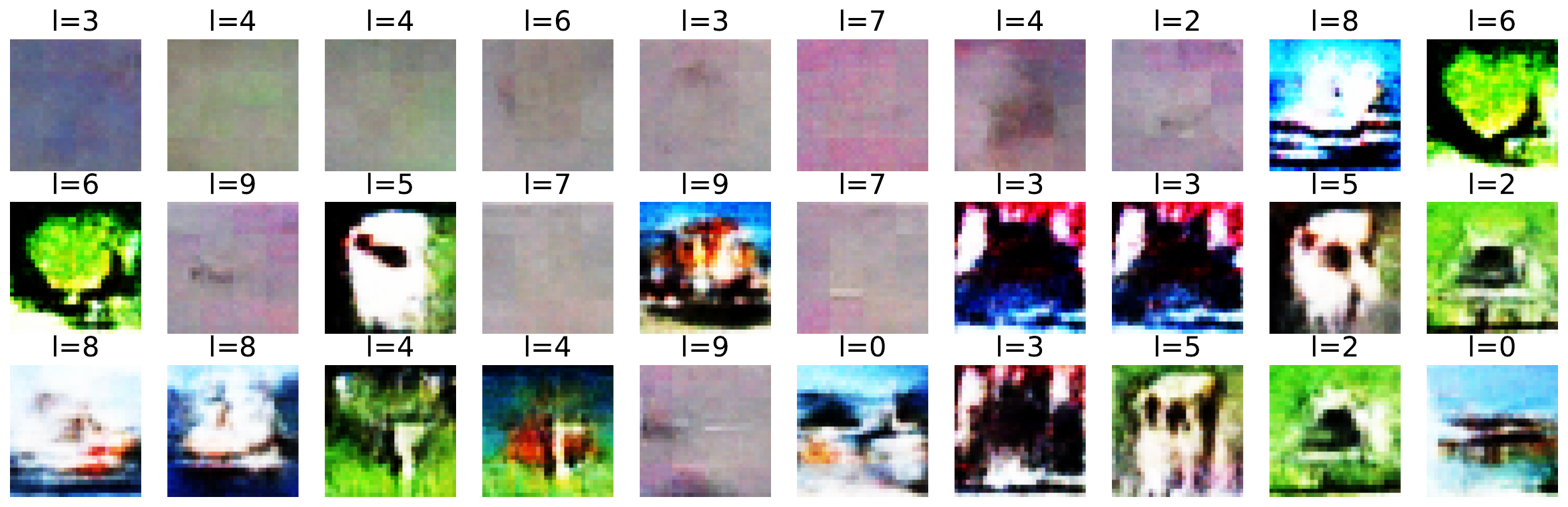}
    \caption{Reconstructed images by GGL in FL}
    \label{fig:ggl_result_fedavg}
\end{figure}

Different from the previous optimization-based attacks directly recovering dummy images, GGL employs a pretrained generator to search through its noise input, thereby significantly reducing the search space. It should be noticed that GGL reconstructs only a single dummy image, regardless of the local batch size or data size, and infers the label directly from the model difference using iDLG's method in Eq. \eqref{eq:idlg}. We suspect that this label inference approach, in this context, is equivalent to random label class sampling. Nevertheless, we maintain the same batch size and local data size as in the previous simulations for consistency. Moreover, gradient-free optimizer like CMA-ES is applied here to achieve more stable attack performance. The reconstructed images by GGL for 30 communication rounds are shown in Fig. \ref{fig:ggl_result_fedavg}, with some images appearing to be successfully recovered. However, we suspect that the reconstruction process is heavily influenced by the training data of the GAN generator, which may limit the accuracy of the recovered images.

\begin{figure}[!htbp]
    \centering
    \includegraphics[width=0.46\textwidth]{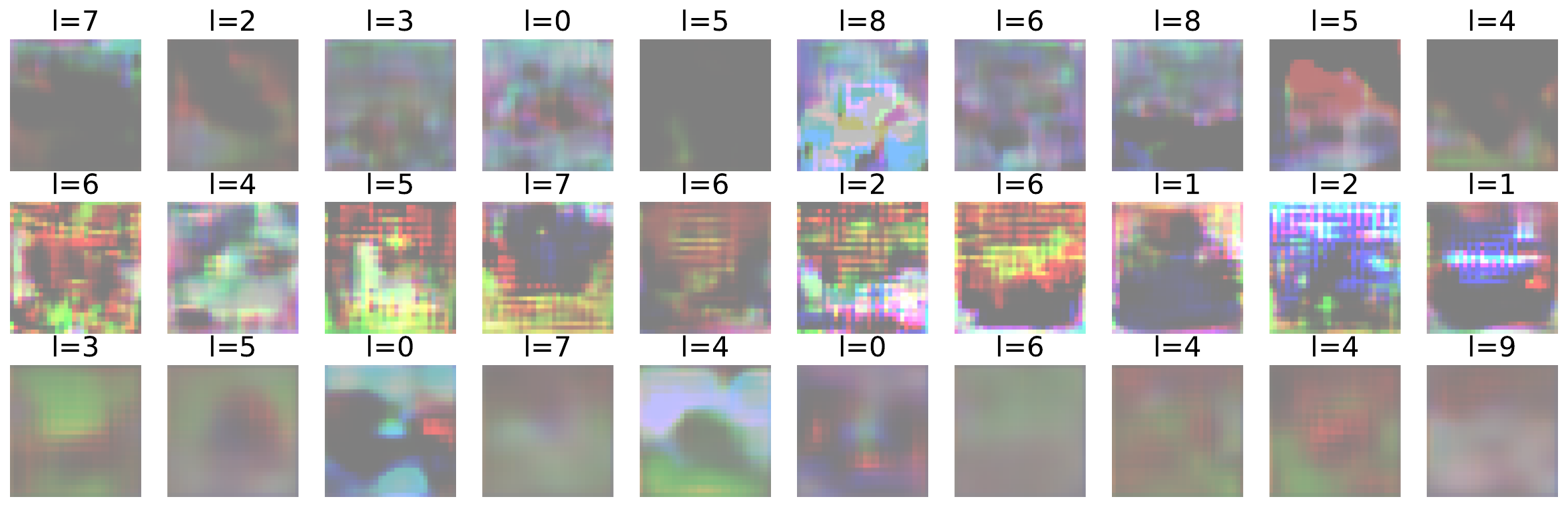}
    \caption{Reconstructed images by GRNN in FL}
    \label{fig:grnn_result_fedavg}
\end{figure}

GRNN, in contrast, does not rely on a pretrained generator. Instead of optimizing the input noise, GRNN trains the model parameters of its GAN generator directly during the reconstruction process. It is worth noting that in GRNN, dummy labels are reconstructed through the outputs of the generator, where a Gated Linear Unit (GLU) module \cite{dauphin2017language} is used in place of a commonly employed activation function. The authors argue that GLU offers greater stability compared to ReLU and learns faster than the Sigmoid activation function. The simulation results of GRNN for 10 batch size and 3 communication rounds are shown in Fig. \ref{fig:grnn_result_fedavg}, where each row represents 10 reconstructed dummy images for each round. The overall quality of the dummy images appears acceptable, however, it remains challenging to discern the semantic meaning from these reconstructed images. This is likely due to GRNN focusing solely on batch gradient attacks, which limits its performance in more complex FL environments.

\begin{figure}[!htbp]
    \centering
    \includegraphics[width=0.46\textwidth]{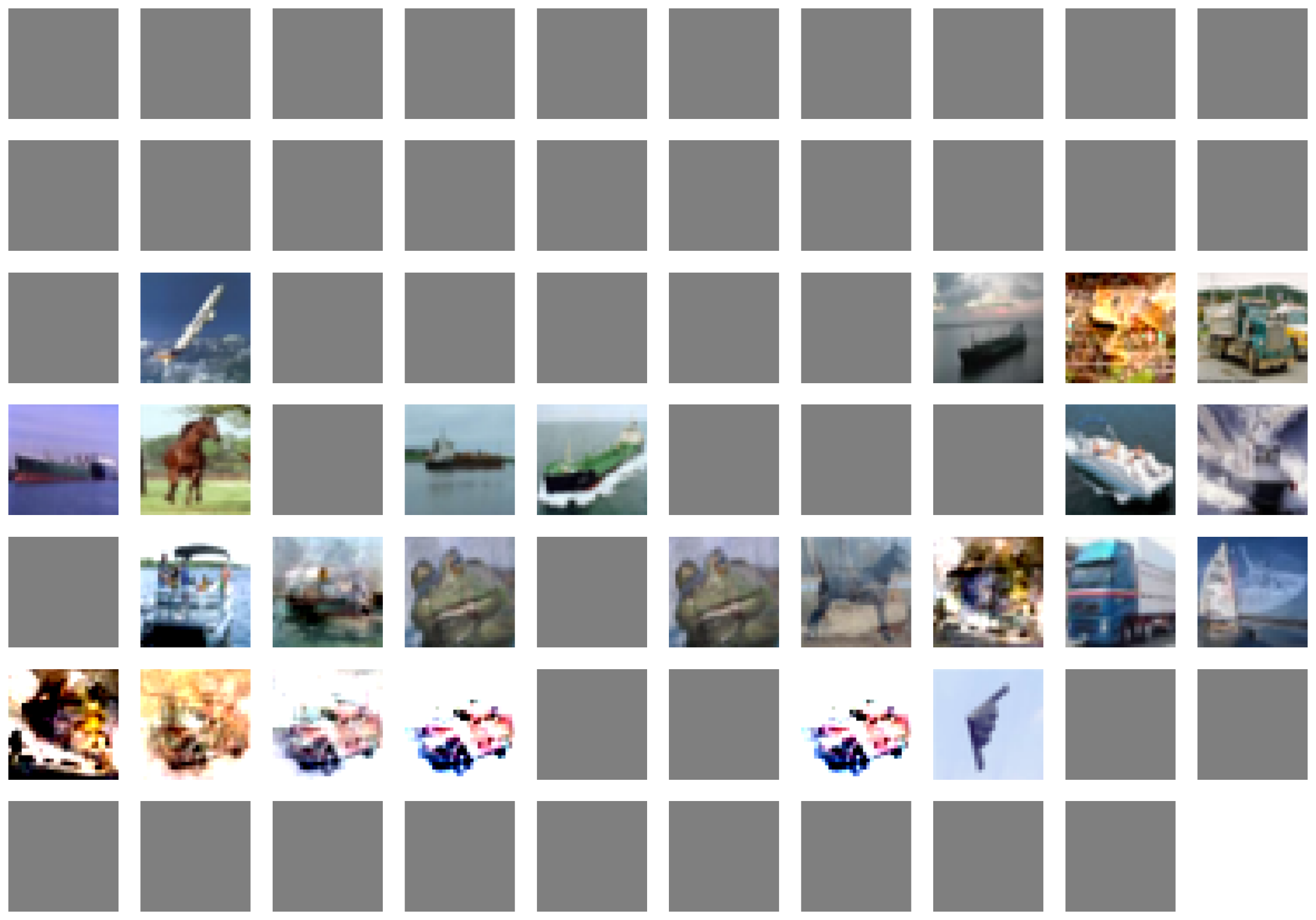}
    \caption{Reconstructed images by RTF in FL}
    \label{fig:rtf_result_fedavg}
\end{figure}

Finally, the experimental results of RTF attack are presented. RTF, being an analytical server-side attack method, reconstructs images independently of the reconstruction batch size, thus eliminating the need for setting a learning rate. However, RTF requires a modification of the original learning model by introducing an Imprint block before it. The Imprint block is essentially a MLP neural network with one hidden layer consisting of 100 neurons (bins). The input and output dimensions of this block are equivalent to the image size. The reconstructed images by RTF attack for one communication round are shown in Fig. \ref{fig:rtf_result_fedavg}. Notably, some of the recovered images closely resemble the original ones, while others fail to be accurately reconstructed. This is because the brightness $h(\mathbf{X}_{i})$ of certain image is not successfully captured within the interval between $-\mathbf{b}_{l}$ and $-\mathbf{b}_{l+1}$.

RTF appears to effectively recover private images from the model parameters uploaded by the client to some extent. However, further validation experiments are necessary to assess the effectiveness of RTF by increasing the local batch size, the number of local epochs, and the local data size. It is also important to note that the Imprint block in RTF may cause some degradation in model performance. Therefore, we also conduct standard FL training to validate the learning .

\subsection{Client-Side Privacy Attack}
In this section, one of the most representative client-side attack algorithms, DMGAN, is tested and evaluated. Other methods, such as mGAN-AI, are not included in this experimental study, as the focus is specifically on input reconstruction privacy attacks rather than on membership inference. Unlike the aforementioned server-side approaches, DMGAN sets the total number of clients to 10, thereby increasing the local data size for more sufficient local GAN training. The local training epochs and batch size are set to 2 and 50, respectively. In practice, DMGAN is not particularly sensitive to variations in these parameters. 

\begin{figure}[!htbp]
    \centering
    \includegraphics[width=0.46\textwidth]{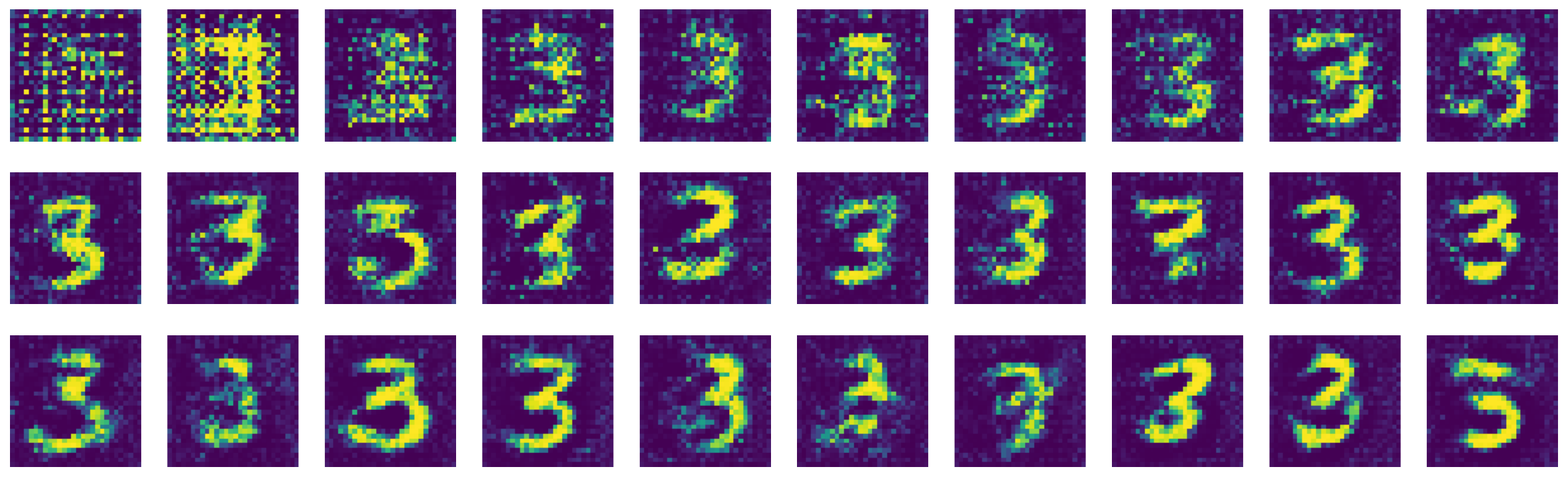}
    \caption{Reconstructed images by DMGAN in FL}
    \label{fig:dmgan_result_fedavg}
\end{figure}

The tracked MNIST images with label class 3 over 30 communication rounds are shown in Fig. \ref{fig:dmgan_result_fedavg}, where each image is recovered by the local generator of adversarial client for each round. It is obvious to see that the local generator begins to produce clearly recognizable digit images after three rounds of communication. However, as previously discussed, DMGAN requires structural modifications to the global model by adding an extra neuron to the output layer. Since the server in this case is not the attacker, and clients lack the authority to alter the global model's structure, making DMGAN unrealistic applied in real-world FL systems.

\subsection{More Challenging FL Environments}
In this section, we aim to evaluate several algorithms with relatively strong attack performance, such as Inverting Gradients, DLF, GGL, RTF and DMGAN, in more challenging FL environments. Other attack algorithms are excluded from consideration due to their poor performance even in relatively easier FL settings. For server-side attacks, more challenging conditions typically involve larger local data size (fewer total clients), larger local batch sizes, and more local training epochs. In contrast, for the client-side attack DMGAN, more difficult FL settings are characterized by smaller local data size (more total clients). It should be noticed that among these testing algorithms, among these testing algorithms only Inverting Gradients and DLF do not require violations of FL protocols, such as modifying the global architecture or using auxiliary server data for GAN pretraining.

\subsubsection{Local Dataset}
Increasing the amount of local training data also increases the number of batch iterations within a single training epoch. This, in turn, heightens the complexity of the model differences between the global model and the updated local model, making it easier to obscure the gradient information of individual batch images within the model. Specifically for server-side attacks, with the total dataset size fixed, we reduce the number of clients to 10 (20 for DLF, since the label count inference would raise running errors for 10), resulting in an approximate tenfold increase in the local data size per client while keeping all other FL settings unchanged.

\begin{figure}[!htbp]
    \centering
    \includegraphics[width=0.46\textwidth]{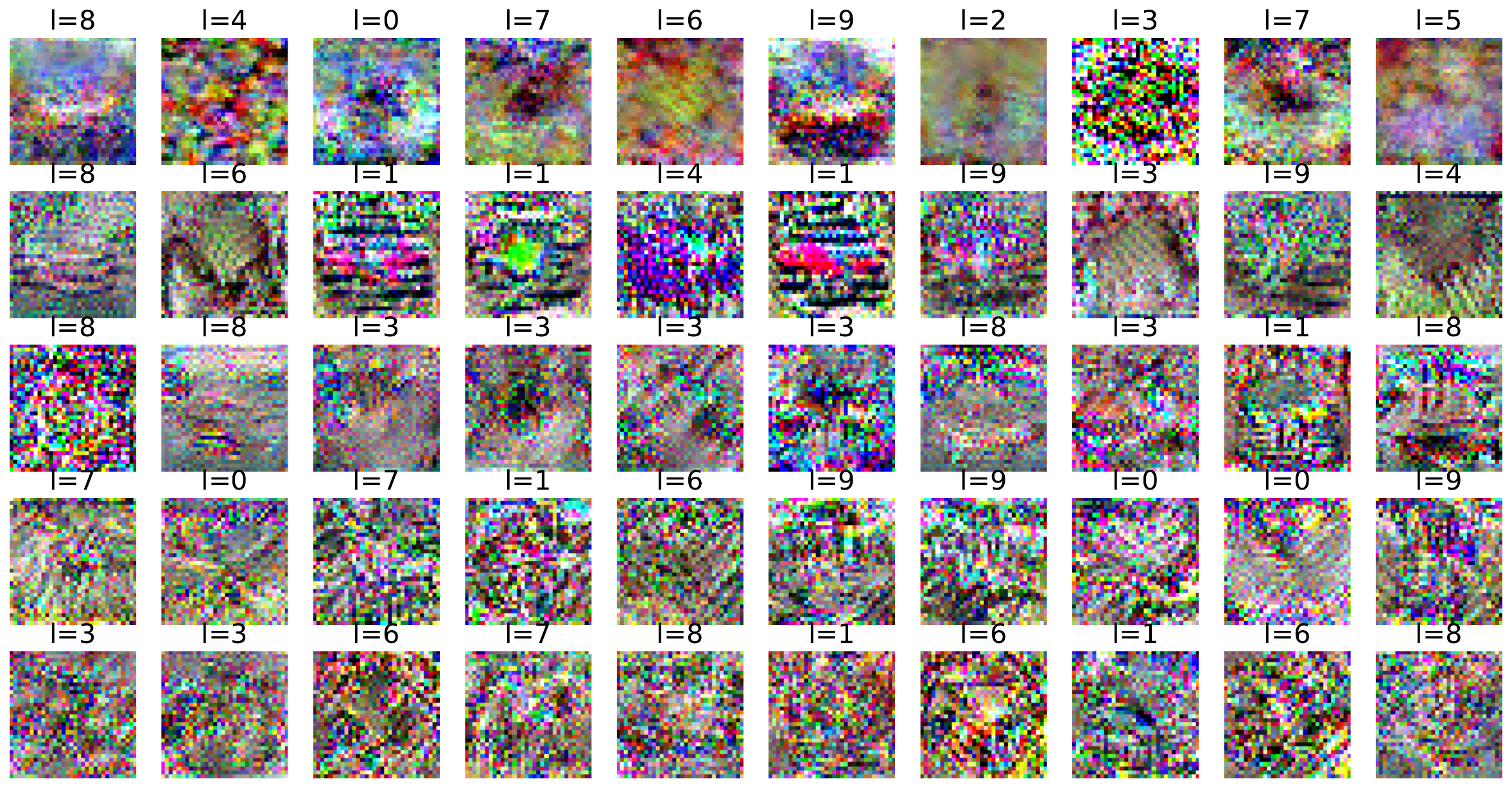}
    \caption{Reconstructed images by Inverting Gradients for 10 clients}
    \label{fig:ig_10c}
\end{figure}

The reconstructed images by Inverting Gradients over 5 communication rounds are shown in Fig. \ref{fig:ig_10c}. it is evident that the quality of the recovered images has significantly degraded, with only chaotic color blocks visible and no useful information discernible. This degradation occurs because the true gradients of the ground-truth local images are much harder to be captured through gradient inversion attacks.

\begin{figure}[!htbp]
    \centering
    \includegraphics[width=0.46\textwidth]{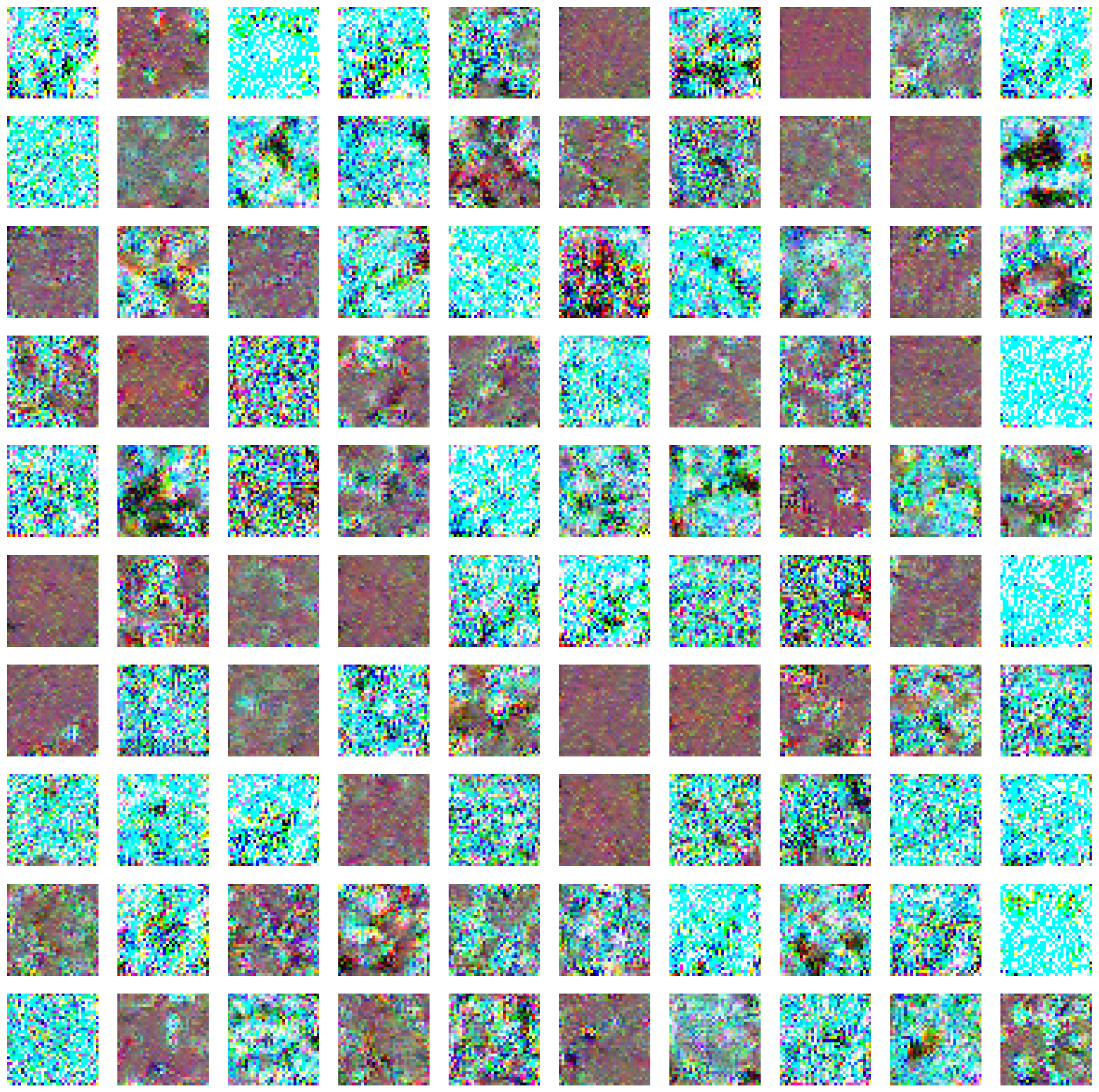}
    \caption{Reconstructed images by DLF for 20 clients}
    \label{fig:dlf_10c}
\end{figure}

\begin{figure}[!htbp]
    \centering
    \includegraphics[width=0.46\textwidth]{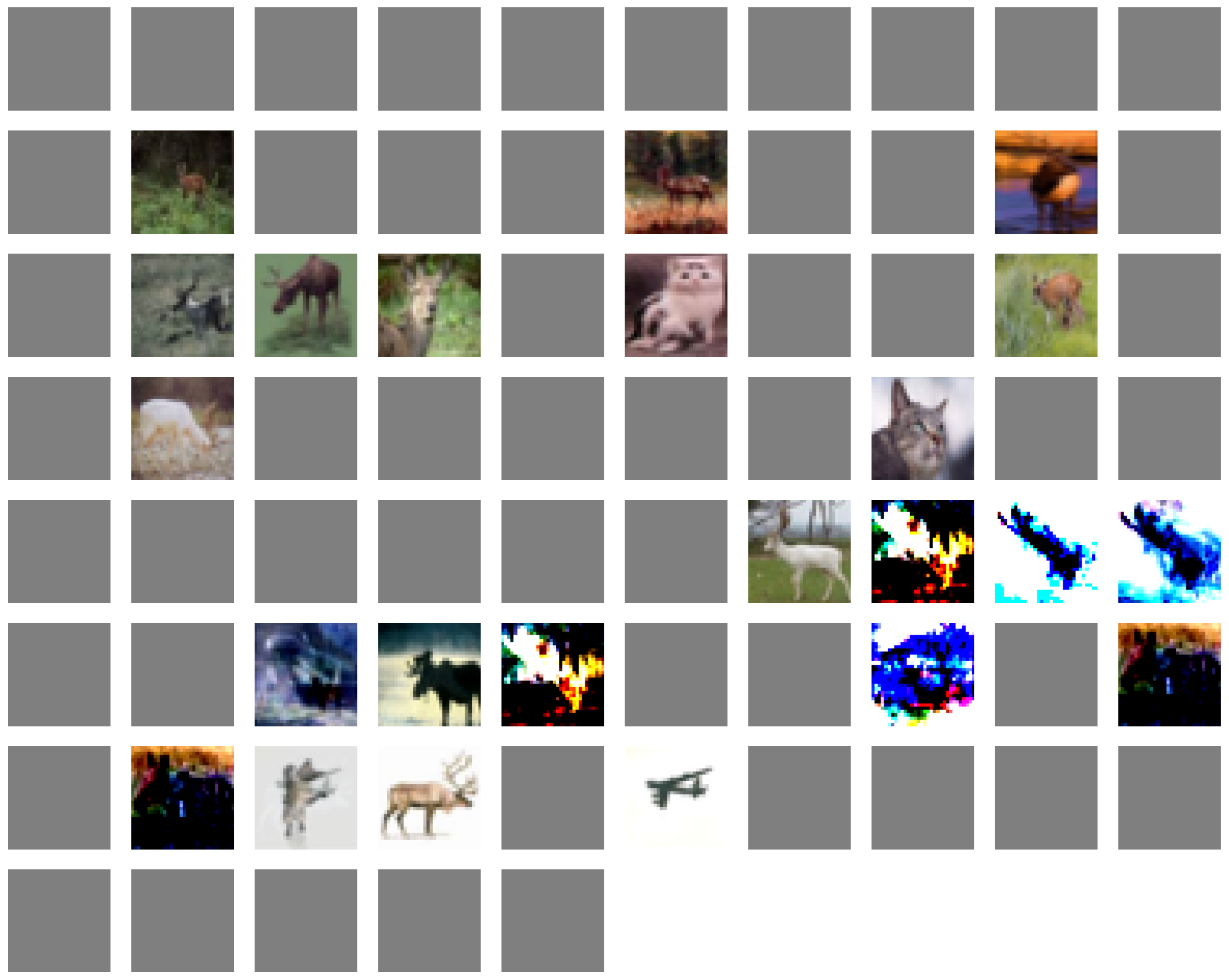}
    \caption{Reconstructed images by RTF for 10 clients}
    \label{fig:rtf_10c}
\end{figure}

\begin{figure}[!htbp]
    \centering
    \includegraphics[width=0.46\textwidth]{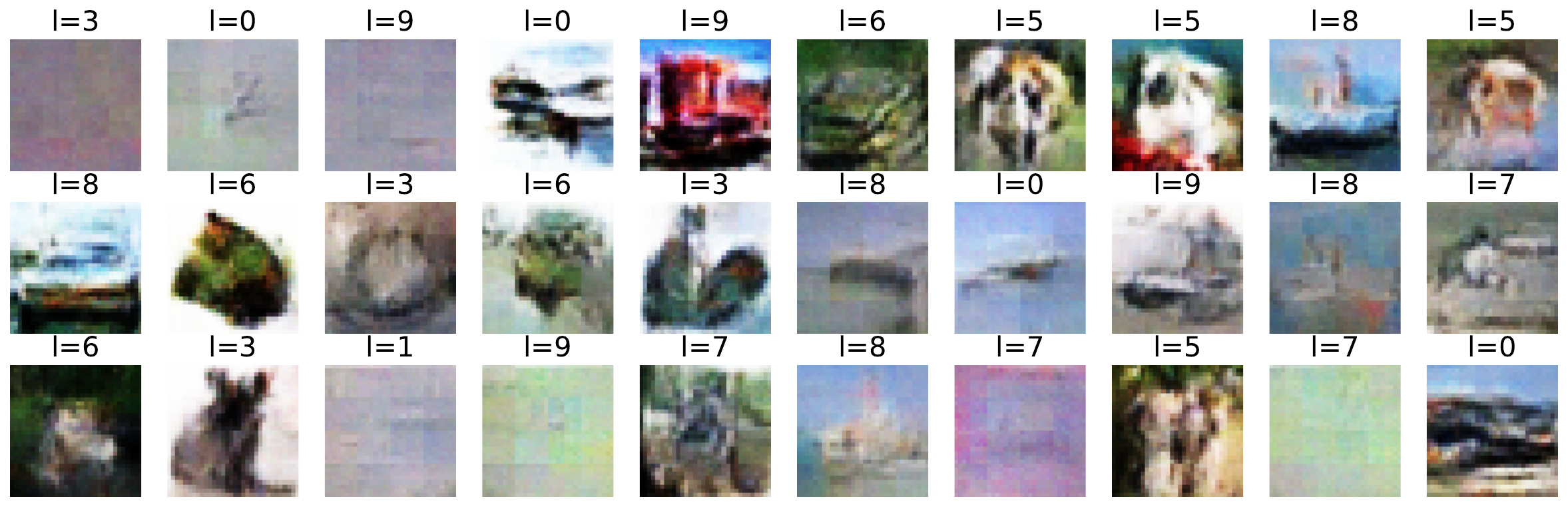}
    \caption{Reconstructed images by GGL for 10 clients}
    \label{fig:ggl_10c}
\end{figure}

The reconstructed results by DLF algorithm with 20 clients are presented in Fig. \ref{fig:dlf_10c}, where 100 dummy images are randomly selected from the entire set of reconstructed images. This aligns with previous findings that a larger local data size significantly increases the error in approximating label counts during DLF attacks, thereby misleading the direction of subsequent image reconstructions. And the reconstructed images by RTF for one communication round are shown in Fig. \ref{fig:rtf_10c}. It appears that a larger local data size does not pose challenges for RTF, as it is still able to generate recognizable high-quality dummy images. Additionally, the reconstructed images by GGL for 10 clients are depicted in Fig. \ref{fig:ggl_10c}. Surprisingly, much better quality images with clear semantic information were recovered compared to the results with 100 clients in Fig. \ref{fig:ggl_result_fedavg}. This may be due to the fact that a larger local data size reduces the influence of gradients from any individual local data point. As a result, the generated dummy images are more likely to be dominated by the generator's pretrained mechanism. To further verify this hypothesis, we will conduct experiments using an untrained generator in GGL.

\begin{figure}[!htbp]
    \centering
    \includegraphics[width=0.46\textwidth]{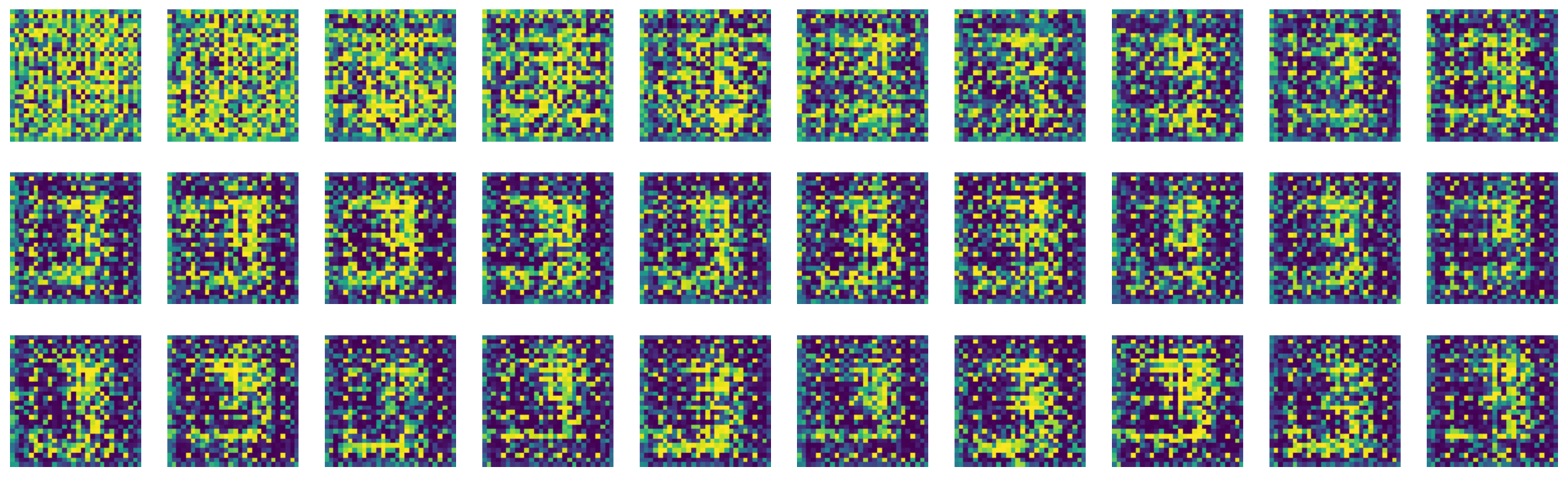}
    \caption{Reconstructed images by RTF for 100 clients}
    \label{fig:dmgan_100c}
\end{figure}

For the client-side DMGAN attack, since it is not based on gradient inversion optimization, the local data size primarily influences the training effectiveness of the local GAN generator. Therefore, reducing the local data size creates a more challenging FL setting for DMGAN. The reconstructed images by DMGAN for 100 clients are shown in Fig. \ref{fig:dmgan_100c}. It is clear that the quality of the recovered images is significantly worse than those generated with 10 clients (in Fig. \ref{fig:dmgan_result_fedavg}) that had relatively sufficient local training data. However, the dummy images remain recognizable as the digit 3.

\subsubsection{Larger Batch Size} \label{subsec:bs}
\begin{figure}[!htbp]
    \centering
    \includegraphics[width=0.46\textwidth]{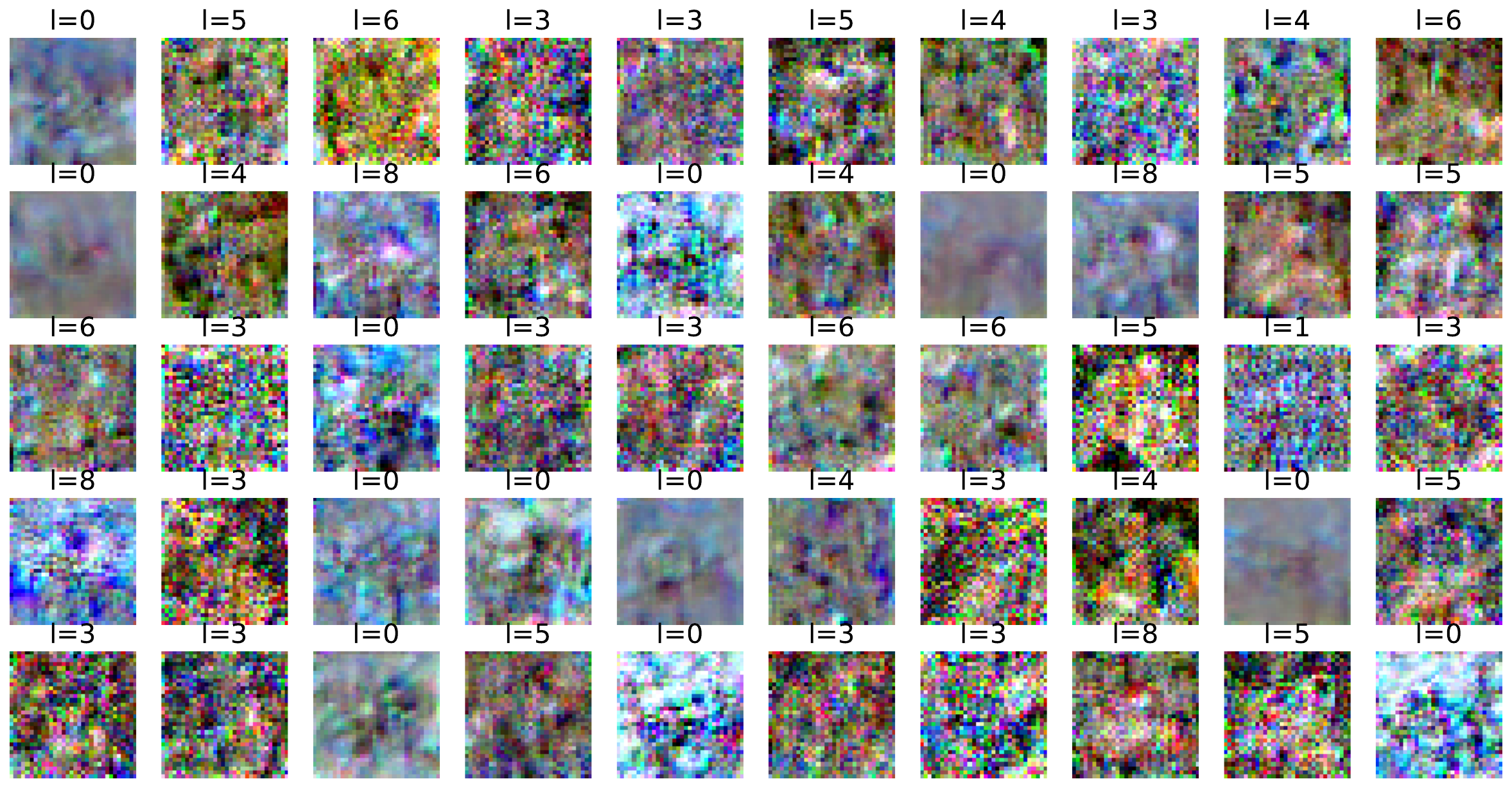}
    \caption{Reconstructed images by Inverting Gradients for 50 batch size}
    \label{fig:ig_50b}
\end{figure}

\begin{figure}[!htbp]
    \centering
    \includegraphics[width=0.46\textwidth]{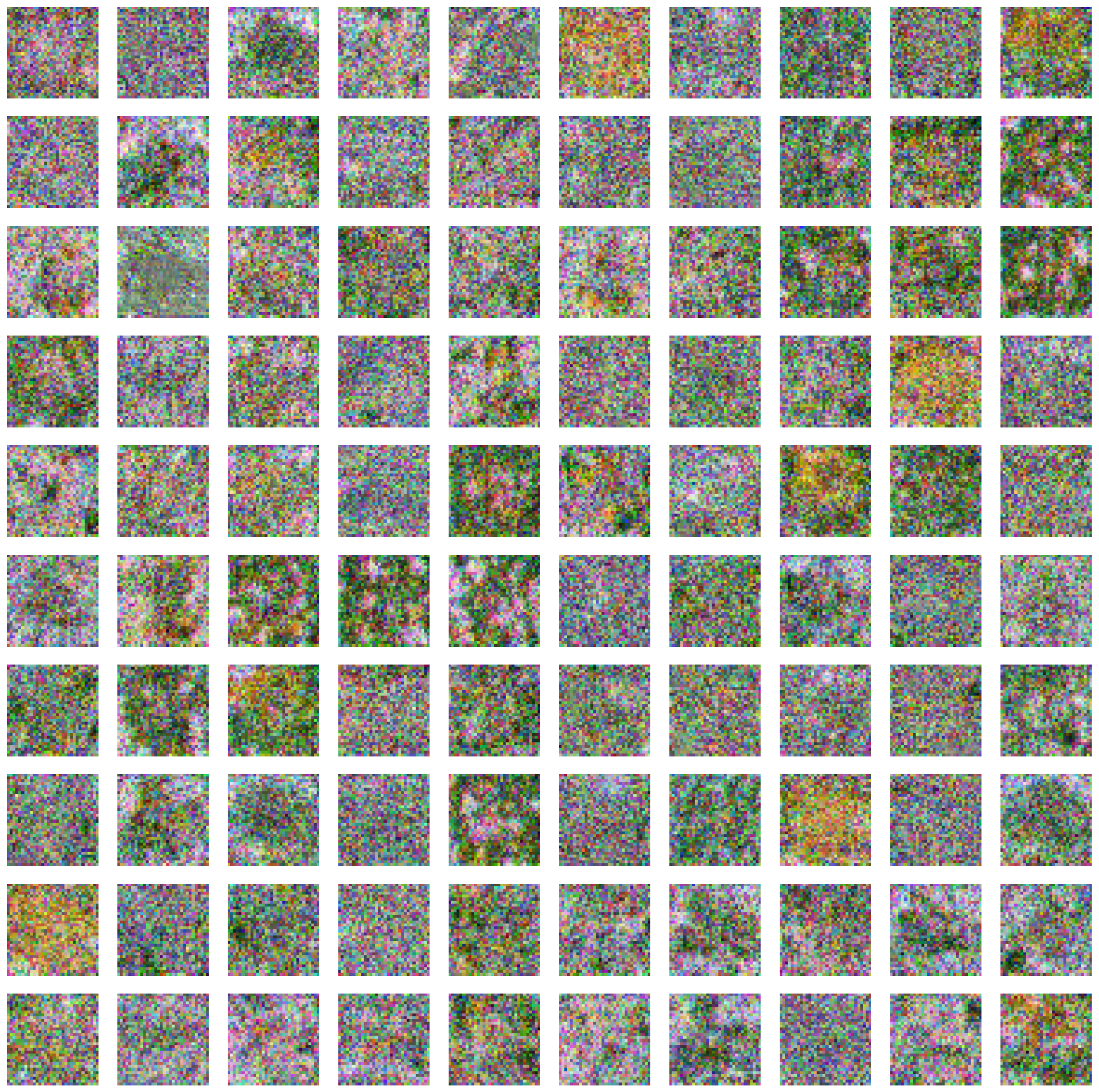}
    \caption{Reconstructed images by DLF for 50 batch size}
    \label{fig:dlf_50b}
\end{figure}

\begin{figure}[!htbp]
    \centering
    \includegraphics[width=0.46\textwidth]{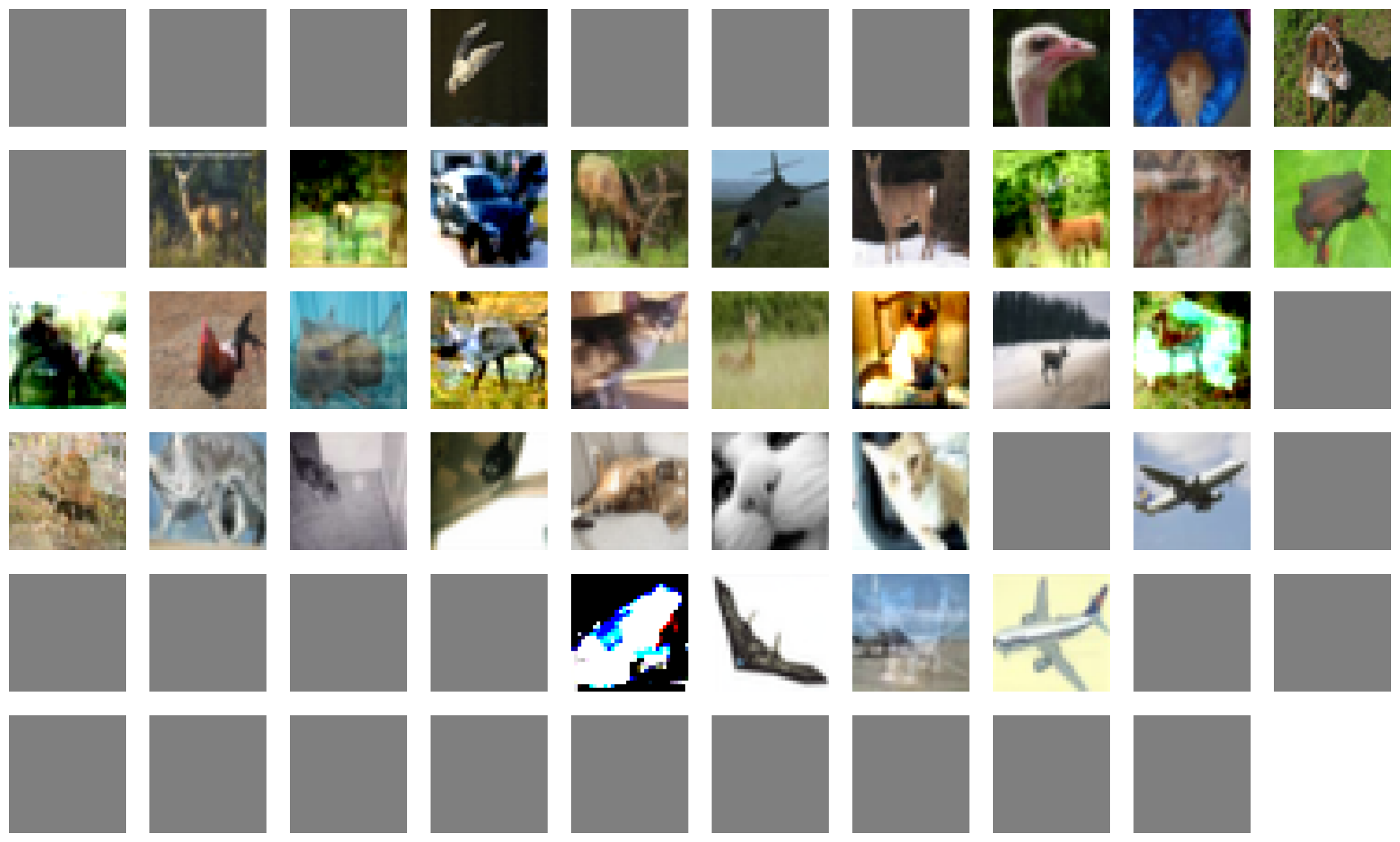}
    \caption{Reconstructed images by RTF for 50 batch size}
    \label{fig:rtf_50b}
\end{figure}

\begin{figure}[!htbp]
    \centering
    \includegraphics[width=0.46\textwidth]{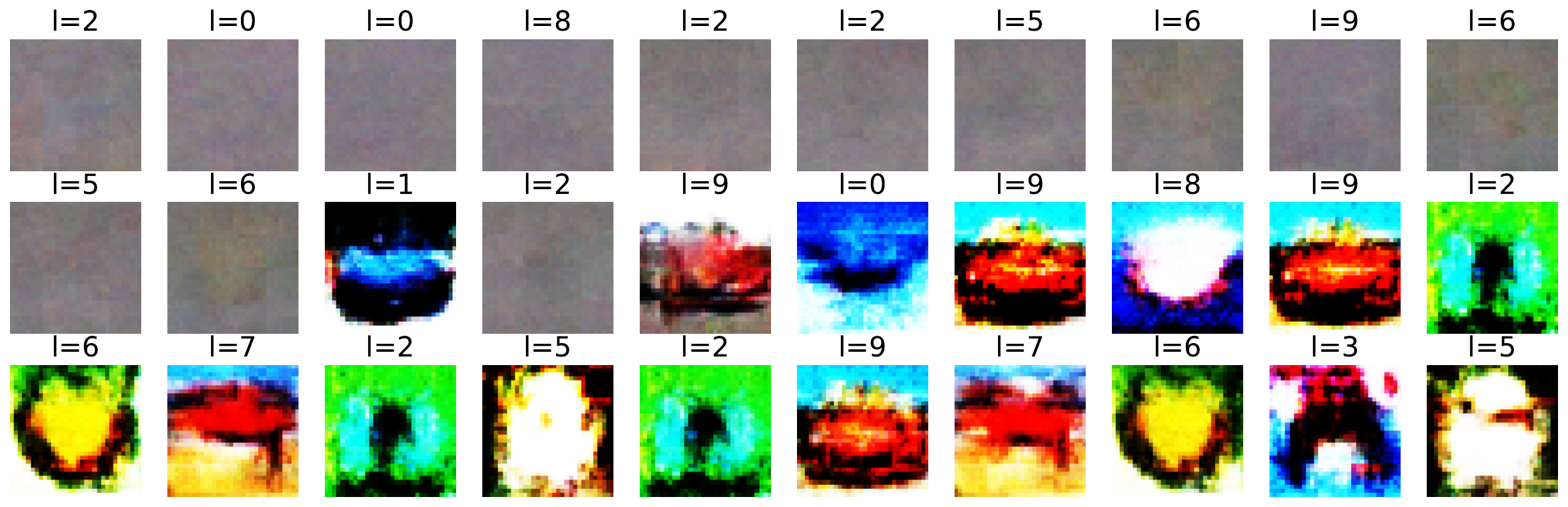}
    \caption{Reconstructed images by GGL for 50 batch size}
    \label{fig:ggl_50b}
\end{figure}

To evaluate the robustness of the algorithms to batch size, we increase the local training batch size from 10 to 50. This change makes it more difficult for gradient inversion algorithms to separate and recover the aggregated model parameters effectively. All other hyperparameters, such as the total number of clients and local training epochs, are kept the same as in Section \ref{sec:serverattack} to provide a more comprehensive ablation study. The reconstructed images by Inverting Gradients, DLF, RTF, GGL for 50 batch size are shown in Fig. \ref{fig:ig_50b}, Fig. \ref{fig:dlf_50b}, Fig. \ref{fig:rtf_50b}, and Fig. \ref{fig:ggl_50b}, respectively.

It is unsurprising to observe that both Inverting Gradients and DLF attacks perform poorly with larger local batch sizes. This suggests that these gradient inversion approaches struggle to extract useful image information from aggregated gradients when dealing with a larger number of elements. While GGL demonstrates similar attack performance compared to the recovered dummy images in Fig. \ref{fig:ggl_10c}, empirically proving that the reconstruction performance of GGL is primarily determined by the quality of the pretrained GAN generator. And the FL environment for reconstruction process is comparatively less important. Finally, RTF continues to exhibit superior attack performance, even with large local training batch size.

\subsubsection{Training Epochs}
\begin{figure}[!htbp]
    \centering
    \includegraphics[width=0.46\textwidth]{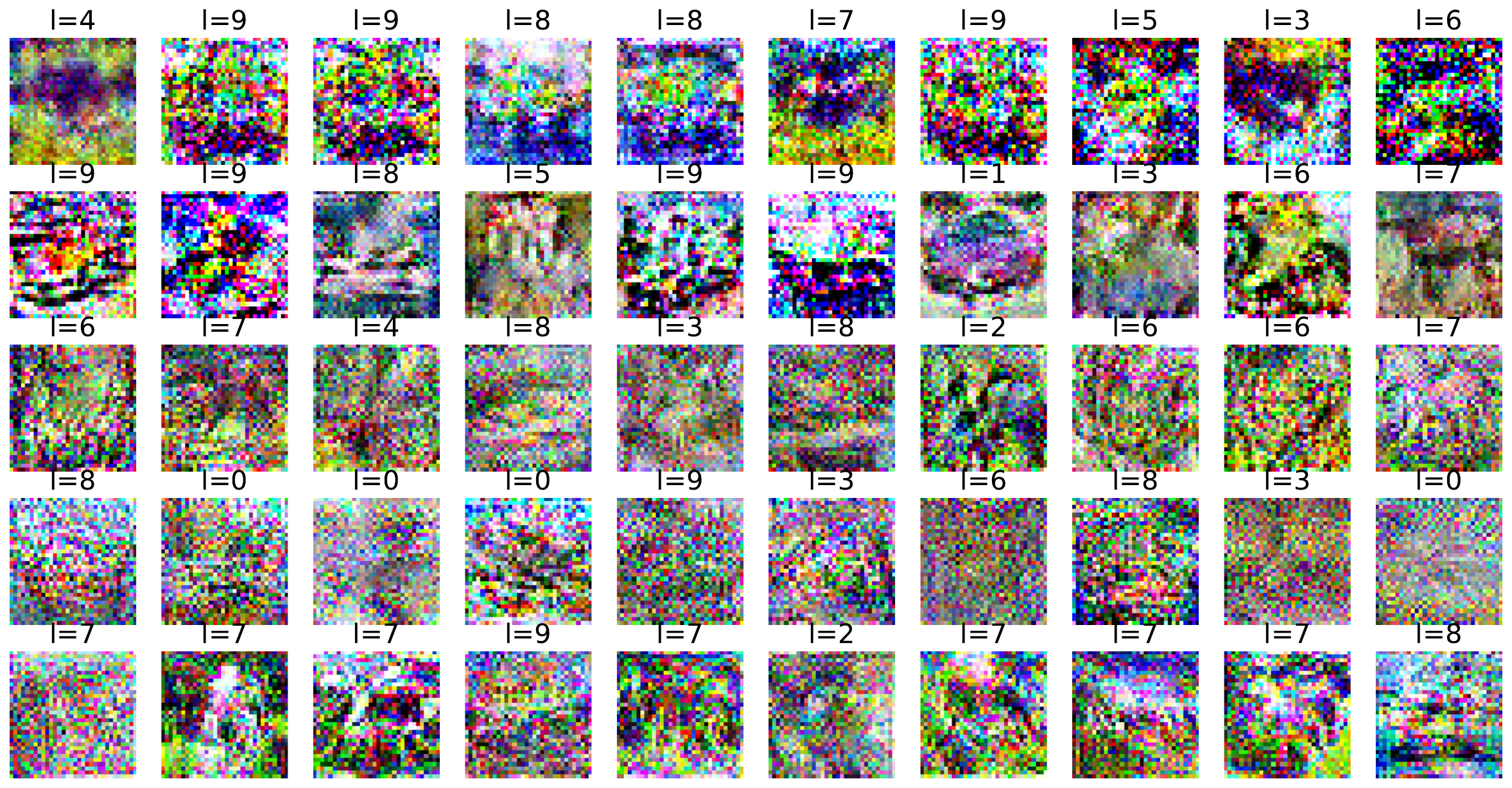}
    \caption{Reconstructed images by Inverting Gradients for 5 local epochs}
    \label{fig:ig_5e}
\end{figure}

\begin{figure}[!htbp]
    \centering
    \includegraphics[width=0.46\textwidth]{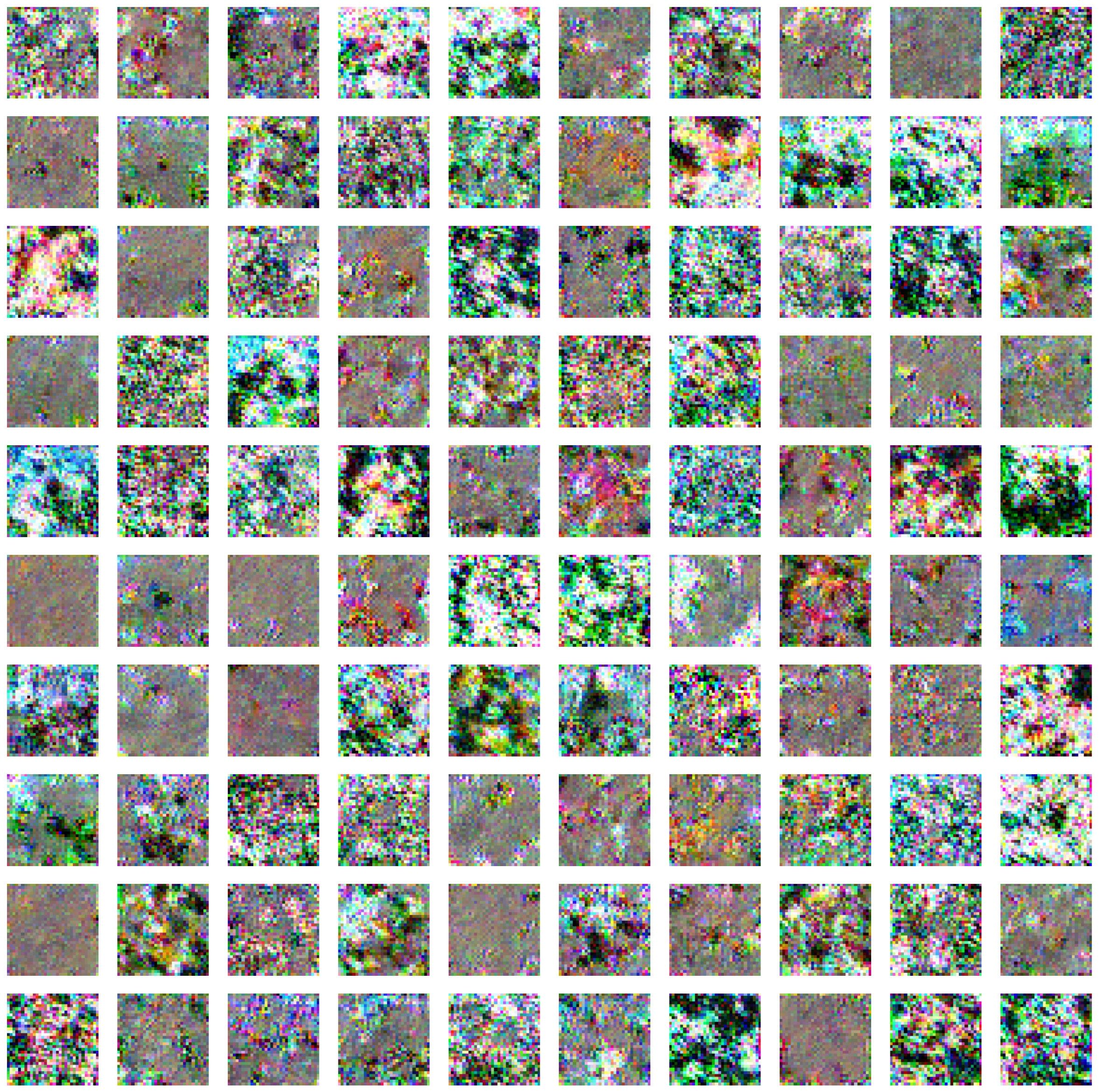}
    \caption{Reconstructed images by DLF for 5 local epochs}
    \label{fig:dlf_5e}
\end{figure}

\begin{figure}[!htbp]
    \centering
    \includegraphics[width=0.46\textwidth]{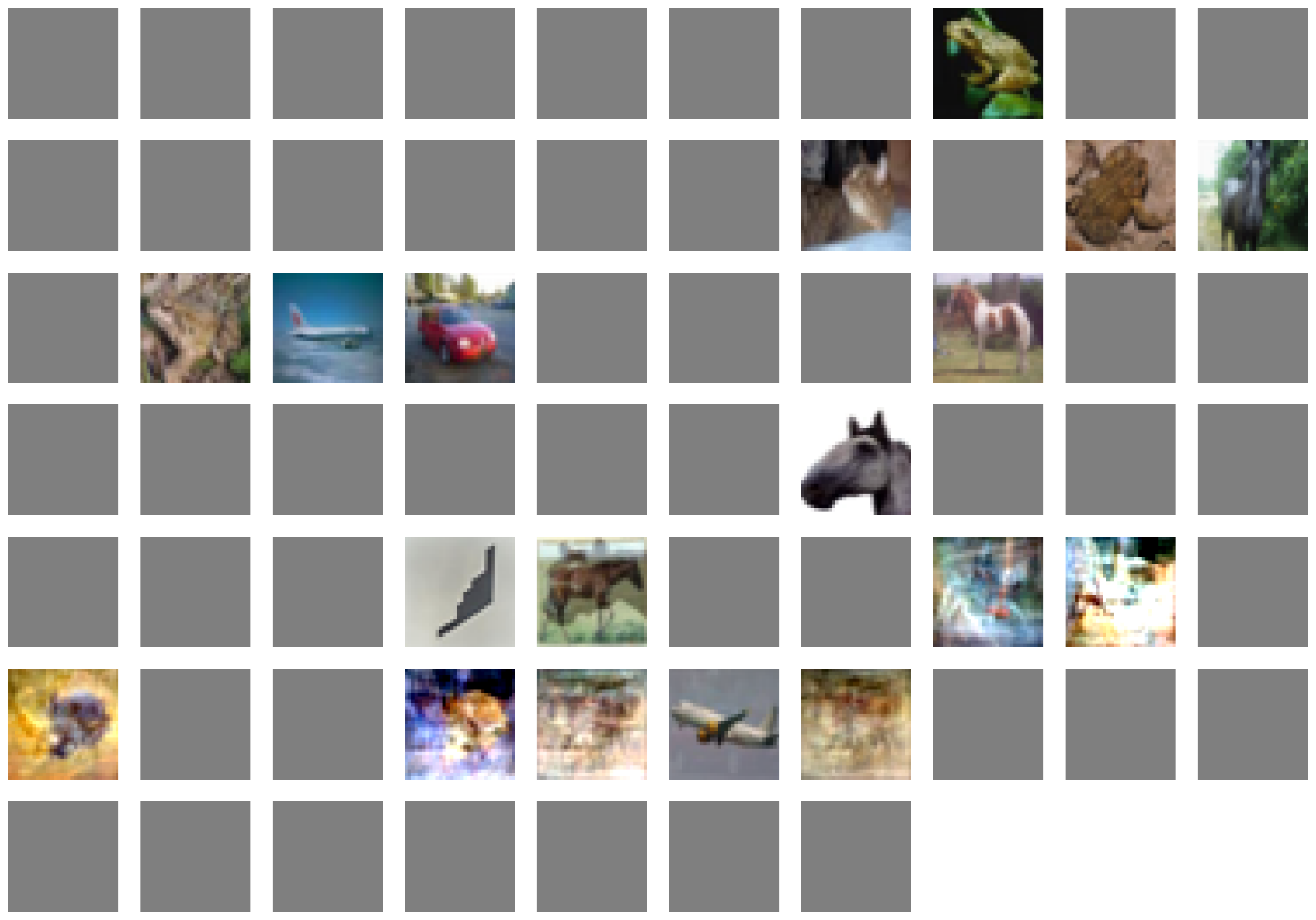}
    \caption{Reconstructed images by RTF for 5 local epochs}
    \label{fig:rtf_5e}
\end{figure}

\begin{figure}[!htbp]
    \centering
    \includegraphics[width=0.46\textwidth]{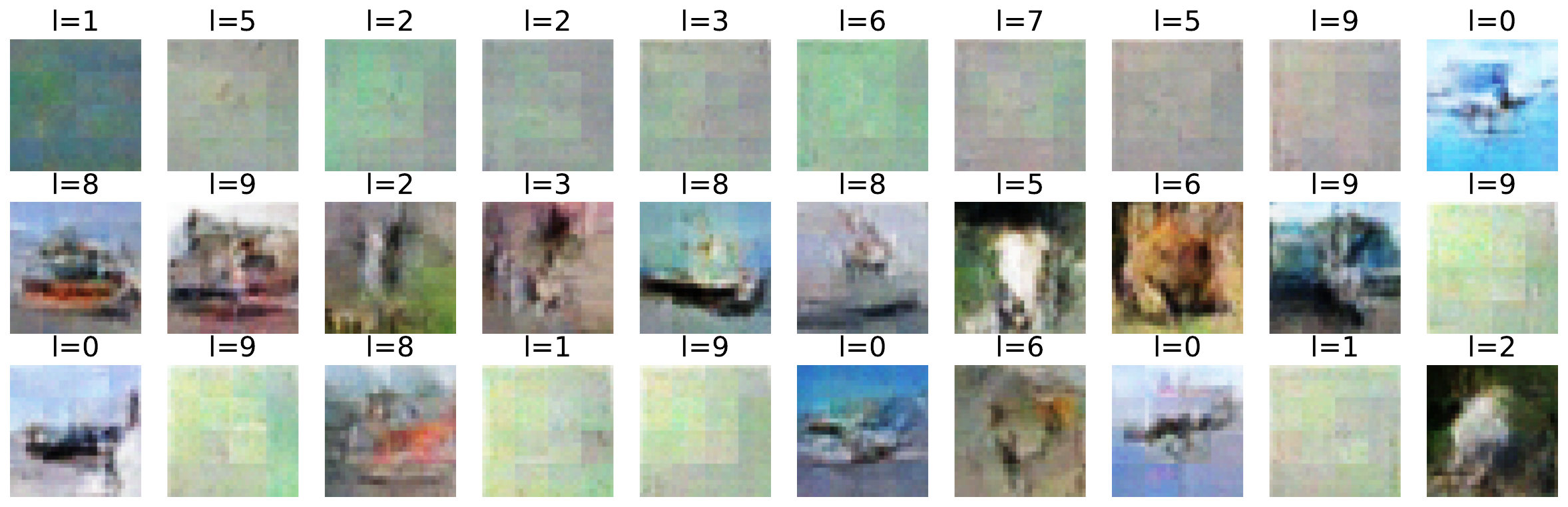}
    \caption{Reconstructed images by GGL for 5 local epochs}
    \label{fig:ggl_5e}
\end{figure}

Increasing the number of local training epochs forces the local data to be repeatedly applied for client model updates. This further widens the gap between the global model and the locally updated model, significantly increasing the complexity and difficulty of performing gradient inversion. To evaluate the impact of local training epochs on attack performance, we increase the number of local training epochs from 1 to 5. And the reconstructed images by Inverting Gradients, DLF, RTF and GGL are shown in Fig. \ref{fig:ig_5e}, Fig. \ref{fig:dlf_5e}, Fig. \ref{fig:rtf_5e}, and Fig. \ref{fig:ggl_5e}, respectively.

Similar to the experimental results in the previous subsection, Inverting Gradients and DLF fail to reconstruct meaningful dummy images with clear semantic information. This is due to the increased difficulty of inferring correct label counts during repeated local batch training with the shuffling process. While GGL shows even better attack performance with multiple local training epochs compared to a single epoch, indirectly supporting our earlier assertion that the quality of reconstructed images is primarily determined by the pretrained generator.  Increasing the local training epochs reduces the influence of gradients from individual image data, further emphasizing the generator's role in reconstruction. 
Finally, as an analytic method, RTF remains robust to larger local epochs. Its success is determined by whether the brightness of the input images can be effectively captured by two adjacent biases ($-\mathbf{b}_{l}$ \& $-\mathbf{b}_{l+1}$), rather than being influenced by the number of training epochs.

\subsection{More Challenging Training Data}
\begin{figure}[!htbp]
    \centering
    \includegraphics[width=0.46\textwidth]{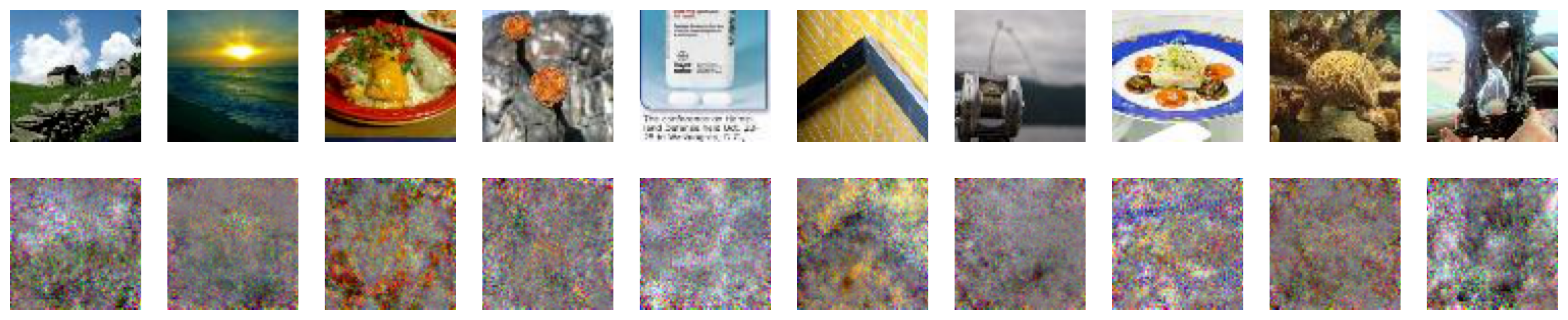}
    \caption{The outcomes of gradient attack by Inverting Gradients on Tiny ImageNet}
    \label{fig:ig_tiny}
\end{figure}

In addition to different FL settings significantly impacting privacy attack performance, more complex datasets with higher-resolution images further increase the difficulty of executing successful attacks. A simple experiment using Inverting Gradients for single-image gradient attack on Tiny ImageNet dataset is presented in Fig. \ref{fig:ig_tiny}, where the second row represents the reconstructed images for 10 repeated simulations. It is evident that the quality of the recovered images in Fig. \ref{fig:ig_tiny} experiences a steep decline compared to the results from CIFAR-10 shown in Fig. \ref{fig:ig_result}. Since Inverting Gradients fails to demonstrate promising attack performance even for simple single gradient attacks on Tiny ImageNet, applying it in more challenging FL environments would be ineffective and impractical. Consequently, current optimization-based attack methods are not only sensitive to FL settings but also struggle to handle complex datasets with larger images effectively.

\begin{figure}[!htbp]
    \centering
    \includegraphics[width=0.46\textwidth]{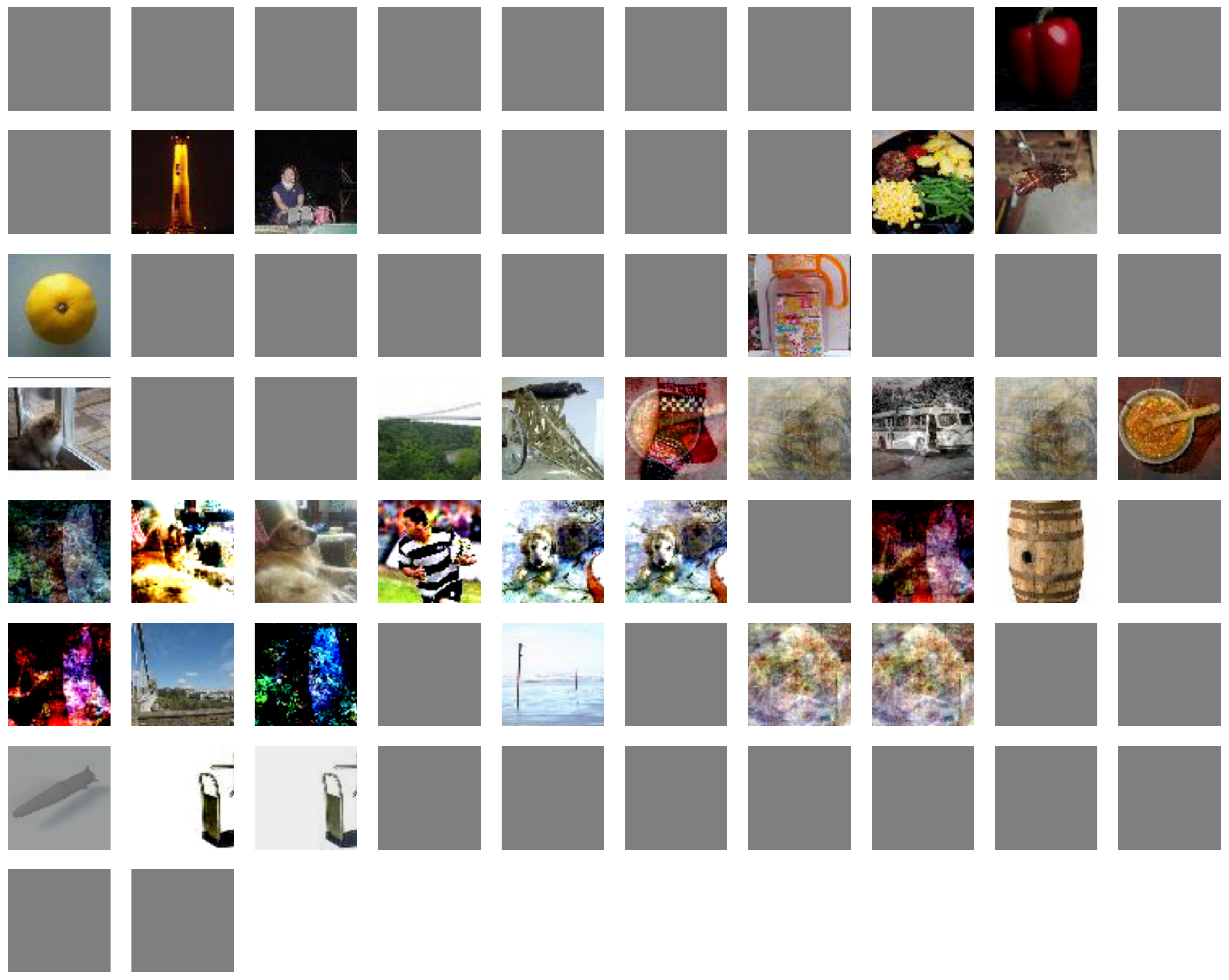}
    \caption{The outcomes of gradient attack by RTF on Tiny ImageNet}
    \label{fig:rtf_tiny}
\end{figure}

While RTF, as a powerful analytic attack method, demonstrates superior reconstruction performance even in more challenging FL environments, it will be interesting to explore whether it can still recover high-quality dummy images on more complex datasets. The reconstructed dummy images by RTF using ResNet18 model are shown in Fig. \ref{fig:rtf_tiny}. It is evident that RTF is still capable of successfully recovering high-quality dummy inputs even on more complex training images. If we set aside the restriction of the imprint block on the global model in FL, RTF is currently the most powerful privacy attack approach. It is also the best suited for realistic FL environments compared to other methods.

\begin{figure}[!htbp]
    \centering
    \includegraphics[width=0.46\textwidth]{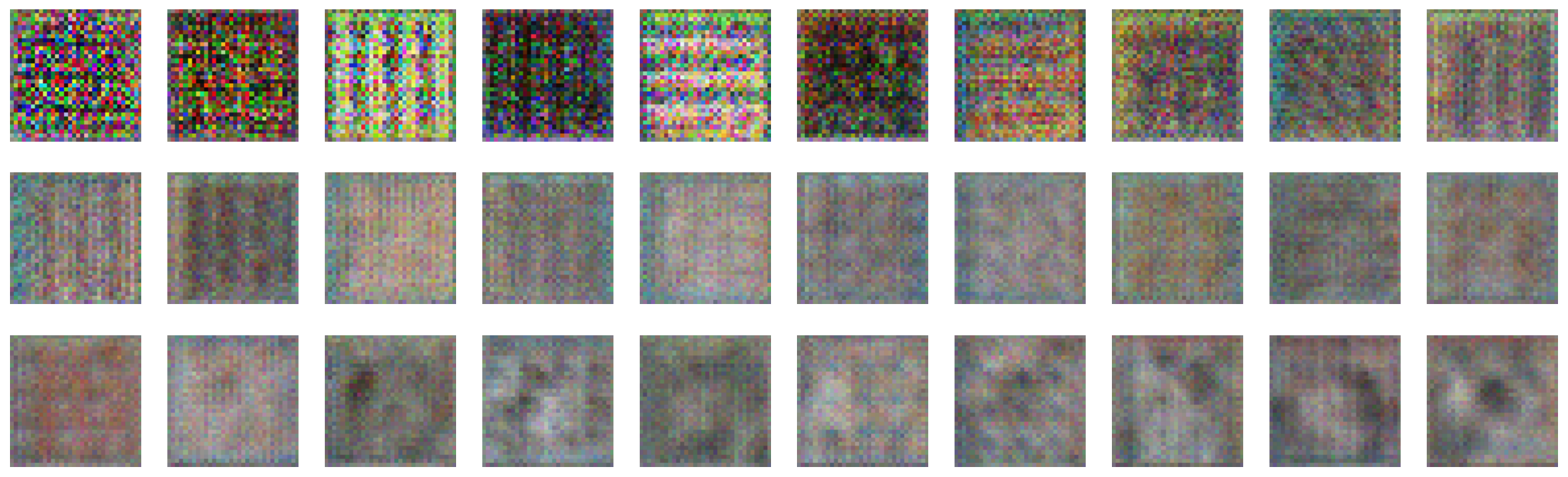}
    \caption{The reconstructed images by DMGAN on CIFAR-10 dataset}
    \label{fig:dmgan_cifar}
\end{figure}

For the client-side attack method DMGAN, the more challenging CIFAR-10 dataset is adopted to replace the simple grayscale handwritten digits MNIST. And the reconstructed images over 30 communication rounds are illustrated in Fig. \ref{fig:dmgan_cifar}. It is surprising to observe that the generator of the adversarial client fails to reconstruct private CIFAR-10 images from other participants. Although the hyperparameters of DMGAN for optimal attack performance were not fine-tuned, we can at least conclude that DMGAN lacks robustness when dealing with more complex training images.

\subsection{Different Models}
Different types of training models undoubtedly influence the overall learning performance of FL. Additionally, model variations can significantly impact the effectiveness of privacy attack methods applied within FL systems. Therefore, this experimental study in this subsection is divided into two main parts: 1) For optimization-based attacks, we will use a relatively simple model, such as MLP, to reduce the reconstruction difficulty. 2) For the analytic attack approach, RTF, we will modify its backbone model behind the imprint module to further explore its robustness.

\begin{figure}[!htbp]
    \centering
    \includegraphics[width=0.46\textwidth]{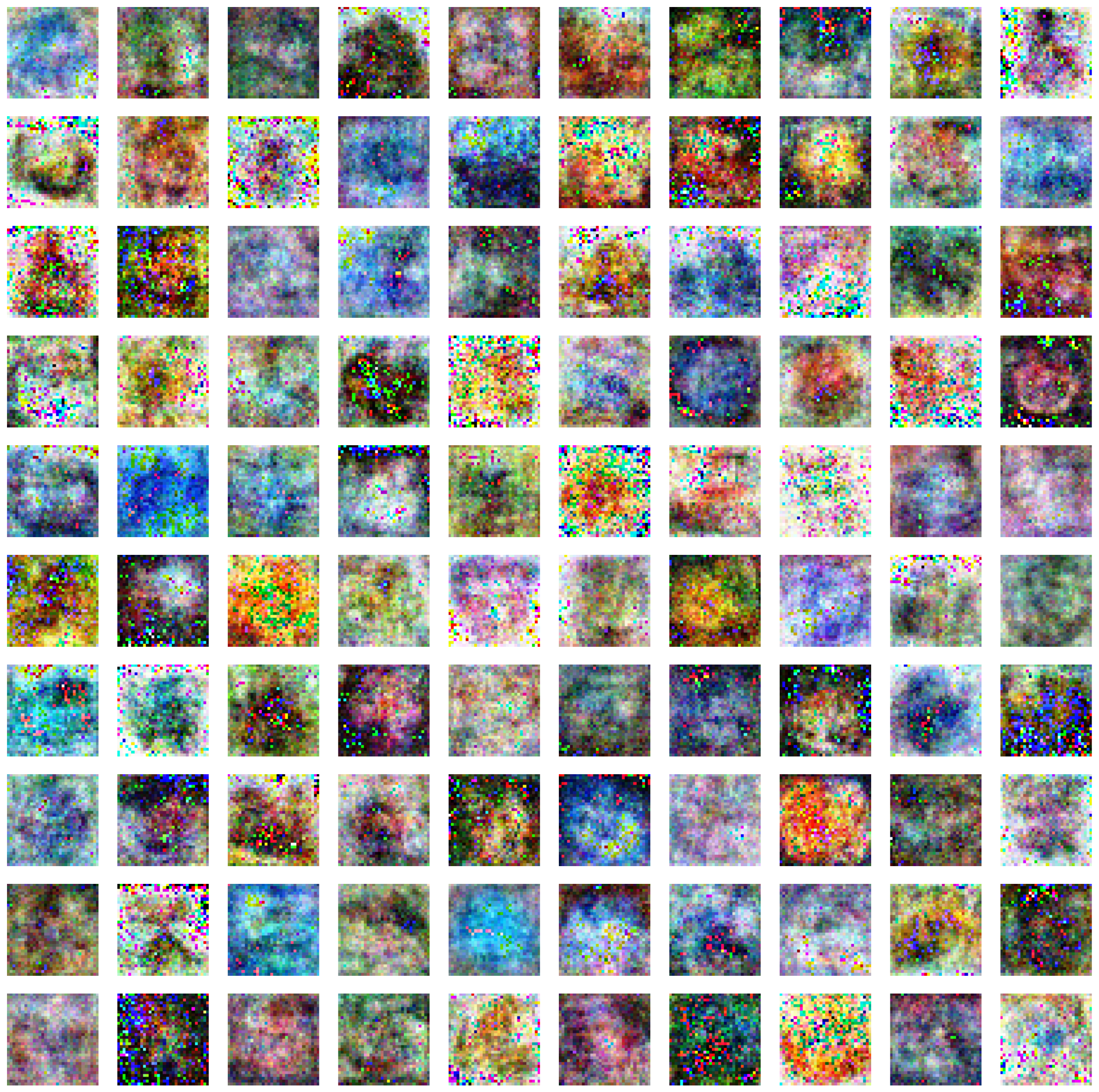}
    \caption{The reconstructed images by DLF on MLP}
    \label{fig:dlf_mlp}
\end{figure}

Recall that CPA uses a simple MLP with just one hidden layer of 256 neurons to successfully separate and recover dummy images from aggregated gradients (in Fig. \ref{fig:cpafc2}). Similarly, we will incorporate this simple MLP into optimization-based attack methods, such as DLF, to reduce the attack difficulty. The reconstructed images by DLF (100 total clients, 10 batch size, and 2 local epochs) on CIFAR-100 dataset are shown in Fig. \ref{fig:dlf_mlp}. It is evident that the reconstruction quality is better than that in Fig. \ref{fig:dlf_result_fedavg} using CNN model, although the objects in the images are still not highly recognizable.

\begin{figure}[!htbp]
    \centering
    \includegraphics[width=0.46\textwidth]{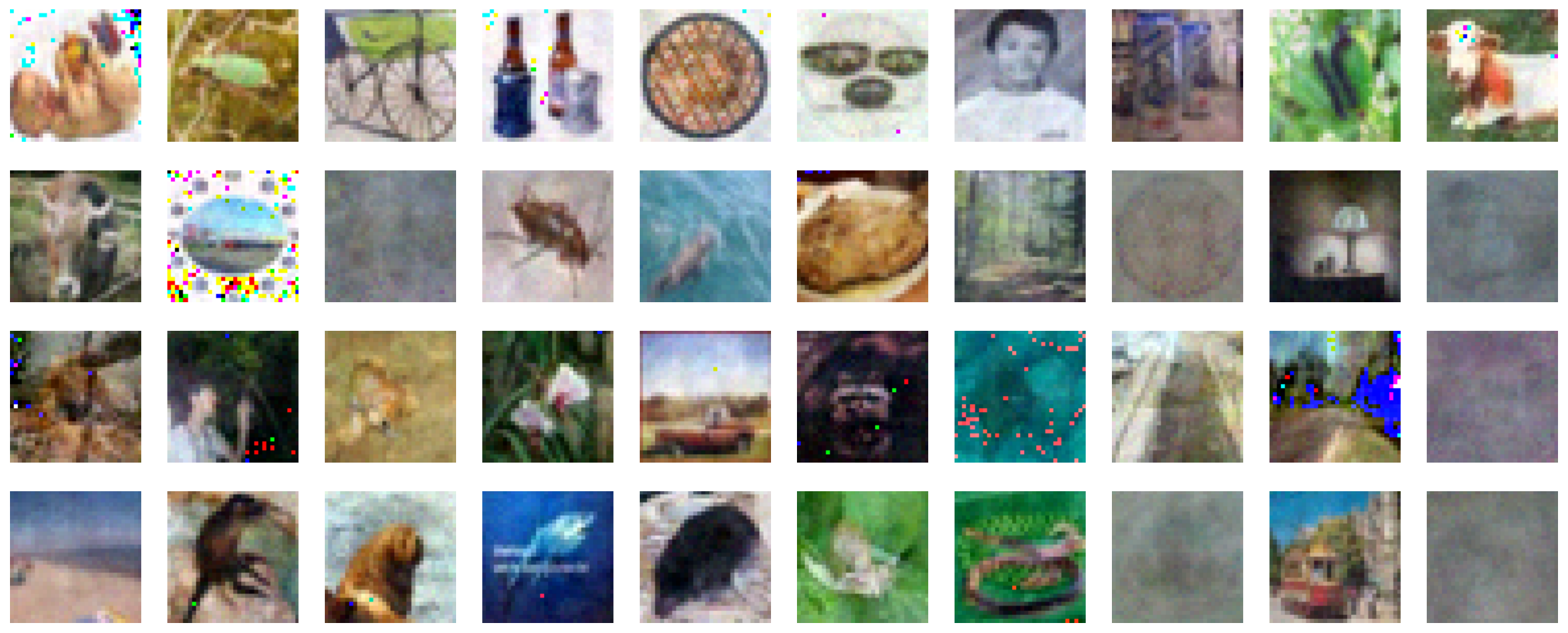}
    \caption{The reconstructed images by DLF on MLP with 1000 clients}
    \label{fig:dlf_mlp_1000c}
\end{figure}

To further reduce the difficulty of privacy attacks, we increase the total number of clients to 1,000, resulting in an approximate local data size of 50 per client, while keeping all other FL settings unchanged. The corresponding recovered images are shown in Fig. \ref{fig:dlf_mlp_1000c}. It is encouraging to see that, with a small local data size and a relatively simple training model, the optimization-based DLF can successfully recover private images with high quality.

\begin{figure}[!htbp]
    \centering
    \includegraphics[width=0.46\textwidth]{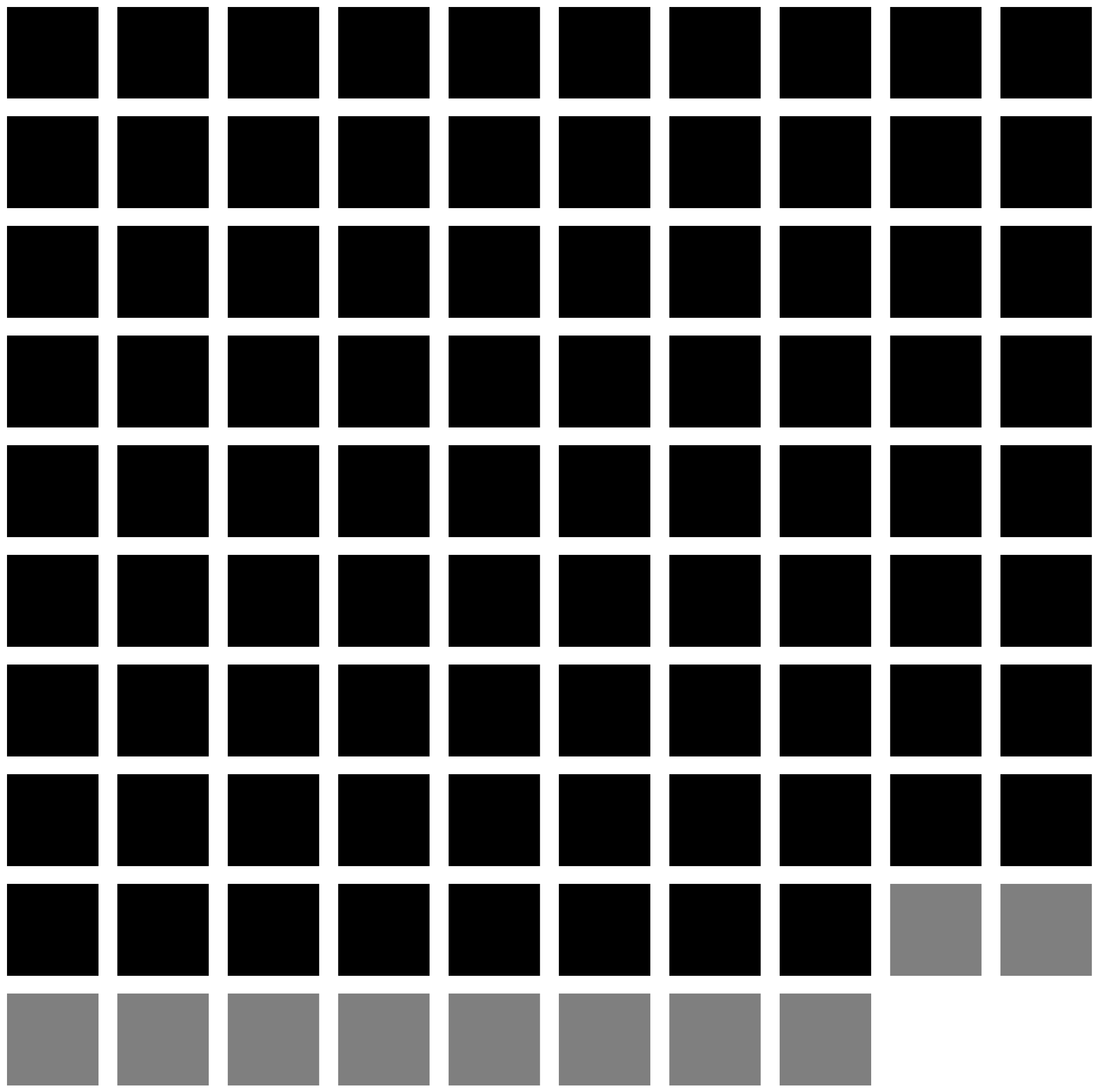}
    \caption{The reconstructed images by RTF on CNN}
    \label{fig:rtf_cnn}
\end{figure}

For RTF, we modify the backbone behind the Imprint model from ResNet18 to a CNN, and the results are presented in Fig. \ref{fig:rtf_cnn}. It is surprising to see that RTF fails to recover private images when the backbone is changed to a simpler CNN! This is likely due to the initialized parameters of the second linear layer in the Imprint module being too large, which can lead to severe gradient explosion during client local training. However, using alternative initialization methods or scaling the parameters to smaller values may negatively impact the quality of the inverted images. 

\subsection{Untrained Generator for GGL}
\begin{figure}[!htbp]
    \centering
    \includegraphics[width=0.46\textwidth]{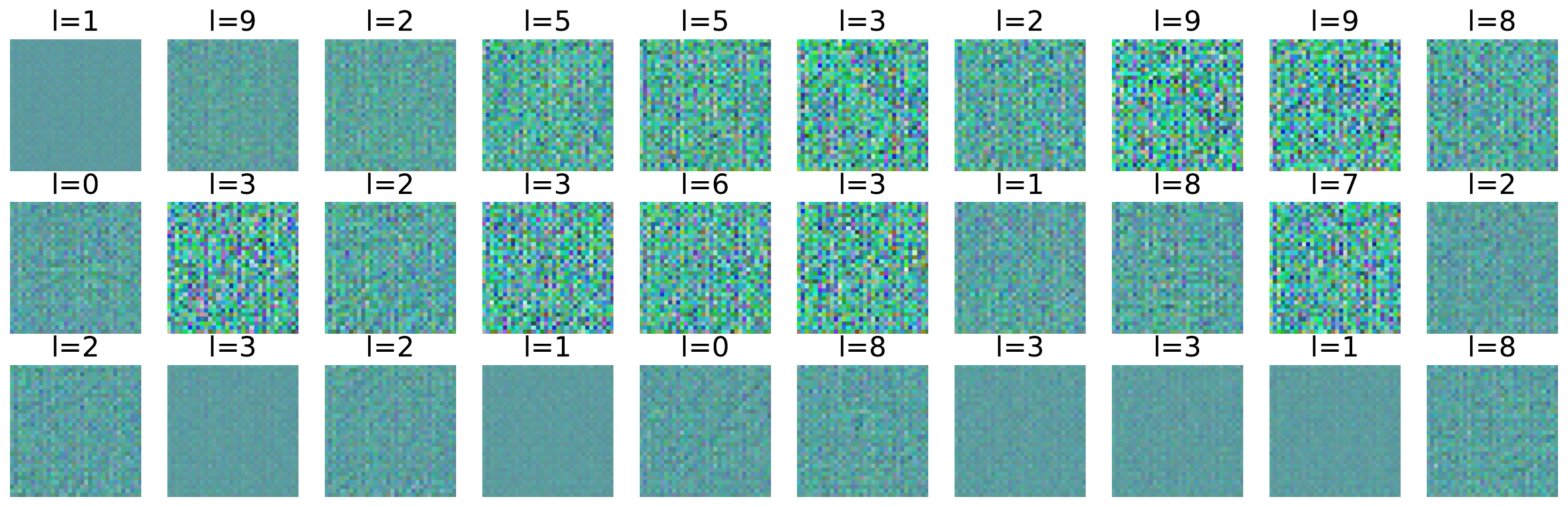}
    \caption{The reconstructed images by GGL with untrained generator}
    \label{fig:ggl_ut}
\end{figure}

In this section, we aim to investigate the previous reconstruction phenomenon of GGL by using an untrained generator. This approach will help determine whether the reconstruction process is dominated by the pretrained generator or influenced more by the client’s local training data. The reconstructed images by GGL using an untrained generator are shown in Fig. \ref{fig:ggl_ut}, which confirms our hypothesis that the recovered private images are primarily dominated by the training data of the pretrained generator. With the untrained generator, the gradient-free optimization loses its directional search capability.

\subsection{Federated Learning Performance}
\begin{figure}[!htbp]
    \centering
    \includegraphics[width=0.46\textwidth]{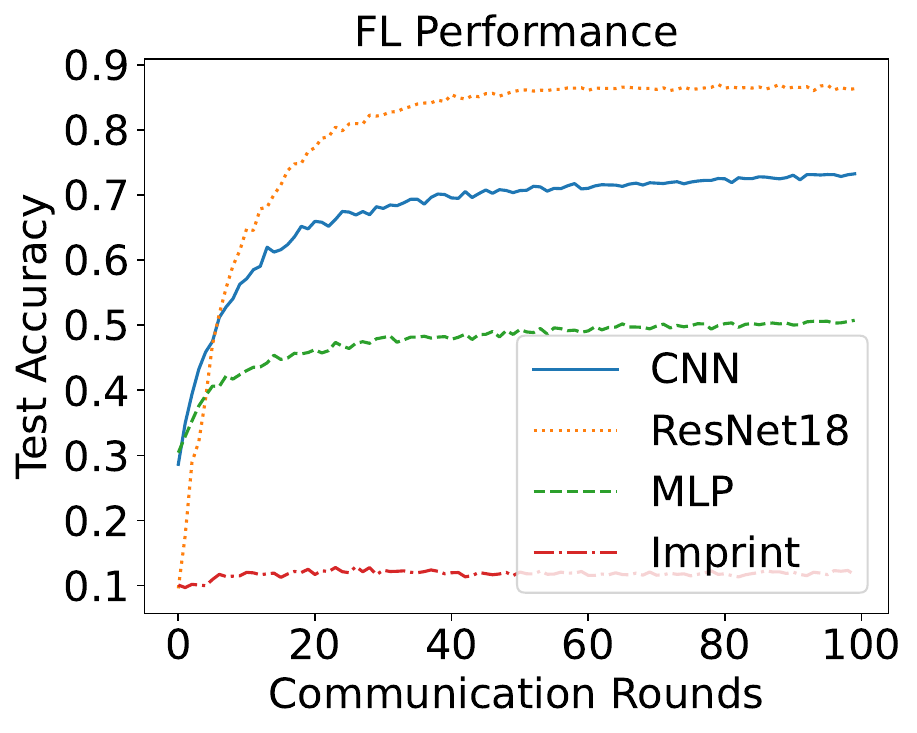}
    \caption{Test accuracy for 4 neural network models}
    \label{fig:fedavg_test_acc}
\end{figure}

To assess the applicability of the models used in privacy attacks, it is crucial to verify their learning performance under the FL environment. The corresponding test accuracy of CNN, ResNet18, MLP with a hidden layer, and ResNet18 with Imprint block for 100 communication rounds on CIFAR-10 dataset is shown in Fig. \ref{fig:fedavg_test_acc}, where both CNN and ResNet18 demonstrate promising FL performance with test accuracy of 73\% and 86\%, respectively. While the simple MLP achieves a relatively low convergence performance with 51\% test accuracy. And the Imprint block combined with ResNet18 appears to fail in achieving convergence, which significantly limits its applicability in real-world FL applications.

Consequently, neural network models with MLP-like structures (only containing fully connected layers) are more susceptible to privacy attacks. However, due to their poor learning performance, FL system builders are unlikely to use MLP models for training. As a result, attackers have fewer opportunities to exploit MLP-based privacy vulnerabilities. In addition, although RTF demonstrates superior attack performance even in real FL environments, the Imprint block significantly degrades overall FL performance. Since it is unlikely that the Imprint module would be incorporated into federated training when the goal is to achieve a high-quality global model, the RTF attack becomes unrealistic for application in FL systems.

 
\subsection{Remaining Challenges and Future Directions}
Although numerous privacy attack methods in FL have been proposed, many challenges remain to be addressed in the future.

\begin{enumerate}
    \item Most privacy attack algorithms are initially developed for gradient inversion scenarios, which differ significantly from the challenges posed by FL environments. As a result, these methods often struggle to effectively recover private images within the more complex context of FL, highlighting that privacy attacks in this setting are inherently challenging.
    \item Optimization-based attack approaches exhibit increased sensitivity to varying federated learning (FL) settings. Factors such as larger local data sizes, larger batch sizes, and an increased number of local epochs complicate the inference of data labels and the separation of individual data points. These complexities can misguide the optimization direction for generating dummy data, ultimately reducing the effectiveness of the attack.
    \item The analytic method RTF is the only evaluated server-side attack that demonstrates promising attack performance in FL environment. However, it necessitates modifying the global model by adding an additional Imprint block. This modification hinders the convergence of the global model, rendering it unsuitable for federated training.
    \item The client-side attack method DMGAN demonstrates good performance on the MNIST dataset, however, like RTF, it necessitates modifying the global model structure by adding an extra output class. Additionally, DMGAN struggles to perform effectively on more complex datasets, such as CIFAR-10.
    \item Future directions could focus on two main aspects: (1) developing analytic attacks against standard federated learning models, such as CNNs, and (2) exploring more advanced label count inference attacks that can effectively handle client model updates derived from averaged batch gradients computed over multiple local iterations.
    
\end{enumerate}


\section{Conclusion}\label{conclusion}
In this paper, we summarize various privacy attacks in FL and discuss their potential principle in detail. To evaluate their actual attack performance, we conducted an experimental study on nine representative attack algorithms, including eight server-side attacks and one client-side attack. Rather than relying on simple gradient inversion attacks, we applied these methods in more realistic FL environments. This approach highlights the significant gap between model gradients and the model differences between the global model and updated local models, making privacy attacks in these scenarios considerably more challenging.

The experimental results empirically demonstrate that most server-side optimization-based attacks fail to recover recognizable images containing clear sensitive private information. This failure is primarily due to incorrect label count approximations that mislead the optimization direction of the input search. Additionally, the task of separating and recovering individual images from client model parameters, which are computed using multiple averaged gradients from randomly sampled batch training data, is inherently complex and challenging. Nonetheless, the analytic method RTF exhibits superior image recovery performance by effectively capturing image brightness through model weights of linear layers. However, the RTF attack is limited to FC layers of the MLP neural network, and its carefully fine-tuned Imprint module encounters convergence issues with the global model, making it unsuitable for real FL systems. Similar situations also arise in the client-side attack DMGAN, where an additional output neuron is incorporated into the global model to function as the discriminator.

In summary, none of the existing privacy attack algorithms can effectively compromise private client images in FL without violating FL protocols or making inappropriate modifications to the global model, even in the absence of defense strategies. There remains significant work to be done in exploring privacy attack issues within FL in the future research work.



%



\section*{Acknowledgment}
This work was supported in part by the National Science Foundation of China (NSFC) under Grant 6240074545 and 62272201, in part by the Wuxi Science and Technology Development Fund Project under Grant K20231012, and in part by Basic Scientific Research Business Fund Project at Central Universities under Grant 22520231033.

\ifCLASSOPTIONcaptionsoff
  \newpage
\fi


\bibliographystyle{IEEEtran}
\bibliography{reference}

\begin{thebibliography}{100}
\providecommand{\url}[1]{#1}
\csname url@samestyle\endcsname
\providecommand{\newblock}{\relax}
\providecommand{\bibinfo}[2]{#2}
\providecommand{\BIBentrySTDinterwordspacing}{\spaceskip=0pt\relax}
\providecommand{\BIBentryALTinterwordstretchfactor}{4}
\providecommand{\BIBentryALTinterwordspacing}{\spaceskip=\fontdimen2\font plus
\BIBentryALTinterwordstretchfactor\fontdimen3\font minus \fontdimen4\font\relax}
\providecommand{\BIBforeignlanguage}[2]{{%
\expandafter\ifx\csname l@#1\endcsname\relax
\typeout{** WARNING: IEEEtran.bst: No hyphenation pattern has been}%
\typeout{** loaded for the language `#1'. Using the pattern for}%
\typeout{** the default language instead.}%
\else
\language=\csname l@#1\endcsname
\fi
#2}}
\providecommand{\BIBdecl}{\relax}
\BIBdecl

\bibitem{voulodimos2018deep}
A.~Voulodimos, N.~Doulamis, A.~Doulamis, and E.~Protopapadakis, ``Deep learning for computer vision: A brief review,'' \emph{Computational intelligence and neuroscience}, vol. 2018, 2018.

\bibitem{chowdhary2020natural}
K.~Chowdhary and K.~Chowdhary, ``Natural language processing,'' \emph{Fundamentals of artificial intelligence}, pp. 603--649, 2020.

\bibitem{lecun2015deep}
Y.~LeCun, Y.~Bengio, and G.~Hinton, ``Deep learning,'' \emph{nature}, vol. 521, no. 7553, pp. 436--444, 2015.

\bibitem{kaissis2020secure}
G.~A. Kaissis, M.~R. Makowski, D.~R{\"u}ckert, and R.~F. Braren, ``Secure, privacy-preserving and federated machine learning in medical imaging,'' \emph{Nature Machine Intelligence}, vol.~2, no.~6, pp. 305--311, 2020.

\bibitem{fredrikson2014privacy}
M.~Fredrikson, E.~Lantz, S.~Jha, S.~Lin, D.~Page, and T.~Ristenpart, ``Privacy in pharmacogenetics: An $\{$End-to-End$\}$ case study of personalized warfarin dosing,'' in \emph{23rd USENIX security symposium (USENIX Security 14)}, 2014, pp. 17--32.

\bibitem{mcpherson2016defeating}
R.~McPherson, R.~Shokri, and V.~Shmatikov, ``Defeating image obfuscation with deep learning,'' \emph{arXiv preprint arXiv:1609.00408}, 2016.

\bibitem{regulation2016regulation}
P.~Regulation, ``Regulation (eu) 2016/679 of the european parliament and of the council,'' \emph{Regulation (eu)}, vol. 679, p. 2016, 2016.

\bibitem{mcmahan2017communication}
B.~McMahan, E.~Moore, D.~Ramage, S.~Hampson, and B.~A. y~Arcas, ``Communication-efficient learning of deep networks from decentralized data,'' in \emph{Artificial intelligence and statistics}.\hskip 1em plus 0.5em minus 0.4em\relax PMLR, 2017, pp. 1273--1282.

\bibitem{yang2019federated}
Q.~Yang, Y.~Liu, T.~Chen, and Y.~Tong, ``Federated machine learning: Concept and applications,'' \emph{ACM Transactions on Intelligent Systems and Technology (TIST)}, vol.~10, no.~2, pp. 1--19, 2019.

\bibitem{fang2020local}
M.~Fang, X.~Cao, J.~Jia, and N.~Gong, ``Local model poisoning attacks to $\{$Byzantine-Robust$\}$ federated learning,'' in \emph{29th USENIX security symposium (USENIX Security 20)}, 2020, pp. 1605--1622.

\bibitem{bhagoji2019analyzing}
A.~N. Bhagoji, S.~Chakraborty, P.~Mittal, and S.~Calo, ``Analyzing federated learning through an adversarial lens,'' in \emph{International Conference on Machine Learning}.\hskip 1em plus 0.5em minus 0.4em\relax PMLR, 2019, pp. 634--643.

\bibitem{Cao_2022_CVPR}
X.~Cao and N.~Z. Gong, ``Mpaf: Model poisoning attacks to federated learning based on fake clients,'' in \emph{Proceedings of the IEEE/CVF Conference on Computer Vision and Pattern Recognition (CVPR) Workshops}, June 2022, pp. 3396--3404.

\bibitem{fi13030073}
\BIBentryALTinterwordspacing
X.~Zhou, M.~Xu, Y.~Wu, and N.~Zheng, ``Deep model poisoning attack on federated learning,'' \emph{Future Internet}, vol.~13, no.~3, 2021. [Online]. Available: \url{https://www.mdpi.com/1999-5903/13/3/73}
\BIBentrySTDinterwordspacing

\bibitem{8975792}
D.~Cao, S.~Chang, Z.~Lin, G.~Liu, and D.~Sun, ``Understanding distributed poisoning attack in federated learning,'' in \emph{2019 IEEE 25th International Conference on Parallel and Distributed Systems (ICPADS)}, 2019, pp. 233--239.

\bibitem{NEURIPS2021_692baebe}
J.~Sun, A.~Li, L.~DiValentin, A.~Hassanzadeh, Y.~Chen, and H.~Li, ``Fl-wbc: Enhancing robustness against model poisoning attacks in federated learning from a client perspective,'' in \emph{Advances in Neural Information Processing Systems}, M.~Ranzato, A.~Beygelzimer, Y.~Dauphin, P.~Liang, and J.~W. Vaughan, Eds., vol.~34.\hskip 1em plus 0.5em minus 0.4em\relax Curran Associates, Inc., 2021, pp. 12\,613--12\,624.

\bibitem{pmlr-v151-panda22a}
\BIBentryALTinterwordspacing
A.~Panda, S.~Mahloujifar, A.~Nitin~Bhagoji, S.~Chakraborty, and P.~Mittal, ``Sparsefed: Mitigating model poisoning attacks in federated learning with sparsification,'' in \emph{Proceedings of The 25th International Conference on Artificial Intelligence and Statistics}, ser. Proceedings of Machine Learning Research, G.~Camps-Valls, F.~J.~R. Ruiz, and I.~Valera, Eds., vol. 151.\hskip 1em plus 0.5em minus 0.4em\relax PMLR, 28-30 Mar 2022, pp. 7587--7624. [Online]. Available: \url{https://proceedings.mlr.press/v151/panda22a.html}
\BIBentrySTDinterwordspacing

\bibitem{Shejwalkar2021ManipulatingTB}
\BIBentryALTinterwordspacing
V.~Shejwalkar and A.~Houmansadr, ``Manipulating the byzantine: Optimizing model poisoning attacks and defenses for federated learning,'' \emph{Proceedings 2021 Network and Distributed System Security Symposium}, 2021. [Online]. Available: \url{https://api.semanticscholar.org/CorpusID:231861235}
\BIBentrySTDinterwordspacing

\bibitem{wang2020attack}
H.~Wang, K.~Sreenivasan, S.~Rajput, H.~Vishwakarma, S.~Agarwal, J.-y. Sohn, K.~Lee, and D.~Papailiopoulos, ``Attack of the tails: Yes, you really can backdoor federated learning,'' \emph{Advances in Neural Information Processing Systems}, vol.~33, pp. 16\,070--16\,084, 2020.

\bibitem{bagdasaryan2020backdoor}
E.~Bagdasaryan, A.~Veit, Y.~Hua, D.~Estrin, and V.~Shmatikov, ``How to backdoor federated learning,'' in \emph{International conference on artificial intelligence and statistics}.\hskip 1em plus 0.5em minus 0.4em\relax PMLR, 2020, pp. 2938--2948.

\bibitem{orekondy2018gradient}
T.~Orekondy, S.~J. Oh, Y.~Zhang, B.~Schiele, and M.~Fritz, ``Gradient-leaks: Understanding and controlling deanonymization in federated learning,'' \emph{arXiv preprint arXiv:1805.05838}, 2018.

\bibitem{nasr2019comprehensive}
M.~Nasr, R.~Shokri, and A.~Houmansadr, ``Comprehensive privacy analysis of deep learning: Passive and active white-box inference attacks against centralized and federated learning,'' in \emph{2019 IEEE symposium on security and privacy (SP)}.\hskip 1em plus 0.5em minus 0.4em\relax IEEE, 2019, pp. 739--753.

\bibitem{luo2021feature}
X.~Luo, Y.~Wu, X.~Xiao, and B.~C. Ooi, ``Feature inference attack on model predictions in vertical federated learning,'' in \emph{2021 IEEE 37th International Conference on Data Engineering (ICDE)}.\hskip 1em plus 0.5em minus 0.4em\relax IEEE, 2021, pp. 181--192.

\bibitem{pustozerova2020information}
A.~Pustozerova and R.~Mayer, ``Information leaks in federated learning,'' in \emph{Proceedings of the network and distributed system security symposium}, vol.~10, 2020, p. 122.

\bibitem{carlini2021extracting}
N.~Carlini, F.~Tramer, E.~Wallace, M.~Jagielski, A.~Herbert-Voss, K.~Lee, A.~Roberts, T.~Brown, D.~Song, U.~Erlingsson \emph{et~al.}, ``Extracting training data from large language models,'' in \emph{30th USENIX Security Symposium (USENIX Security 21)}, 2021, pp. 2633--2650.

\bibitem{jin2021cafe}
X.~Jin, P.-Y. Chen, C.-Y. Hsu, C.-M. Yu, and T.~Chen, ``Cafe: Catastrophic data leakage in vertical federated learning,'' \emph{Advances in Neural Information Processing Systems}, vol.~34, pp. 994--1006, 2021.

\bibitem{wu2023learning}
R.~Wu, X.~Chen, C.~Guo, and K.~Q. Weinberger, ``Learning to invert: Simple adaptive attacks for gradient inversion in federated learning,'' in \emph{Uncertainty in Artificial Intelligence}.\hskip 1em plus 0.5em minus 0.4em\relax PMLR, 2023, pp. 2293--2303.

\bibitem{zhu2019deep}
L.~Zhu, Z.~Liu, and S.~Han, ``Deep leakage from gradients,'' \emph{Advances in neural information processing systems}, vol.~32, 2019.

\bibitem{zhao2020idlg}
B.~Zhao, K.~R. Mopuri, and H.~Bilen, ``idlg: Improved deep leakage from gradients,'' \emph{arXiv preprint arXiv:2001.02610}, 2020.

\bibitem{geiping2020inverting}
J.~Geiping, H.~Bauermeister, H.~Dr{\"o}ge, and M.~Moeller, ``Inverting gradients-how easy is it to break privacy in federated learning?'' \emph{Advances in Neural Information Processing Systems}, vol.~33, pp. 16\,937--16\,947, 2020.

\bibitem{hitaj2017deep}
B.~Hitaj, G.~Ateniese, and F.~Perez-Cruz, ``Deep models under the gan: information leakage from collaborative deep learning,'' in \emph{Proceedings of the 2017 ACM SIGSAC conference on computer and communications security}, 2017, pp. 603--618.

\bibitem{kariyappa2023cocktail}
S.~Kariyappa, C.~Guo, K.~Maeng, W.~Xiong, G.~E. Suh, M.~K. Qureshi, and H.-H.~S. Lee, ``Cocktail party attack: Breaking aggregation-based privacy in federated learning using independent component analysis,'' in \emph{International Conference on Machine Learning}.\hskip 1em plus 0.5em minus 0.4em\relax PMLR, 2023, pp. 15\,884--15\,899.

\bibitem{yin2021see}
H.~Yin, A.~Mallya, A.~Vahdat, J.~M. Alvarez, J.~Kautz, and P.~Molchanov, ``See through gradients: Image batch recovery via gradinversion,'' in \emph{Proceedings of the IEEE/CVF Conference on Computer Vision and Pattern Recognition}, 2021, pp. 16\,337--16\,346.

\bibitem{zhao2023secure}
J.~C. Zhao, A.~Sharma, A.~R. Elkordy, Y.~H. Ezzeldin, S.~Avestimehr, and S.~Bagchi, ``Secure aggregation in federated learning is not private: Leaking user data at large scale through model modification,'' \emph{arXiv preprint arXiv:2303.12233}, 2023.

\bibitem{li2022auditing}
Z.~Li, J.~Zhang, L.~Liu, and J.~Liu, ``Auditing privacy defenses in federated learning via generative gradient leakage,'' in \emph{Proceedings of the IEEE/CVF Conference on Computer Vision and Pattern Recognition}, 2022, pp. 10\,132--10\,142.

\bibitem{ren2022grnn}
H.~Ren, J.~Deng, and X.~Xie, ``Grnn: Generative regression neural network—a data leakage attack for federated learning,'' \emph{ACM Transactions on Intelligent Systems and Technology (TIST)}, vol.~13, no.~4, pp. 1--24, 2022.

\bibitem{geng2021towards}
J.~Geng, Y.~Mou, F.~Li, Q.~Li, O.~Beyan, S.~Decker, and C.~Rong, ``Towards general deep leakage in federated learning,'' \emph{arXiv preprint arXiv:2110.09074}, 2021.

\bibitem{fowl2022robbing}
\BIBentryALTinterwordspacing
L.~H. Fowl, J.~Geiping, W.~Czaja, M.~Goldblum, and T.~Goldstein, ``Robbing the fed: Directly obtaining private data in federated learning with modified models,'' in \emph{International Conference on Learning Representations}, 2022. [Online]. Available: \url{https://openreview.net/forum?id=fwzUgo0FM9v}
\BIBentrySTDinterwordspacing

\bibitem{9076003}
Y.~Liu, Y.~Kang, C.~Xing, T.~Chen, and Q.~Yang, ``A secure federated transfer learning framework,'' \emph{IEEE Intelligent Systems}, vol.~35, no.~4, pp. 70--82, 2020.

\bibitem{saha2021federated}
S.~Saha and T.~Ahmad, ``Federated transfer learning: Concept and applications,'' \emph{Intelligenza Artificiale}, vol.~15, no.~1, pp. 35--44, 2021.

\bibitem{hard2018federated}
A.~Hard, K.~Rao, R.~Mathews, S.~Ramaswamy, F.~Beaufays, S.~Augenstein, H.~Eichner, C.~Kiddon, and D.~Ramage, ``Federated learning for mobile keyboard prediction,'' \emph{arXiv preprint arXiv:1811.03604}, 2018.

\bibitem{liu2020fedvision}
Y.~Liu, A.~Huang, Y.~Luo, H.~Huang, Y.~Liu, Y.~Chen, L.~Feng, T.~Chen, H.~Yu, and Q.~Yang, ``Fedvision: An online visual object detection platform powered by federated learning,'' in \emph{Proceedings of the AAAI conference on artificial intelligence}, vol.~34, no.~08, 2020, pp. 13\,172--13\,179.

\bibitem{li2021survey}
Q.~Li, Z.~Wen, Z.~Wu, S.~Hu, N.~Wang, Y.~Li, X.~Liu, and B.~He, ``A survey on federated learning systems: Vision, hype and reality for data privacy and protection,'' \emph{IEEE Transactions on Knowledge and Data Engineering}, 2021.

\bibitem{kang2022privacy}
Y.~Kang, Y.~He, J.~Luo, T.~Fan, Y.~Liu, and Q.~Yang, ``Privacy-preserving federated adversarial domain adaptation over feature groups for interpretability,'' \emph{IEEE Transactions on Big Data}, 2022.

\bibitem{liang2021Self}
X.~Liang, Y.~Liu, J.~Luo, Y.~He, T.~Chen, and Q.~Yang, ``Self-supervised cross-silo federated neural architecture search,'' \emph{arXiv preprint arXiv:2101.11896}, 2021.

\bibitem{li2021label}
O.~Li, J.~Sun, X.~Yang, W.~Gao, H.~Zhang, J.~Xie, V.~Smith, and C.~Wang, ``Label leakage and protection in two-party split learning,'' \emph{arXiv preprint arXiv:2102.08504}, 2021.

\bibitem{wainakh2022user}
A.~Wainakh, F.~Ventola, T.~M{\"u}{\ss}ig, J.~Keim, C.~G. Cordero, E.~Zimmer, T.~Grube, K.~Kersting, and M.~M{\"u}hlh{\"a}user, ``User-level label leakage from gradients in federated learning,'' \emph{Proceedings on Privacy Enhancing Technologies}, vol.~2, pp. 227--244, 2022.

\bibitem{dimitrov2022data}
D.~I. Dimitrov, M.~Balunovic, N.~Konstantinov, and M.~Vechev, ``Data leakage in federated averaging,'' \emph{Transactions on Machine Learning Research}, 2022.

\bibitem{ma2022instance}
K.~Ma, Y.~Sun, J.~Cui, D.~Li, Z.~Guan, and J.~Liu, ``Instance-wise batch label restoration via gradients in federated learning,'' in \emph{The Eleventh International Conference on Learning Representations}, 2022.

\bibitem{chen2024recovering}
H.~Chen and H.~Vikalo, ``Recovering labels from local updates in federated learning,'' \emph{arXiv preprint arXiv:2405.00955}, 2024.

\bibitem{ganju2018property}
K.~Ganju, Q.~Wang, W.~Yang, C.~A. Gunter, and N.~Borisov, ``Property inference attacks on fully connected neural networks using permutation invariant representations,'' in \emph{Proceedings of the 2018 ACM SIGSAC conference on computer and communications security}, 2018, pp. 619--633.

\bibitem{CARTUYVELS2021143}
R.~Cartuyvels, G.~Spinks, and M.-F. Moens, ``Discrete and continuous representations and processing in deep learning: Looking forward,'' \emph{AI Open}, vol.~2, pp. 143--159, 2021.

\bibitem{melis2019exploiting}
L.~Melis, C.~Song, E.~De~Cristofaro, and V.~Shmatikov, ``Exploiting unintended feature leakage in collaborative learning,'' in \emph{2019 IEEE symposium on security and privacy (SP)}.\hskip 1em plus 0.5em minus 0.4em\relax IEEE, 2019, pp. 691--706.

\bibitem{gupta2022recovering}
S.~Gupta, Y.~Huang, Z.~Zhong, T.~Gao, K.~Li, and D.~Chen, ``Recovering private text in federated learning of language models,'' \emph{Advances in Neural Information Processing Systems}, vol.~35, pp. 8130--8143, 2022.

\bibitem{song2020information}
C.~Song and A.~Raghunathan, ``Information leakage in embedding models,'' in \emph{Proceedings of the 2020 ACM SIGSAC conference on computer and communications security}, 2020, pp. 377--390.

\bibitem{lyu2021novel}
L.~Lyu and C.~Chen, ``A novel attribute reconstruction attack in federated learning,'' \emph{arXiv preprint arXiv:2108.06910}, 2021.

\bibitem{goodfellow2014generative}
I.~J. Goodfellow, J.~Pouget-Abadie, M.~Mirza, B.~Xu, D.~Warde-Farley, S.~Ozair, A.~Courville, and Y.~Bengio, ``Generative adversarial networks,'' 2014.

\bibitem{deng-etal-2021-tag-gradient}
\BIBentryALTinterwordspacing
J.~Deng, Y.~Wang, J.~Li, C.~Wang, C.~Shang, H.~Liu, S.~Rajasekaran, and C.~Ding, ``{TAG}: Gradient attack on transformer-based language models,'' in \emph{Findings of the Association for Computational Linguistics: EMNLP 2021}, M.-F. Moens, X.~Huang, L.~Specia, and S.~W.-t. Yih, Eds.\hskip 1em plus 0.5em minus 0.4em\relax Punta Cana, Dominican Republic: Association for Computational Linguistics, Nov. 2021, pp. 3600--3610. [Online]. Available: \url{https://aclanthology.org/2021.findings-emnlp.305}
\BIBentrySTDinterwordspacing

\bibitem{pan2020theory}
X.~Pan, M.~Zhang, Y.~Yan, J.~Zhu, and M.~Yang, ``Theory-oriented deep leakage from gradients via linear equation solver,'' \emph{arXiv preprint arXiv:2010.13356}, vol.~1, 2020.

\bibitem{li2021deep}
Z.~Li, M.~Hubchak, and Y.~Zhu, ``Deep leakage from gradients in multiple-label medical image classification,'' in \emph{2021 IEEE 9th International Conference on Healthcare Informatics (ICHI)}.\hskip 1em plus 0.5em minus 0.4em\relax IEEE, 2021, pp. 447--448.

\bibitem{luo2022effective}
Z.~Luo, C.~Zhu, L.~Fang, G.~Kou, R.~Hou, and X.~Wang, ``An effective and practical gradient inversion attack,'' \emph{International Journal of Intelligent Systems}, vol.~37, no.~11, pp. 9373--9389, 2022.

\bibitem{qian2020can}
J.~Qian and L.~K. Hansen, ``What can we learn from gradients?'' 2020.

\bibitem{krizhevsky2009learning}
A.~Krizhevsky, G.~Hinton \emph{et~al.}, ``Learning multiple layers of features from tiny images,'' 2009.

\bibitem{wang2020sapag}
Y.~Wang, J.~Deng, D.~Guo, C.~Wang, X.~Meng, H.~Liu, C.~Ding, and S.~Rajasekaran, ``Sapag: A self-adaptive privacy attack from gradients,'' \emph{arXiv preprint arXiv:2009.06228}, 2020.

\bibitem{RUDIN1992259}
L.~I. Rudin, S.~Osher, and E.~Fatemi, ``Nonlinear total variation based noise removal algorithms,'' \emph{Physica D: Nonlinear Phenomena}, vol.~60, no.~1, pp. 259--268, 1992.

\bibitem{he2023fast}
X.~He, C.~Peng, W.~Tan \emph{et~al.}, ``Fast and accurate deep leakage from gradients based on wasserstein distance,'' \emph{International Journal of Intelligent Systems}, vol. 2023, 2023.

\bibitem{vaserstein1969markov}
L.~N. Vaserstein, ``Markov processes over denumerable products of spaces, describing large systems of automata,'' \emph{Problemy Peredachi Informatsii}, vol.~5, no.~3, pp. 64--72, 1969.

\bibitem{jeon2021gradient}
J.~Jeon, K.~Lee, S.~Oh, J.~Ok \emph{et~al.}, ``Gradient inversion with generative image prior,'' \emph{Advances in neural information processing systems}, vol.~34, pp. 29\,898--29\,908, 2021.

\bibitem{hatamizadeh2022gradvit}
A.~Hatamizadeh, H.~Yin, H.~R. Roth, W.~Li, J.~Kautz, D.~Xu, and P.~Molchanov, ``Gradvit: Gradient inversion of vision transformers,'' in \emph{Proceedings of the IEEE/CVF Conference on Computer Vision and Pattern Recognition}, 2022, pp. 10\,021--10\,030.

\bibitem{10.5555/2969033.2969202}
P.~Bachman, O.~Alsharif, and D.~Precup, ``Learning with pseudo-ensembles,'' in \emph{Proceedings of the 27th International Conference on Neural Information Processing Systems - Volume 2}, ser. NIPS'14.\hskip 1em plus 0.5em minus 0.4em\relax Cambridge, MA, USA: MIT Press, 2014, p. 3365–3373.

\bibitem{yang2022using}
H.~Yang, M.~Ge, K.~Xiang, and J.~Li, ``Using highly compressed gradients in federated learning for data reconstruction attacks,'' \emph{IEEE Transactions on Information Forensics and Security}, vol.~18, pp. 818--830, 2022.

\bibitem{wei2020framework}
W.~Wei, L.~Liu, M.~Loper, K.-H. Chow, M.~E. Gursoy, S.~Truex, and Y.~Wu, ``A framework for evaluating gradient leakage attacks in federated learning,'' \emph{arXiv preprint arXiv:2004.10397}, 2020.

\bibitem{pmlr-v234-zhao24b}
\BIBentryALTinterwordspacing
Z.~Zhao, M.~Luo, and W.~Ding, ``Deep leakage from model in federated learning,'' in \emph{Conference on Parsimony and Learning}, ser. Proceedings of Machine Learning Research, Y.~Chi, G.~K. Dziugaite, Q.~Qu, A.~W. Wang, and Z.~Zhu, Eds., vol. 234.\hskip 1em plus 0.5em minus 0.4em\relax PMLR, 03--06 Jan 2024, pp. 324--340. [Online]. Available: \url{https://proceedings.mlr.press/v234/zhao24b.html}
\BIBentrySTDinterwordspacing

\bibitem{10445924}
Y.~Sun, G.~Xiong, X.~Yao, K.~Ma, and J.~Cui, ``Gi-pip: Do we require impractical auxiliary dataset for gradient inversion attacks?'' in \emph{ICASSP 2024 - 2024 IEEE International Conference on Acoustics, Speech and Signal Processing (ICASSP)}, 2024, pp. 4675--4679.

\bibitem{yin2020dreaming}
H.~Yin, P.~Molchanov, J.~M. Alvarez, Z.~Li, A.~Mallya, D.~Hoiem, N.~K. Jha, and J.~Kautz, ``Dreaming to distill: Data-free knowledge transfer via deepinversion,'' in \emph{Proceedings of the IEEE/CVF Conference on Computer Vision and Pattern Recognition}, 2020, pp. 8715--8724.

\bibitem{HYVARINEN2000411}
A.~Hyvärinen and E.~Oja, ``Independent component analysis: algorithms and applications,'' \emph{Neural Networks}, vol.~13, no.~4, pp. 411--430, 2000.

\bibitem{le2015tiny}
Y.~Le and X.~Yang, ``Tiny imagenet visual recognition challenge,'' \emph{CS 231N}, vol.~7, no.~7, p.~3, 2015.

\bibitem{mahendran2015understanding}
A.~Mahendran and A.~Vedaldi, ``Understanding deep image representations by inverting them,'' in \emph{Proceedings of the IEEE conference on computer vision and pattern recognition}, 2015, pp. 5188--5196.

\bibitem{ulyanov2018deep}
D.~Ulyanov, A.~Vedaldi, and V.~Lempitsky, ``Deep image prior,'' in \emph{Proceedings of the IEEE conference on computer vision and pattern recognition}, 2018, pp. 9446--9454.

\bibitem{simonyan2014very}
K.~Simonyan and A.~Zisserman, ``Very deep convolutional networks for large-scale image recognition,'' \emph{arXiv preprint arXiv:1409.1556}, 2014.

\bibitem{phong2017privacy}
L.~T. Phong, Y.~Aono, T.~Hayashi, L.~Wang, and S.~Moriai, ``Privacy-preserving deep learning: Revisited and enhanced,'' in \emph{Applications and Techniques in Information Security: 8th International Conference, ATIS 2017, Auckland, New Zealand, July 6--7, 2017, Proceedings}.\hskip 1em plus 0.5em minus 0.4em\relax Springer, 2017, pp. 100--110.

\bibitem{enthoven2022fidel}
D.~Enthoven and Z.~Al-Ars, ``Fidel: Reconstructing private training samples from weight updates in federated learning,'' in \emph{2022 9th International Conference on Internet of Things: Systems, Management and Security (IOTSMS)}.\hskip 1em plus 0.5em minus 0.4em\relax IEEE, 2022, pp. 1--8.

\bibitem{qian2020minimal}
J.~Qian, H.~Nassar, and L.~K. Hansen, ``Minimal model structure analysis for input reconstruction in federated learning,'' \emph{arXiv preprint arXiv:2010.15718}, 2020.

\bibitem{fan2020rethinking}
L.~Fan, K.~W. Ng, C.~Ju, T.~Zhang, C.~Liu, C.~S. Chan, and Q.~Yang, ``Rethinking privacy preserving deep learning: How to evaluate and thwart privacy attacks,'' \emph{Federated Learning: Privacy and Incentive}, pp. 32--50, 2020.

\bibitem{zhu2020r}
J.~Zhu and M.~Blaschko, ``R-gap: Recursive gradient attack on privacy,'' \emph{arXiv preprint arXiv:2010.07733}, 2020.

\bibitem{wen2022fishing}
Y.~Wen, J.~A. Geiping, L.~Fowl, M.~Goldblum, and T.~Goldstein, ``Fishing for user data in large-batch federated learning via gradient magnification,'' in \emph{International Conference on Machine Learning}.\hskip 1em plus 0.5em minus 0.4em\relax PMLR, 2022, pp. 23\,668--23\,684.

\bibitem{fowl2022decepticons}
L.~Fowl, J.~Geiping, S.~Reich, Y.~Wen, W.~Czaja, M.~Goldblum, and T.~Goldstein, ``Decepticons: Corrupted transformers breach privacy in federated learning for language models,'' \emph{arXiv preprint arXiv:2201.12675}, 2022.

\bibitem{zhao2023loki}
J.~C. Zhao, A.~Sharma, A.~R. Elkordy, Y.~H. Ezzeldin, S.~Avestimehr, and S.~Bagchi, ``Loki: Large-scale data reconstruction attack against federated learning through model manipulation,'' in \emph{2024 IEEE Symposium on Security and Privacy (SP)}.\hskip 1em plus 0.5em minus 0.4em\relax IEEE Computer Society, 2023, pp. 30--30.

\bibitem{boenisch2023curious}
F.~Boenisch, A.~Dziedzic, R.~Schuster, A.~S. Shamsabadi, I.~Shumailov, and N.~Papernot, ``When the curious abandon honesty: Federated learning is not private,'' in \emph{2023 IEEE 8th European Symposium on Security and Privacy (EuroS\&P)}.\hskip 1em plus 0.5em minus 0.4em\relax IEEE, 2023, pp. 175--199.

\bibitem{lam2021gradient}
M.~Lam, G.-Y. Wei, D.~Brooks, V.~J. Reddi, and M.~Mitzenmacher, ``Gradient disaggregation: Breaking privacy in federated learning by reconstructing the user participant matrix,'' in \emph{International Conference on Machine Learning}.\hskip 1em plus 0.5em minus 0.4em\relax PMLR, 2021, pp. 5959--5968.

\bibitem{pasquini2022eluding}
D.~Pasquini, D.~Francati, and G.~Ateniese, ``Eluding secure aggregation in federated learning via model inconsistency,'' in \emph{Proceedings of the 2022 ACM SIGSAC Conference on Computer and Communications Security}, 2022, pp. 2429--2443.

\bibitem{hayes2017logan}
J.~Hayes, L.~Melis, G.~Danezis, and E.~De~Cristofaro, ``Logan: evaluating privacy leakage of generative models using generative adversarial networks,'' \emph{arXiv preprint arXiv:1705.07663}, pp. 506--519, 2017.

\bibitem{ha2022inference}
T.~Ha and T.~K. Dang, ``Inference attacks based on gan in federated learning,'' \emph{International Journal of Web Information Systems}, vol.~18, no. 2/3, pp. 117--136, 2022.

\bibitem{yang2019adversarial}
Z.~Yang, E.-C. Chang, and Z.~Liang, ``Adversarial neural network inversion via auxiliary knowledge alignment,'' \emph{arXiv preprint arXiv:1902.08552}, 2019.

\bibitem{zhang2020secret}
Y.~Zhang, R.~Jia, H.~Pei, W.~Wang, B.~Li, and D.~Song, ``The secret revealer: Generative model-inversion attacks against deep neural networks,'' in \emph{Proceedings of the IEEE/CVF conference on computer vision and pattern recognition}, 2020, pp. 253--261.

\bibitem{salem2020updates}
A.~Salem, A.~Bhattacharya, M.~Backes, M.~Fritz, and Y.~Zhang, ``$\{$Updates-Leak$\}$: Data set inference and reconstruction attacks in online learning,'' in \emph{29th USENIX security symposium (USENIX Security 20)}, 2020, pp. 1291--1308.

\bibitem{wang2019beyond}
Z.~Wang, M.~Song, Z.~Zhang, Y.~Song, Q.~Wang, and H.~Qi, ``Beyond inferring class representatives: User-level privacy leakage from federated learning,'' in \emph{IEEE INFOCOM 2019-IEEE conference on computer communications}.\hskip 1em plus 0.5em minus 0.4em\relax IEEE, 2019, pp. 2512--2520.

\bibitem{hansen2016cma}
N.~Hansen, ``The cma evolution strategy: A tutorial,'' \emph{arXiv preprint arXiv:1604.00772}, 2016.

\bibitem{eriksson2019scalable}
D.~Eriksson, M.~Pearce, J.~Gardner, R.~D. Turner, and M.~Poloczek, ``Scalable global optimization via local bayesian optimization,'' \emph{Advances in neural information processing systems}, vol.~32, 2019.

\bibitem{deng2012mnist}
L.~Deng, ``The mnist database of handwritten digit images for machine learning research [best of the web],'' \emph{IEEE signal processing magazine}, vol.~29, no.~6, pp. 141--142, 2012.

\bibitem{ILSVRC15}
O.~Russakovsky, J.~Deng, H.~Su, J.~Krause, S.~Satheesh, S.~Ma, Z.~Huang, A.~Karpathy, A.~Khosla, M.~Bernstein, A.~C. Berg, and L.~Fei-Fei, ``{ImageNet Large Scale Visual Recognition Challenge},'' \emph{International Journal of Computer Vision (IJCV)}, vol. 115, no.~3, pp. 211--252, 2015.

\bibitem{fletcher2000practical}
R.~Fletcher, \emph{Practical methods of optimization}.\hskip 1em plus 0.5em minus 0.4em\relax John Wiley \& Sons, 2000.

\bibitem{dauphin2017language}
Y.~N. Dauphin, A.~Fan, M.~Auli, and D.~Grangier, ``Language modeling with gated convolutional networks,'' in \emph{International conference on machine learning}.\hskip 1em plus 0.5em minus 0.4em\relax PMLR, 2017, pp. 933--941.

\end{thebibliography}

\end{document}